\documentclass[aip, amsmath,amssymb, reprint]{revtex4-1}

\usepackage{graphicx}
\usepackage{dcolumn}
\usepackage{bm}

\usepackage[utf8]{inputenc}
\usepackage[T1]{fontenc}
\usepackage{mathptmx}
\usepackage{multirow}
\usepackage{color}
\usepackage{float}
\usepackage[caption=false]{subfig}
\usepackage{caption}

\draft 

\begin{document}
	
	\title[Modelling the Fluctuating Torque]{A Statistical Analysis Towards Modelling the Fluctuating Torque on Particles in Particle-laden Turbulent Shear Flow }
	
	\author{S.Ghosh}
	\altaffiliation[Also at ]{Department of Chemical Engineering,
		\\Indian Institute of Technology, Bombay}
	\email{133020003@iitb.ac.in}
	\author{P.S.Goswami}%
	\affiliation{Department of Chemical Engineering,
		\\Indian Institute of Technology, Bombay
		\\Powai, Mumbai-400-076, India
	}%
	
	\date{\today}
	
	\begin{abstract}
		Dynamics of the particle phase in a particle laden turbulent flow is highly influenced by the fluctuating velocity and vorticity field  of the fluid phase. 	The present work mainly focuses on exploring the possibility of applying a Langevin type of random torque model to predict the rotational dynamics of the particle phase. Towards this objective, direct numerical simulations (DNS) have been carried out  for particle laden turbulent shear flow with Reynolds number, $Re_{\delta}=750$  in presence of sub-Kolmogorov sized inertial particles (Stokes number $>>1$).  The inter-particle and wall-particle interactions have also been considered to be elastic. From the particle equation of rotational motion, we arrive at the expression where the fluctuating angular acceleration $\alpha_i'$ of the particle is expressed as the ratio of a linear combination of fluctuating rotational velocities of particle ($\omega_i'$)  and fluid angular velocity ($\varOmega_i'$) to the particle rotational relaxation time $\tau_r$. The analysis was done using p.d.f plots  and \textbf{Jensen-Shannon Divergence} based method to assess the similarity  of the  particle net rotational acceleration distribution $f(\alpha_i')$,  with (i) the distributions of particle acceleration component arising from fluctuating fluid angular velocity computed in the particle-Largrangian frame $f((\varOmega'_i/\tau_r)_{pl})$, (ii) fluctuating particle angular velocity $f(\omega'_i/\tau_r)$, and (iii) the fluid angular velocity  $f((\varOmega'_i/\tau_r)_{e})$ computed in the fluid Eulerian grids.  The analysis leads to the conclusion that $f(\alpha_i')$ can be modeled with a Gaussian white noise using a pre-estimated strength which can be calculated from  the temporal correlation of $(\varOmega'_i/\tau_r)_e$.

	\end{abstract}
	
	
	\maketitle 
	
	\section{Introduction and Background}
	The intricacies of particle-laden turbulent flows arise due to the presence of multiple length and time-scales and also due to the coupling between the
	carrier phase and the dispersed-phase. Understanding and modelling the inter-phase coupling terms are important to accurately predict the two phase dynamics. It is reported in the earlier investigations that at low mass loading, the presence of particle phase does not alter the fluid phase turbulence but the dynamics of particles are strongly influenced by the turbulence fluctuation \cite{soo1960experimental, gore1989effect, elghobashi1993two,louge1991role}. Therefore, a detail understanding of fluctuating force and fluctuating torque exerted on the particles is of utmost need to model the dynamics of the particle phase. Several investigations have been reported to characterize the fluid phase fluctuation through the analysis of probability density function (pdf) of velocity fluctuations. Majority of those studies are for isotropic homogeneous turbulence \cite{andres2019statistics,bandi2009probability, rani2014stochastic, perrin2015relative, meyer2013rotational, pujara2018rotations, mathai2016translational}. Pdf study of the velocity and pressure field to characterize the coherent structure in turbulent channel flow has been done using Direct Numerical Simulations \citep*{lamballais1997probability}.
	
		Non-Gaussian nature of probability density function of longitudinal velocity differences was observed in Couette-Taylor turbulence through a theoritical approach of unifying popular Castaining and Beck-Cohen methods through conditional PDFs. Inverse of granular temperature and dissipation rate were the intensive variables for the two methods respectively.  \citet*{bandi2009probability} investigated power fluctuations in Homogeneous Isotropic Turbulence through joint PDF of velocity and gaussian forcing term. In 3D freely decaying homogeneous turbulence, PDF of velocity and vorticity field increments were computed using DNS \citep*{andres2019statistics}. The incremental velocity PDF was observed to develop exponential and stretched exponential tails with the increase in two-point distance. \\In Particle-laden turbulent flow systems, various works have been done involving computations of particle phase PDF statistics as well.  Joint distribution function of relative velocity and relative separation of inertial particles is studied in the context of particle caustics and clustering \cite{gustavsson2011distribution, gustavsson2014relative, bhatnagar2018relative}. Power-law nature of the joint PDF was derived in various limits. The joint PDF for identical inertial particles in homogeneous isotropic turbulence was found to be independent on the Taylor-based Reynolds number \citep*{perrin2015relative}. \citet*{rani2014stochastic} proposed a closure form of diffusion tensor of heavy inertial particles in isotropic turbulence. From the evolution of the diffusion tensor through Langevin equation, the relative velocity PDF was observed to change its nature from gaussian, at separations of the order of the integral length, to exponential at smaller separations. \citet*{goswami2010particle1}, through DNS of particle laden turbulent channel-flow with low Reynolds’
	Number and high Stokes number, found out that in presence of one-way coupling the
	pdf of acceleration distribution of the particles can be reconstructed using fluid phase velocity
	distribution. Furthermore, it was concluded that the effect of turbulent velocity fluctuation on the particle could be modelled as Gaussian Random White
	Noise.
	\\In the context of fluid-particle one-way coupling it is worth noting that the turbulent
	fluid phase contributes to the rotation or spin of the rough particles through torque coupling \citep*{andersson2012torque}. Rotational statistics of inertial particles in turbulent flows are investigated in the context of non-spherical shaped particles as well. Stereoscopic PIV experiments of spherical and ellipsoidal particles in isotropic turbulence showed that rotational auto-correlation functions decayed exponentially. This decay carried signature of Ornstein-Uhlenbeck process of angular velocity fluctuations of the particles being controlled by the large scales of turbulence \citep*{meyer2013rotational}. \citet*{pujara2018rotations}, performed experiments of freely moving particles of various shapes in isotropic turbulence. It was observed that angular velocity PDF was log-normally distributed and dependent upon only the volume of the particles and not on their shapes. Experiments of large buoyant spheres in isotropic turbulence showed angular acceleration distribution with wide tails along all the directions \citep*{mathai2016translational}.
	\citet*{marchioli2013rotation} studied rotational statistics of fibers in turbulent channel-flow thorugh DNS. The fiber angular velocities, at the center of the channel, were observed to be Gaussian in nature with expoenetially decaying angular velocity auto-correlation function.   
	\\This article encompasses a detailed analysis of fluctuating particle statistics e.g. translational and rotational acceleration and velocity statistics, in the form of probability distribution functions, correlations and moments. All of these statistics are obtained for the particles with high Stokes number. The pdfs of velocity and acceleration are computed at different wall-normal locations of the channel. A Jensen-Shannon divergence based similarity analysis of the pdfs have also been performed to check whether the particle angular acceleration fluctuation can be represented as the ratio of fluid vorticity fluctuation to the angular relaxation time. Autocorrelation coefficient for the particle angular velocity, fluid angular velocity and particle angular acceleration have also been reported. Such a detail analysis helps to explore the possibility to develop a model in which torque exerted on the particle due to fluid vorticity field can be described by a Langevin type random force model. This article, in the subsequent sections, contains the simulation methodology and governing equations (Section-II), the detailed analysis of fluctuating velocity and acceleration statistics of the particle phase for both translational and rotational motion (Section-III), the Jensen-Shannon Divergence test based modelling of fluctuating torque on the particle-phase (section-IV) which is followed by conclusions in the last section (V). 

	\section{Governing Equations and Simulation Methodology}
\begin{figure}
		\centerline{\includegraphics[scale=0.25]{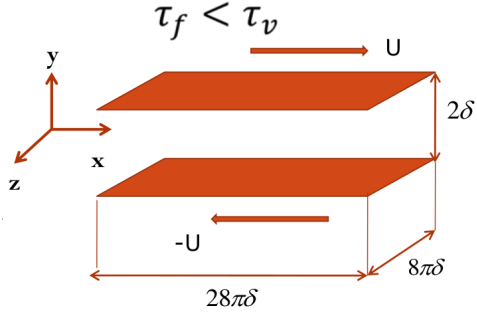}}
		\caption{Schematic of the flow geometry}
		\label{fig:schematic}
	\end{figure}

	An Eulerian-Lagrangian method has been used for the simulation. Since the primary 
	objective of the  present work 	is to explore the possibility and demonstrate a random force based modelling of effect of turbulent fluctuation on the particle dynamics, one way coupled direct numerical simulation is used. Fluid phase is described by continuity  and the Navier Stokes equations (equations \ref{eq_cont} \& \ref{eq_NS}). Both the equations are solved on Eulerian grids.	
	
	\begin{equation}
	\nabla\cdot\bf{u}=0
	\label{eq_cont}
	\end{equation}
	\begin{equation}
	\frac{\partial\bf{u}}{\partial t}+\bf{u}\cdot\nabla \bf{u}=\frac{1}{\rho_{f}}\nabla p+ \nu \nabla^2 \bf{u}
	\label{eq_NS}
	\end{equation}
	Here, $\bf{u}(\bf{x},t)$ represents three dimensional instantaneous fluid velocity field as function of position $\bf{x}$. $p(\bf{x},t)$ denotes the  instantaneous pressure field, $\nu$ and $\rho_f$ are the kinematic viscosity and the density of the fluid respectively. In our isothermal simulations fluid is considered to be incompressible. 
	The flow-field is considered to be periodic along x and z direction, whereas along y-direction is bounded by walls and no-slip boundary condition is applied figure \ref{fig:schematic}. The fluid flow field is  solved using Direct Numerical Simulation (DNS) in a  pseudo-spectral framework.  
    Detail of the numerical scheme, interpolation method for fluid velocity at the particle location, and correction of the velocity field for the calculation of the drag on the particle has been described in detail by our earlier publications \citet{goswami2010particle1} and \citet{muramulla2020disruption}.
	

	The particle-phase is described using Newton's equation of motion with Lagrangian particle tracking algorithm. Acceleration of the particle (p) between two successive collisions can be represented as
	\begin{equation}
	\bf{a}_p=\frac{d\bf{v}_p}{dt}=\frac{\bf{u}(\bf{x}_p)-\bf{v}_p}{\tau_v}.
	\end{equation}
	Similarly, rotational acceleration between two successive collisions, acting on the same particle 
	is described as 
	\begin{equation}
	\bf{\alpha}_p=\frac{d\bf{\omega}_p}{dt}=\frac{\bf{\varOmega}(\bf{x}_p)-\bf{\omega}_p}{\tau_r}.
	\end{equation}
	In the above equations $\bf{a}_p$ and $\bf{\alpha}_p$  denote the instantaneous translational and rotational acceleration of the 'p'th particle with translational velocity $\bf{v}_p$ and rotational velocity $\bf{\omega}_p$. $\bf{u}(\bf{x}_p)$ and $\bf{\varOmega}(\bf{x}_p)$ denote the fluid phase linear and angular velocity interpolated at the particle position $\bf{x}_p$. $\tau_v$ is the viscous relaxation time for particle linear velocity and it is given by 
	\begin{math}
	\tau_v=\frac{\rho_p d_p^2}{18 \mu}
	\end{math}. For spherical particles relaxation time for angular velocity ($\tau_r$) is given by $\tau_r=0.3\tau_v$.
	
	The inter-particle interaction is captured through standard molecular dynamics simulation algorithm \citep{goswami2010particle, goswami2010particle1}. However, a generalized collision rule following the formulation  by \citet{lun1991kinetic} is considered here to capture the effect of inelasticity and effect of roughness using coefficient restitution ($e$) and roughness factor ($\beta$) respectively. 
	\label{Method}
	Here, the generalised collision rule is presented briefly. Let us  consider  $m_1$, $\bf{v_1}$, $\bf{\omega_1}$, $I_1,\sigma_1$  be the  mass, linear velocity, angular velocity, moment of inertia and diameter of the particle 1. The same symbols with a subscript 2 denote the same quantities for the second particle of the colliding pair. Vector $\bf{k}$ represents the unit vector along the line joining centres from particle 1 to 2. The total relative velocity between two particles at the point of contact before and after the collision  is expressed by $\bf{g_{12}}$ and  $\bf{g_{12}^{'}}$  respectively in eq.\ref{eq2.14}. 
	\begin{equation}
	\label{eq2.14}
	\bf{g_{12}}=\bf{v_{12}}-\frac{1}{2}(\bf{k}\times\bf{\xi}), \quad
	\bf{g_{12}}^{'}=\bf{v_{12}}^{'}-\frac{1}{2}(\bf{k}\times\bf{\xi}^{'}).
	\end{equation}
		
	Where 
	\begin{equation}
	\bf{v_{12}}=\bf{v_1}-\bf{v_2},\quad
	\bf{v_{12}}^{'}=\bf{v_1}^{'}-\bf{v_2}^{'}
	\end{equation}
	and 
	\begin{equation}
	\bf{\xi}=\sigma_1\bf{\omega_1}+\sigma_2\bf{\omega_2},\quad 
	\bf{\xi}^{'}=\sigma_1\bf{\omega_1}^{'}+\sigma_2\bf{\omega_2}^{'}.
	\end{equation}
		For inelastic collision the coefficient of restitution $e$ is defined as,
		\begin{equation}
		\label{eq2.16}
		\bf{k}.\bf{g_{12}}^{'}=-e(\bf{k}.\bf{g_{12}})
		\end{equation}	
		For collisions between rough-particles a roughness parameter $\beta$ is introduced and it is defined as,
		\begin{equation}
		\label{eq2.17}
		\bf{k}\times\bf{g_{12}}^{'}=-\beta(\bf{k}\times\bf{g_{12}})
		\end{equation}
		Roughness co-efficient $\beta$ can have a values between $-1$ and $+1$.  $\beta=-1$ refers perfectly smooth collision while for $\beta=1$ indicates that the collisions are perfectly rough.
		In the range of $0<\beta\leq1 $ spin reversal of the colliding particles takes place.
		To estimate the change in linear velocity and the angular velocity of the particle let us define some  parameters $K_1,\eta_1,\eta_2$ as follows.
		\begin{equation}
		\label{eq2.22}
		K_1=\frac{2I_e}{m_{12}\sigma_e^2}
		\end{equation}
		\begin{equation}
		\label{eq2.23}
		\eta_1=1+e
		\end{equation}
		\begin{equation}
		\label{eq2.24}
		\eta_2=(1+\beta)\frac{K_1}{1+K_1}
		\end{equation}
		$K_1$ is non-dimensional moment of inertia. For a sphere having all the mass concentrated at the centre it has a value 0. For a uniform solid sphere it is $2/5$.
		Here,
		\begin{equation}
		\label{eq2.25}
		\frac{\sigma_e^2}{I_e}=\frac{\sigma_1^2}{2I_1}+\frac{\sigma_2^2}{2I_2}
		\end{equation} and 
		\begin{equation}
		\label{eq2.26}
		m_{12}=\frac{m_1m_2}{m_1+m_2}
		\end{equation}
		The change in linear and angular velocities are given by eq. \ref{eq2.27} \& eq. \ref{eq2.28}
		\begin{widetext}
			\begin{equation}
			\label{eq2.27}
			m_1(\bf{v_1}^{'}-\bf{v_1})=m_2(\bf{v_2}-\bf{v_2}^{'})\\=m_{12}\left[-\eta_2\bf{v_{12}}+\frac{1}{2}\eta_2(\bf{k}\times\bf{\xi})-(\eta_1-\eta_2)(\bf{k}.\bf{v_{12}})\bf{k}\right]
			\end{equation}
			\begin{equation}
			\label{eq2.28}
			\frac{I_1}{\sigma_1}(\bf{\omega_1}^{'}-\bf{\omega_1})=\frac{I_2}{\sigma_2}(\bf{\omega_2}^{'}-\bf{\omega_2})=-m_{12}\eta_2[k\times\bf{v_{12}}-\\\frac{1}{2}(\bf{k}.\bf{\xi})\bf{k}+\frac{1}{2}\bf{\xi}]
			\end{equation}
		\end{widetext}
		
		In the present simulations, one-way coupled Direct Numerical Simulation  is used because the particle volume loading ($~10^{-4}$) is such that there is no significant change in turbulence intensity due to the feed back force exerted by the particles \cite{gore1989effect, elghobashi1993two}. All the simulations are performed using 44800 spherical particles of size 39 $\mu$m which is less than the Komogorov scale \cite{goswami2010particle1}. The investigation is conducted in the limit of high Stokes number using particle material density of $2000kg/m^3$. 
		Particle Stokes number based on integral fluid time scale is 
		1.7. Air at ambient condition is used as continuum fluid.\\
		 The dimension of the simulation box is $28\pi\delta$ $\times$ $2\delta$ $\times$ $8\pi\delta$  where $\delta$ is the half-channel-width. Upper and the lower walls of  the Couette move  with positive and negative velocities respectively in the x-direction. The Reynolds' Number based on half channel-width and wall-velocity is 750. The y-axis is along the cross-stream direction and z-axis indicates the span-wise or vorticity  direction as shown figure \ref{fig:schematic}. The number of grids used in x, y, and z directions are 340, 55, and 170 respectively. The grid resolution is 13.64 $\times$ 1.92 $\times$ 7.79 in viscous length scale.  A detail description of the simulation method for smooth particle in turbulent shear flow is presented by \citet{goswami2010particle1}. Here the statistics are presented for perfectly elastic collisions. Two limiting cases of roughness have been considered; for smooth collisions ($\beta=-1.0$) and perfectly  rough collisions ($\beta=+1.0$). 
		\section{Velocity and acceleration statistics for  particle phase}
		The statistical analysis of the particle-phase translational and rotational acceleration and velocity fluctuations are presented here in the presence and absence of particle roughness. A detailed analysis of various factors governing the particle rotational acceleration distribution along with the rotational acceleration correlation and velocity auto-correlation functions is presented here. Jensen-Shannon Divergence test is applied to put the light on modelling the  particle rotational acceleration distributions. 
		
		\subsection{Particle statistics for smooth elastic collisions }
		\subsubsection{Particle Acceleration Distribution}
		Particle acceleration distributions reflect the fluctuating statistics of the net force acting on the particles. Fluctuating acceleration originates from the fluid velocity fluctuation and also due to the particle velocity fluctuations over the local particle mean velocity as described in equation figure \ref{eq_acc_fluc}
			\begin{equation}
		\bf{a'}_p=\frac{\bf{u}(\bf{x}_p)-\overline{\bf{\bf{u}(\bf{x}_p)}}}{\tau_v}-\frac{\bf{v_p}-\bf{\overline{v}}_p}{\tau_v}.
		\label{eq_acc_fluc}
		\end{equation}
		
\begin{figure}[!]
			\centering
			\includegraphics[width=0.5\textwidth]{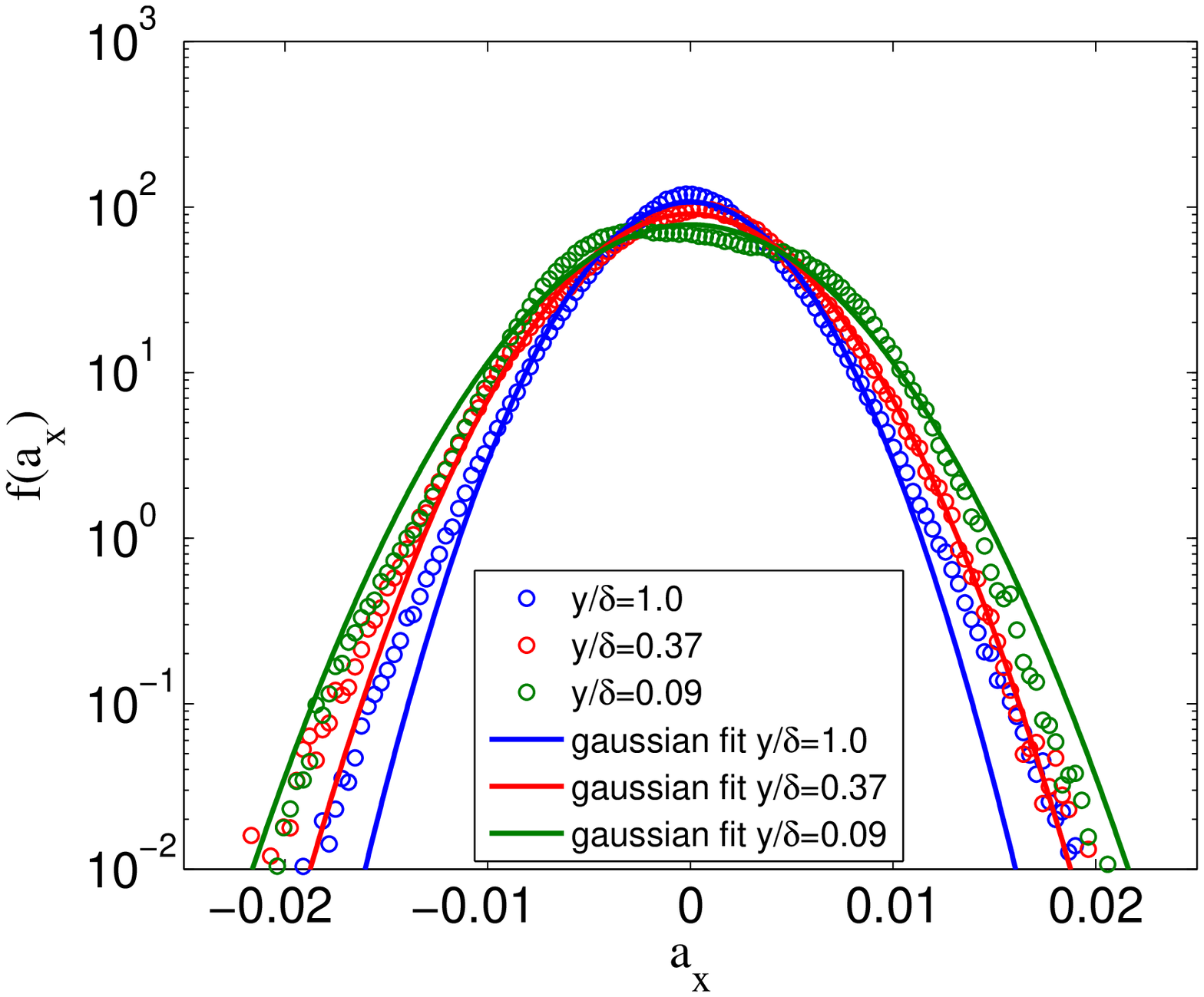}
			\caption*{(a)}
			\includegraphics[width=1.0\linewidth, height=5.5cm]{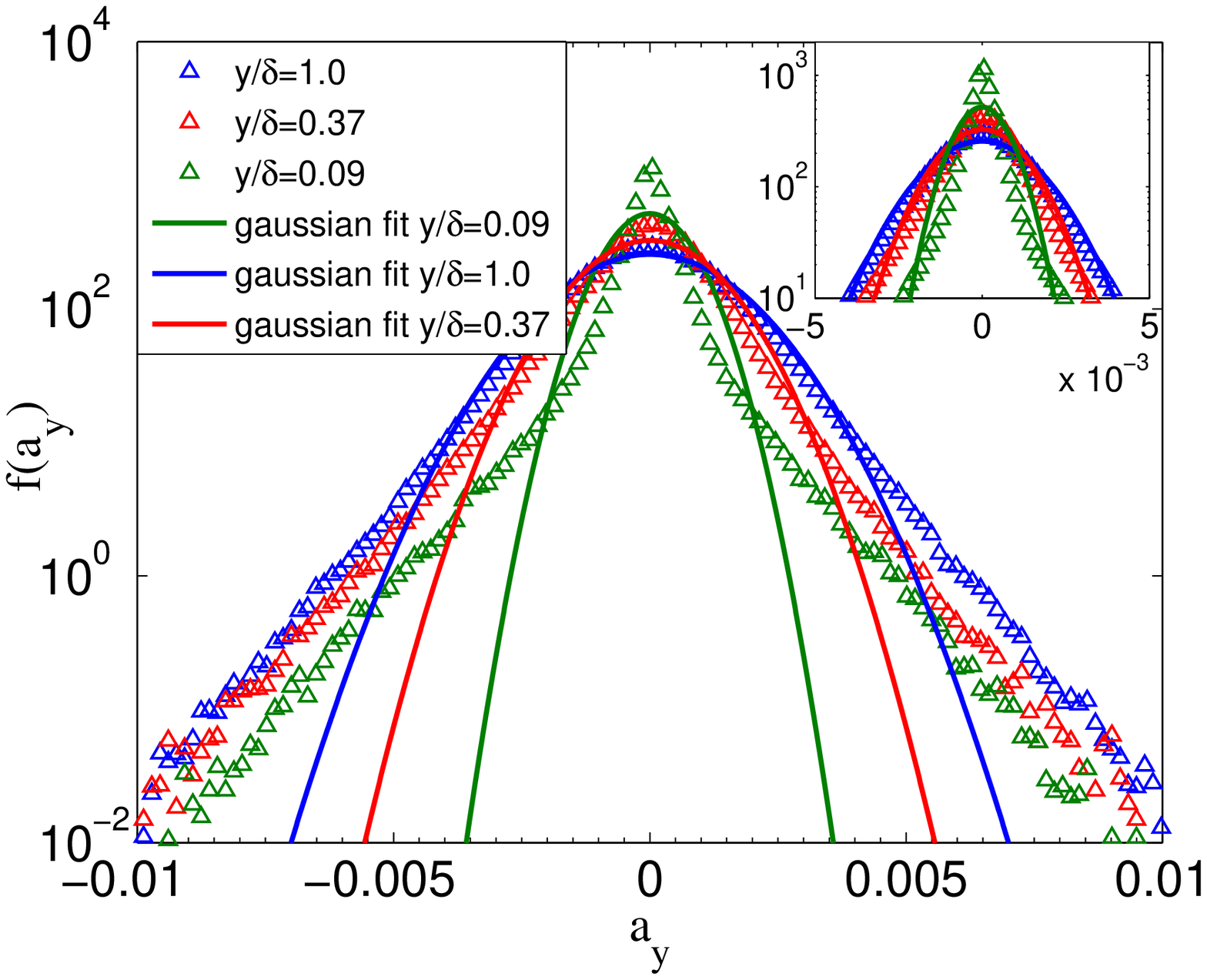}
			\caption*{(b)}
			\includegraphics[width=1.0\linewidth, height= 5.5cm]{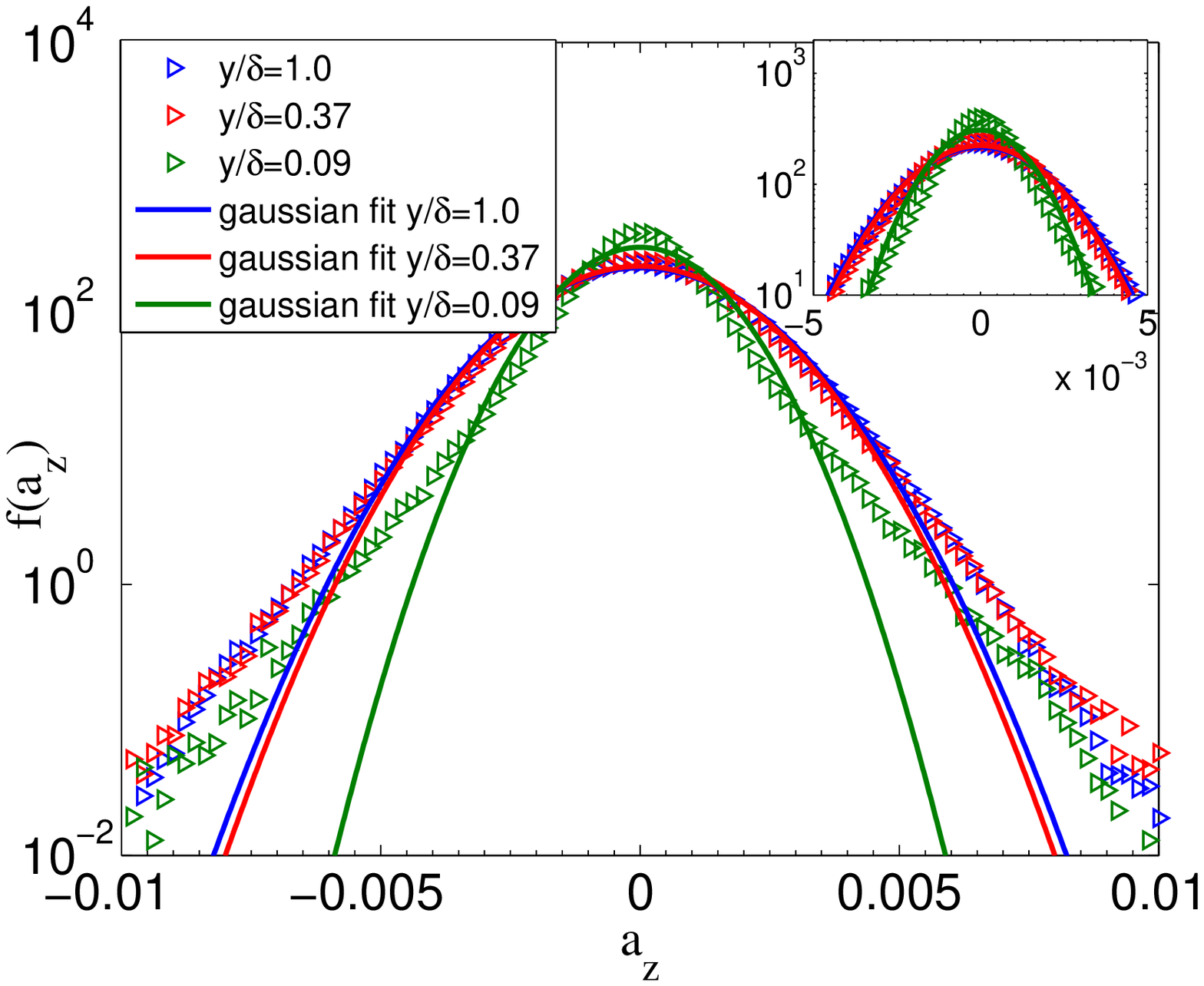}
			\caption*{(c)}
			\caption{Components of particle acceleration distributions (a) $f(a_x)$, (b) $f(a_y)$, and (c) $f(a_z)$ at different wall normal position ($y/\delta$) of the channel.}
			\label{f(a_i)}     
		\end{figure}

\begin{figure*}[!]
			\centering
			\includegraphics[width=1.0\textwidth]{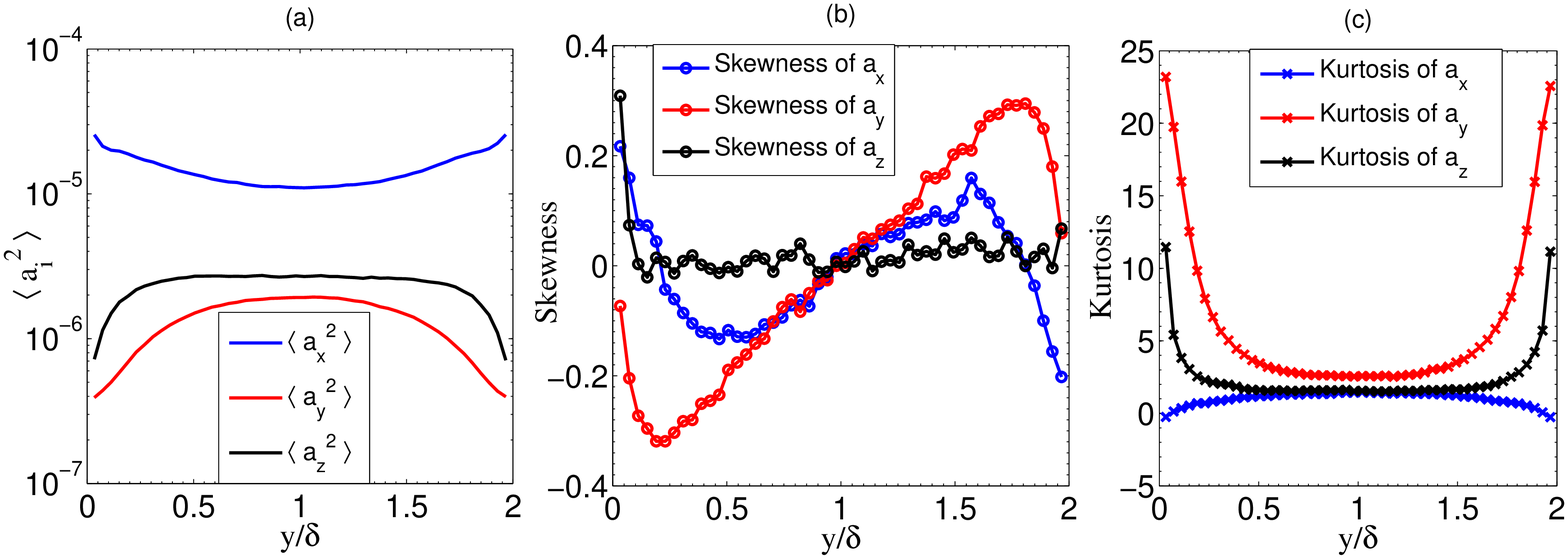}
			\caption{Moments of Acceleration Distributions; (a) second moments of acceleration fluctuations, (b) skewness, and (c) kurtosis}
			\label{acc_moments}
		\end{figure*}

		In this section,  statistics  of overall  acceleration $f(a_i)$ in three different directions at three different wall normal locations ($y/\delta$=0.09. 0.37, and 1.0) is discussed. Hereafter we have omitted the $'$ symbol in the superscript of the 
		fluctuating quantities. Distribution functions for the streamwise, wall normal, and spanwise acceleration fluctuation  are  shown in figure \ref{f(a_i)} (a), (b) and (c) respectively. The variation of second, third, and fourth moments  which are the  mean square, Skewness, and Kurtosis along the wall normal locations are shown in figure \ref{acc_moments}. All the distribution functions have been compared 
		with corresponding Gaussian distribution with same variance and zero mean. The  traits of acceleration fluctuations  are discussed here.	Stream wise acceleration distribution function ($f(a_x)$) can be approximated well with Gaussian distribution. Although, it deviates from the corresponding Gaussian fit at the tail of the distribution function, which is almost two orders of magnitude lower than the peak distribution (figure \ref{f(a_i)} (a)). 
		Distribution function for wall normal ($f(a_y)$) and spanwise accelerations ($f(a_z)$) are shown in  figures \ref{f(a_i)} (b) and (c). In both the cases distributions  significantly deviate from their respective Gaussian distributions with sharper peaks and elongated  exponential tails at higher values of acceleration fluctuations.  
		$f(a_y)$ deviates significantly from the Gaussian distribution at "near wall" region.
		Spanwise acceleration distribution also shows a similar trend of long exponential trend at the higher values of acceleration fluctuations. At the center of the Couette and intermediate locations, both the distributions $f(a_y)$ and $f(a_z)$ follow the 
		Gaussian distributions more than one decade of peak distributions as shown in the 
		insets of the figures. This is due to the fact that the second moment of the fluid velocity and particle velocity fluctuations are homogeneous at those zones. 
		
		The variance of $f(a_x)$ is higher by one order of magnitude than $f(a_y)$ and $f(a_z)$ as shown in figure \ref{acc_moments} (a). Streamwise mean square fluctuation is marginally higher near the wall.  But for wall normal and spanwise fluctuations 
		the variance is two times higher at the center as shown in figure \ref{acc_moments} (a). Figure \ref{acc_moments} (b) shows that the variation of skewness is anti-symmetric with respect to the channel-centre. $f(a_y)$ is the least locally symmetric and  $f(a_z)$ is the most locally symmetric in nature although the  values indicate that the acceleration distributions are very mildly skewed except at the center zone of the Couette.
		For a fluctuating variable $\phi$ the Skewness and excess Kurtosis is defined as:
		\begin{equation}
		Skewness=\frac{\langle{\phi^3}\rangle}{\langle{\phi^2}\rangle^{3/2}}    
		\end{equation}
		\begin{equation}
		 Kurtosis=\frac{\langle{\phi^4}\rangle}{\langle{\phi^2}\rangle^2}-3  
		\end{equation}
		The excess Kurtosis values for all the components of acceleration 
		distribution functions at different wall normal locations are shown  in figure \ref{acc_moments} (c). It is observed that at the center of the Couette, tail of the distribution for all the components  approximately follow Gaussian distribution. Distribution functions near the wall are far from the Gaussian for $f(a_y)$ and $f(a_z)$ and posses  extreme outliers in the form of elongated tails as also shown in  figures \ref{f(a_i)} (b) and (c).

		\subsubsection{Particle Velocity Distribution}
		
\begin{figure}[!]
			\centering
			\includegraphics[width=0.45\textwidth ]{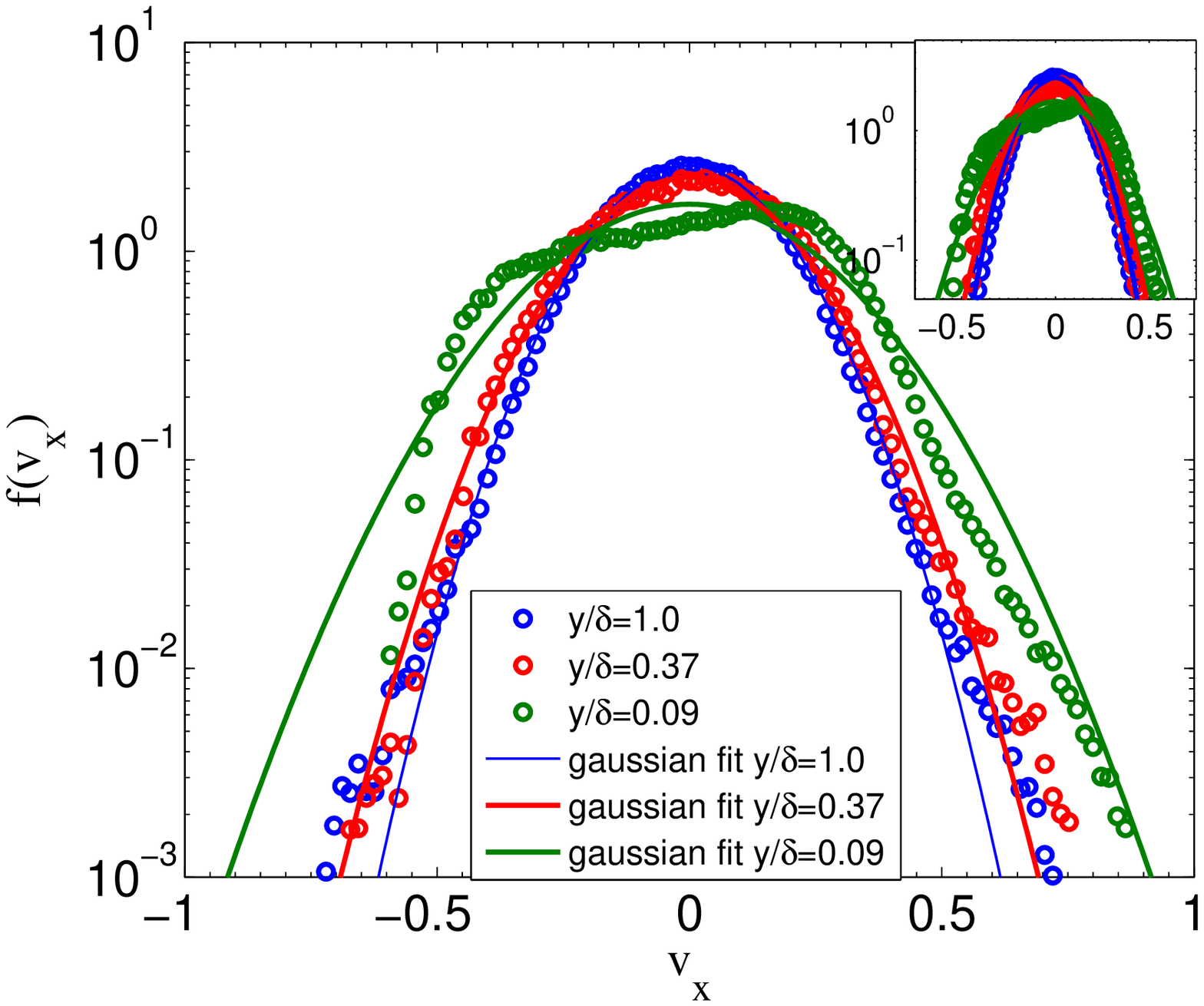}
			\caption*{(a)}
			\includegraphics[width=0.45\textwidth]{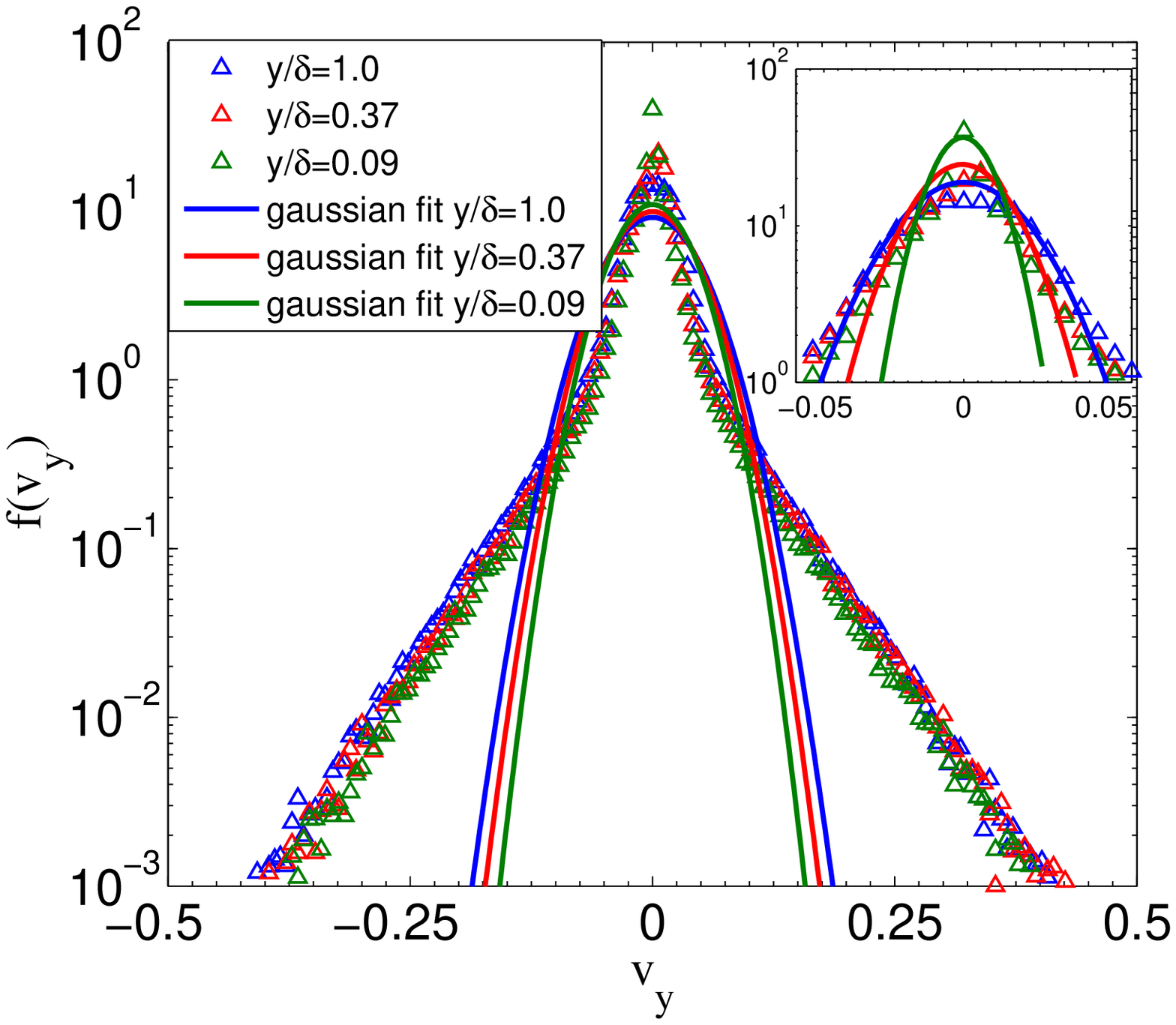}
			\caption*{(b)}
			\includegraphics[width=0.45\textwidth]{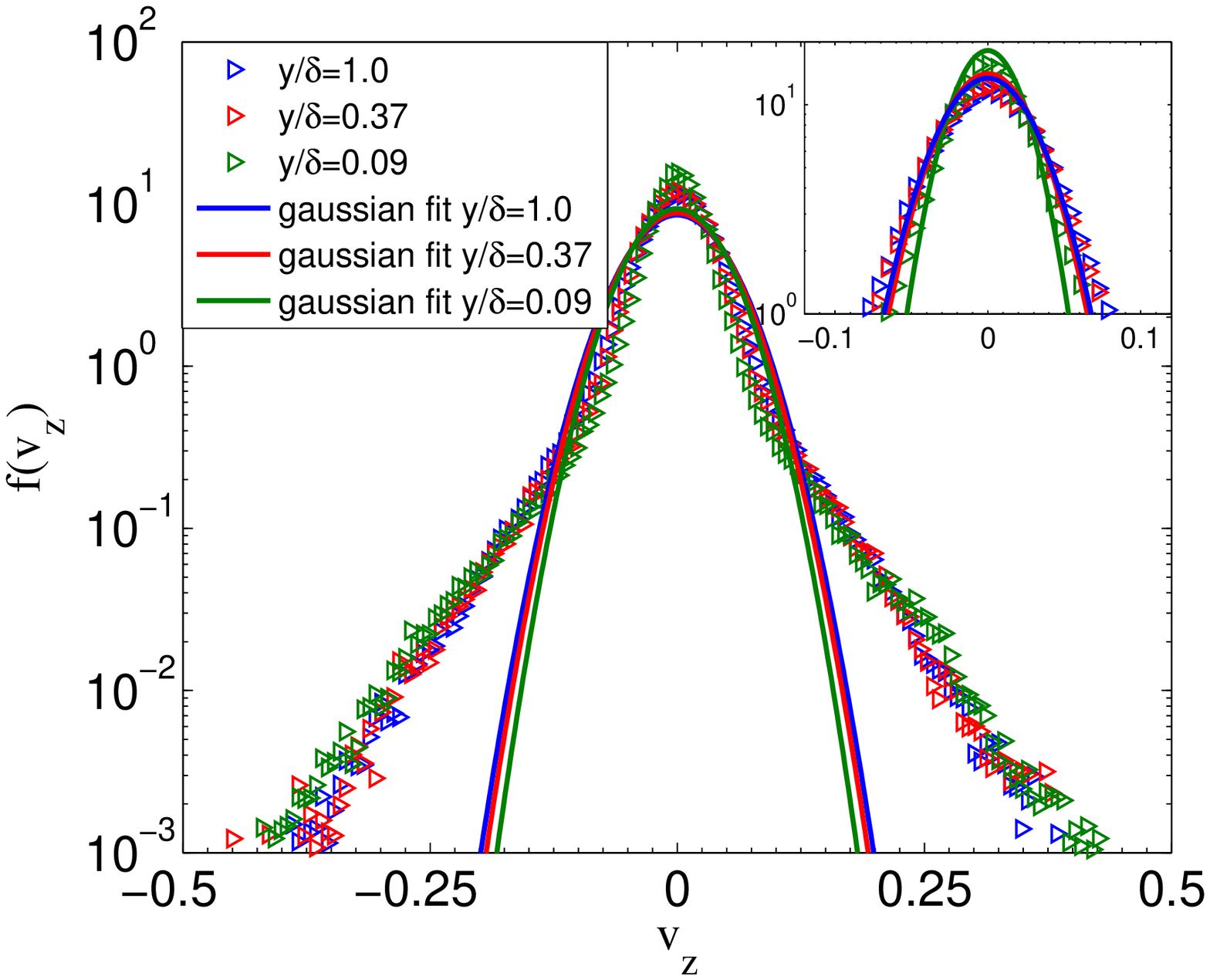}
			\caption*{(c)}
				\caption{Components of particle velocity distributions (a) $f(v_x)$, (b) $f(v_y)$, and (c) $f(v_z)$ at different wall normal position ($y/\delta$) of the channel.}
						 \label{f(v_i)}     
		\end{figure}

\begin{figure*}[!]
			\centering
			\includegraphics[width=1.0\textwidth]{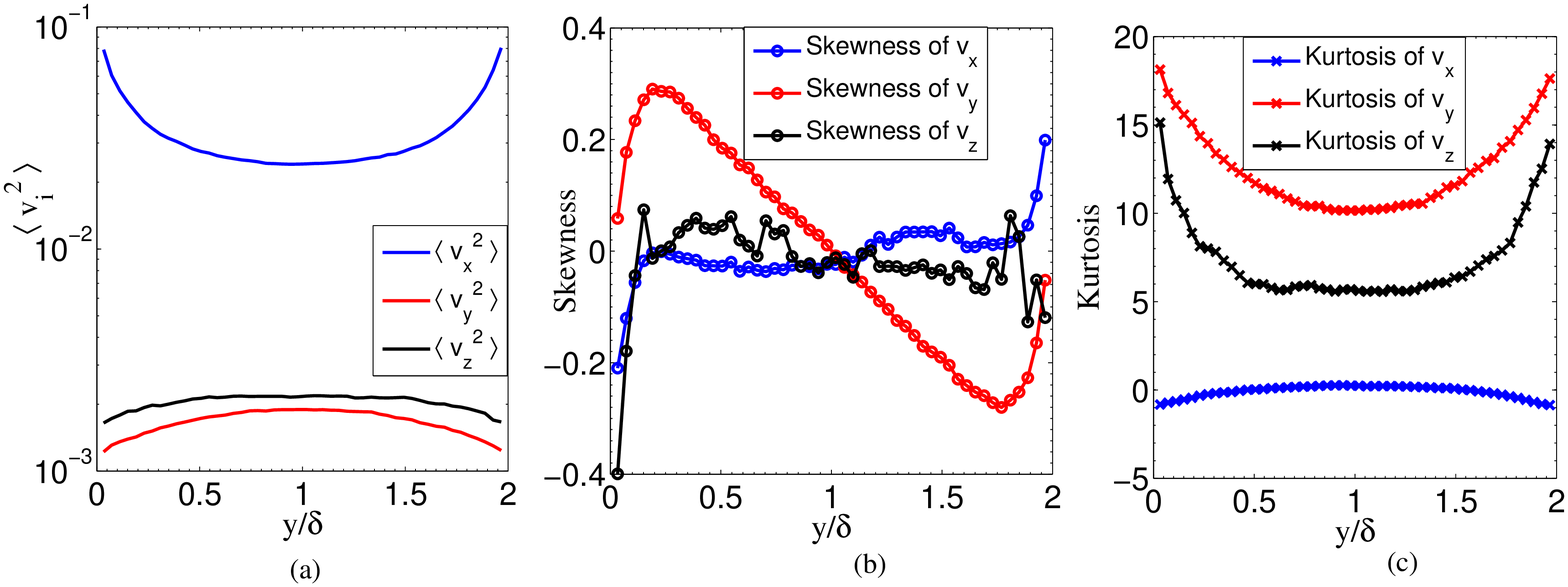}
			\caption{\label{Velocity_moments} Moments of particle velocity distributions; (a) second moments of velocity fluctuations, (b) skewness, and (c) kurtosis.}
			
		\end{figure*}

		In this section, distribution function and the moments of the 
		linear velocity of the particles at different wall normal locations are reported. The distribution functions are shown in figures \ref{f(v_i)} (a) to (c) and the variation of second, third and fourth moments in the form of Mean Square, Skewness and Kurtosis along the channel-width are shown in figure \ref{Velocity_moments}.

	Streamwise particle velocity fluctuation $f(v_x)$ can be well approximated by the Gaussian distribution except near the wall as shown in figure \ref{f(v_i)} (a). The Gaussian fits have the zero mean and same 
	variance as that of the corresponding particle velocity distribution. 
	In the near wall region $f(v_x)$ deviates from the Gaussian distribution. At low negative velocity fluctuation the decay is slower but at higher velocity fluctuation it deacys faster that the Gaussian distribution,  even though the peak distribution appears at the positive velocity fluctuation.  
	Distribution of wall normal ($f(v_y)$) and spanwise velocity field ($f(v_z)$) significantly deviates from the Gaussian distribution as shown in figure \ref{f(v_i)} (b) and (c).  Although  there are  longer exponential tails, the distribution functions can be approximated as Gaussian upto one decade of the peak frequency. Moments of the distribution functions are shown in figure \ref{Velocity_moments}. 
	
	 The variance of $f(v_x)$ is more than one order of magnitude higher than $f(v_y)$ and $f(v_z)$ as shown in figure \ref{Velocity_moments} (a). Since local mean flow is in x direction, cross stream migration of the particles contributes to the stream wise velocity fluctuations. The variance of $f(v_x)$ is higher near the walls and lower at the center of the Couette. An opposite trend is observed for $f(v_y)$ and $f(v_z)$ as shown in figure \ref{Velocity_moments} (a).
	Figure \ref{Velocity_moments} (b) shows that the variation of skewness is anti-symmetric with respect to the channel-centre. It is  to be noted that the outward wall normal vector changes sign for both the walls. Although the magnitude shows the signature of mild asymmetry in the distribution function of wall normal velocity component,  $f(v_y)$ is the least locally symmetric but $f(v_x)$ and $f(v_z)$ is symmetric across the channel except very near the wall. 
	 Figure \ref{acc_moments} (b) and \ref{Velocity_moments} (b) show that the skewness values are of opposite signs along the channel denoting that the positive tailed particle acceleration fluctuations corresponds to negative tailed particle velocity fluctuations and vice versa.
	 The excess Kurtosis values in figure \ref{Velocity_moments} (c) shows that local $f(v_x)$ can be approximated to Gaussian distribution but local $f(v_y)$ and $f(v_z)$ are far from the Gaussian with the existence of  'extreme outliers' in the form of larger tails which are observed in figures \ref{f(v_i)} (a) to (c) as well.

		\subsubsection{Particle Rotational Acceleration Distribution}
\begin{figure}[!]
			\includegraphics[width=0.45\textwidth]{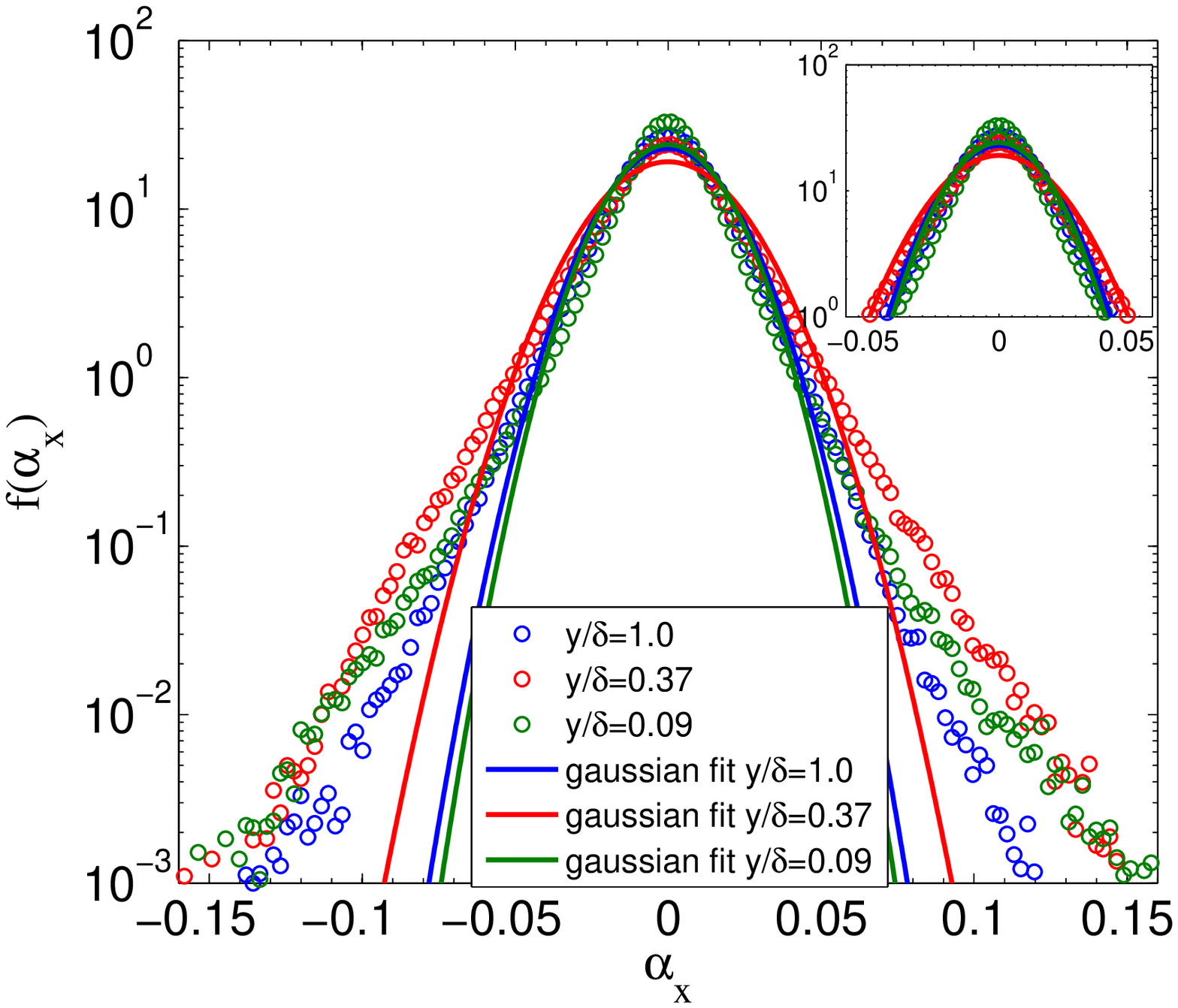}
			\caption*{(a)}
			\includegraphics[width=0.45\textwidth]{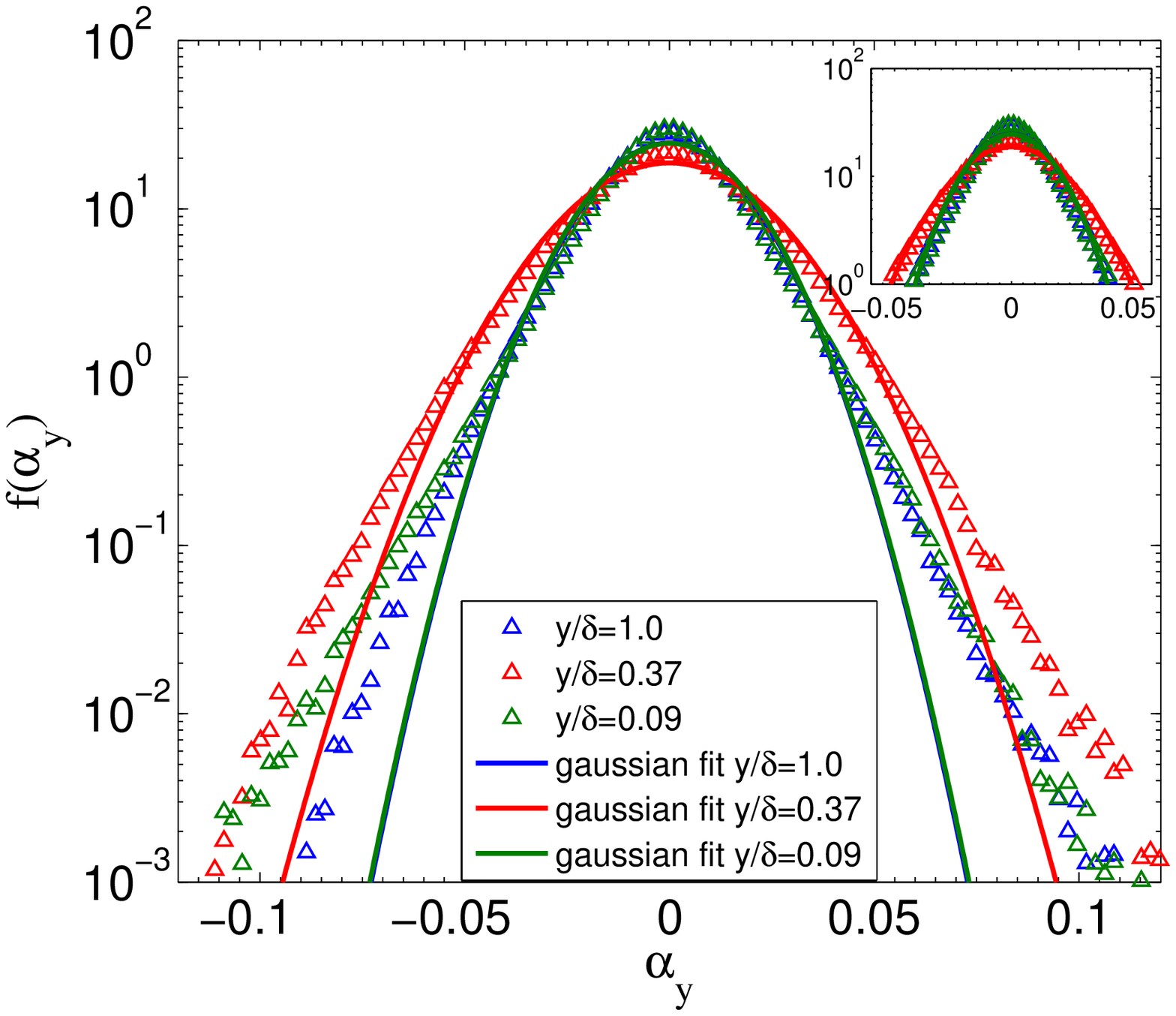}
			\caption*{(b)}
			\centering
			\includegraphics[width=0.45\textwidth]{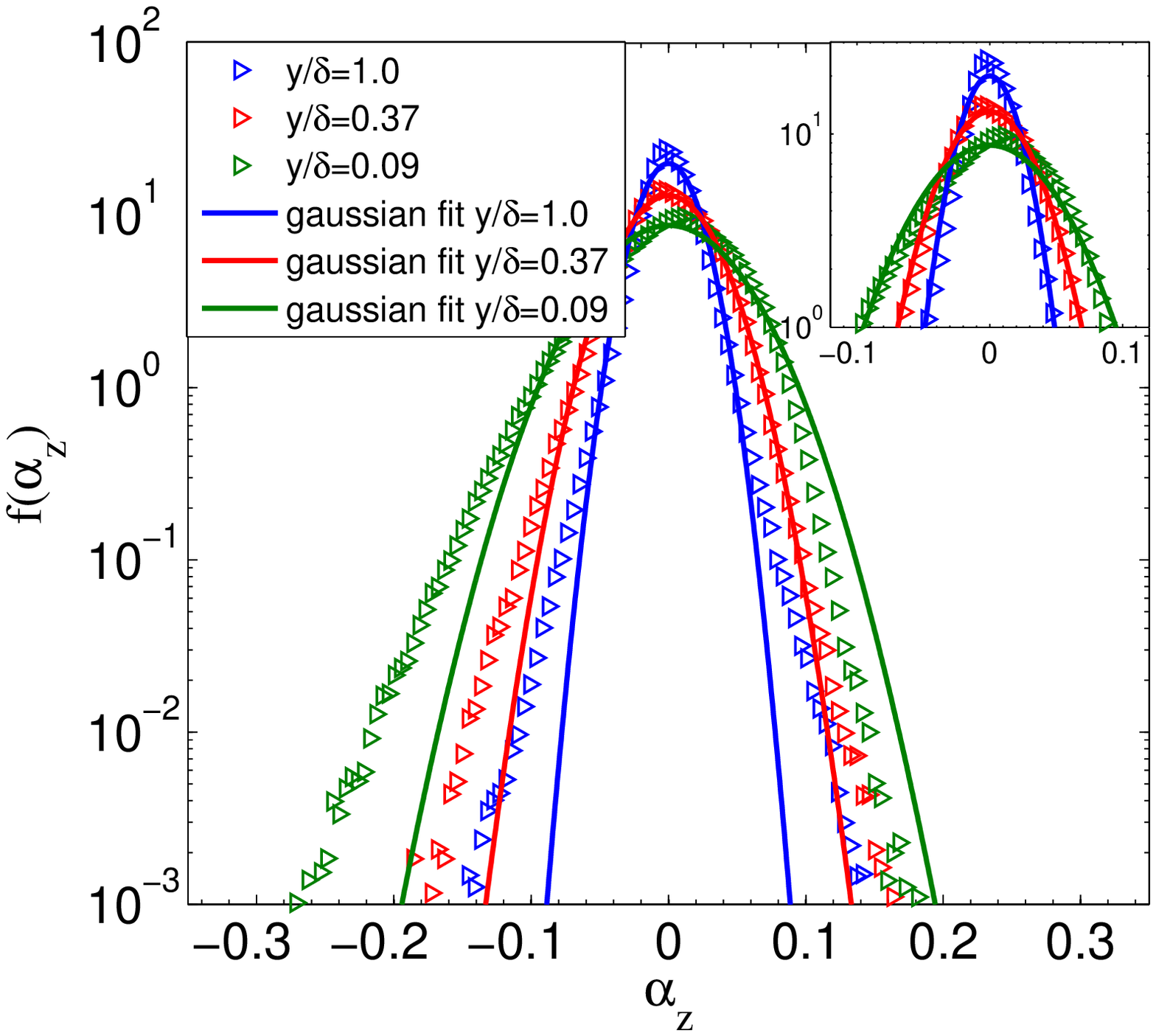}
			\caption*{(c)}
				\caption{Components of particle rotational acceleration distributions (a) $f(\alpha_x)$, (b) $f(\alpha_y)$, and (c) $f(\alpha_z)$ at different wall normal positions ($y/\delta$) of the channel.}
						\label{f(alpha_i)}     
		\end{figure}

\begin{figure*}[!]
			\centering
			\includegraphics[width=1.0\textwidth]{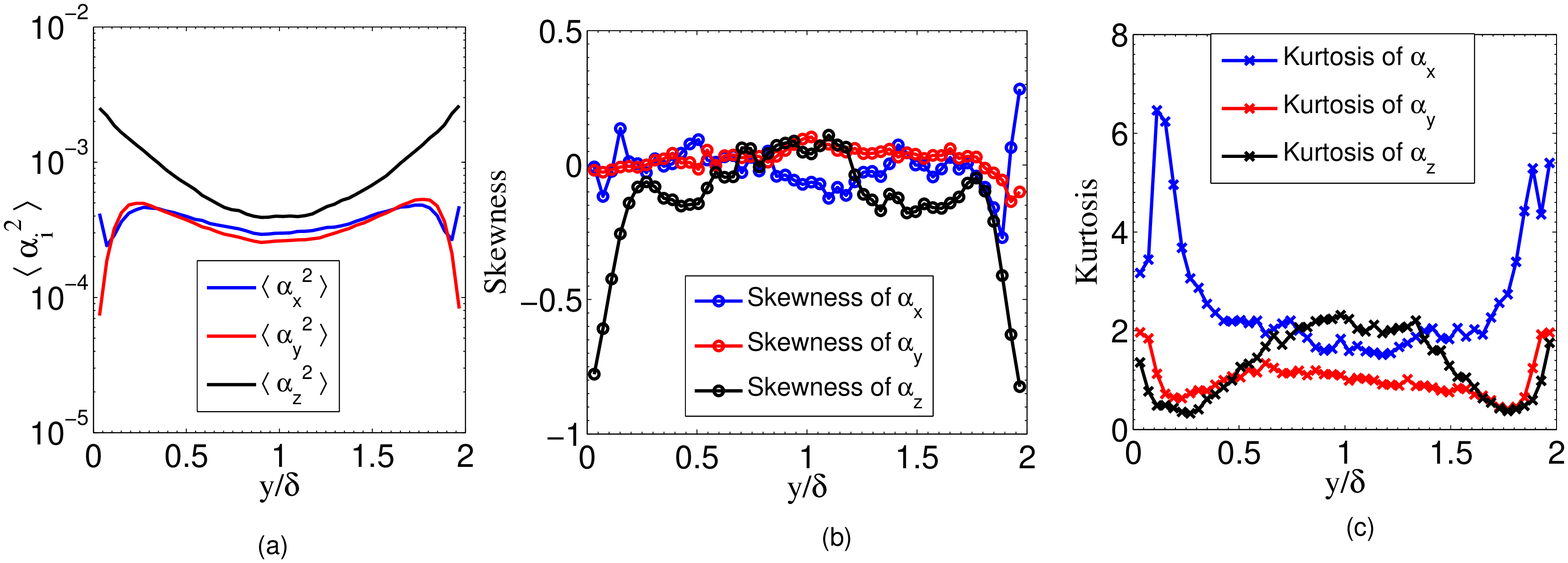}
			\caption{Moments of the rotational acceleration distributions; (a) second moments, (b) skewness, and (c) kurtosis.}
			\label{rot_acc_moments}
		\end{figure*}

		To model the dynamics of particles with surface roughness it is worth to 
		analyze the angular acceleration and angular velocities of the particles. 
		Here, we report the nature of rotational acceleration distribution $f(\alpha_i)$ of the particles in three  directions at different wall normal 
		locations as shown in figures \ref{f(alpha_i)} (a) to (c). Particle rotational acceleration statistics reflect the fluctuating statistics of the net torque acting on the particles. The variation of second, third and fourth moments in the form of Mean Square, Skewness and Kurtosis along the channel-width are shown in figure \ref{rot_acc_moments}. The following traits of the rotational acceleration fluctuation statistics are observed.
		The principal component of the angular acceleration is $\alpha_z$.  $f(\alpha_x)$ and $f(\alpha_y)$ deviate from their respective Gaussian fits only near the tail region with longer tails approximately exponential in nature (figures \ref{f(alpha_i)} (a) to (c). The tails deviate almost  below two decades of the peak distribution. 
		For both the distribution functions, an early deviation from the Gaussian distribution happens at near wall position. $f(\alpha_z)$ follows Gaussian distribution at the center of the channel where  moments of the fluctuations are homogeneous.   However, it deviates from the Gaussian 
		at more than two decades below the peak frequency. This deviation happens earlier in case of distribution at the near wall region. From the  plots in the insets of figures \ref{f(alpha_i)} (a) to (c) it is observed that upto one decade lower from the peak frequency all the components of  $f(\alpha_x)$ and $f(\alpha_y) $ can be approximated by Gaussian function with the same mean and variance as that of the corresponding distribution. But for 
		$f(\alpha_z) $ it is up to two decades. Other properties of $f(\alpha_i)$ e.g. the spread, the skewness, the flatness can be estimated from the higher moments of  rotational acceleration statistics shown in figure \ref{rot_acc_moments}.
		The variance of $f(\alpha_z)$ is approximately one order of magnitude higher than that of  $f(\alpha_x)$ and $f(\alpha_y)$ as shown in figure \ref{rot_acc_moments} (a). $\langle \alpha_z^2\rangle$ is higher near the wall 
		because of high fluid vorticity and decreases monotonically towards the center of the channel. In  contrary $\langle \alpha_x^2\rangle$ and $\langle \alpha_y^2\rangle$ show a non-monotonic decay.  $f(\alpha_x)$ and $f(\alpha_y)$ are symmetric in nature with almost zero skewness across the channel as shown in figure \ref{rot_acc_moments} (b). $f(\alpha_z)$ is  negatively skewed near the wall with a longer negative tail observed in figure
		\ref{f(alpha_i)} (c) and also depicted by the  skewness plot in figure \ref{rot_acc_moments} (b). Such a long tail in negative fluctuation appears due 
		to the contribution of negative fluctuation at higher magnitude due to cross 
		stream migration of the particle. It is worth to note that the the magnitudes of the skewness indicate that in all the cases distributions are mildly skewed. 
		 The positive excess Kurtosis values in figure \ref{rot_acc_moments}(c) reflects  deviation from corresponding normal distributions with more 'outliers' reflected in the longer tails evident as also being reflected in  figures \ref{f(alpha_i)} (a), (b) and (c).

		\subsubsection{Particle Rotational Velocity Distribution}
\begin{figure}[!]
			\includegraphics[width=0.45\textwidth]{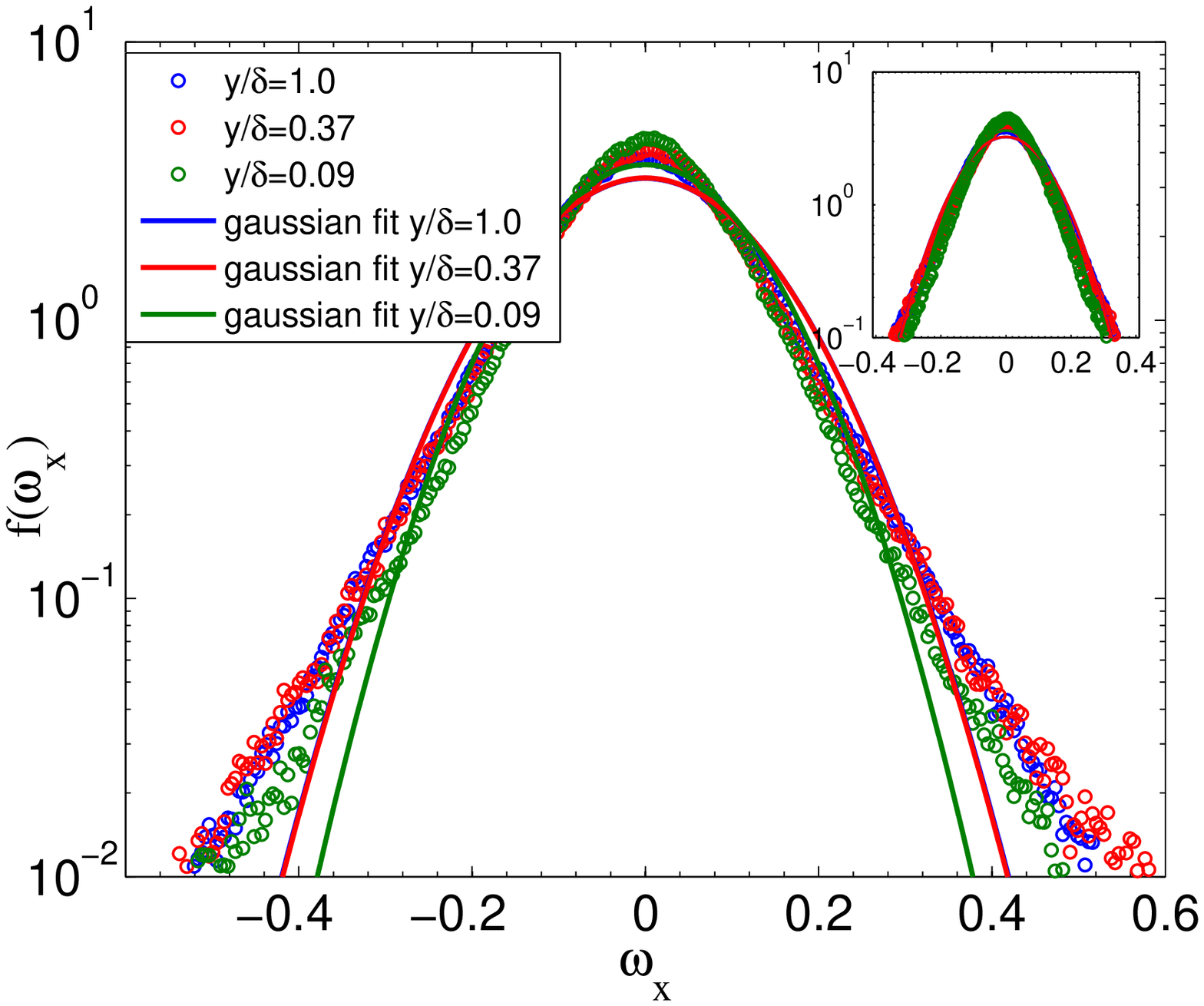}
			\caption*{(a)}
			\includegraphics[width=0.45\textwidth]{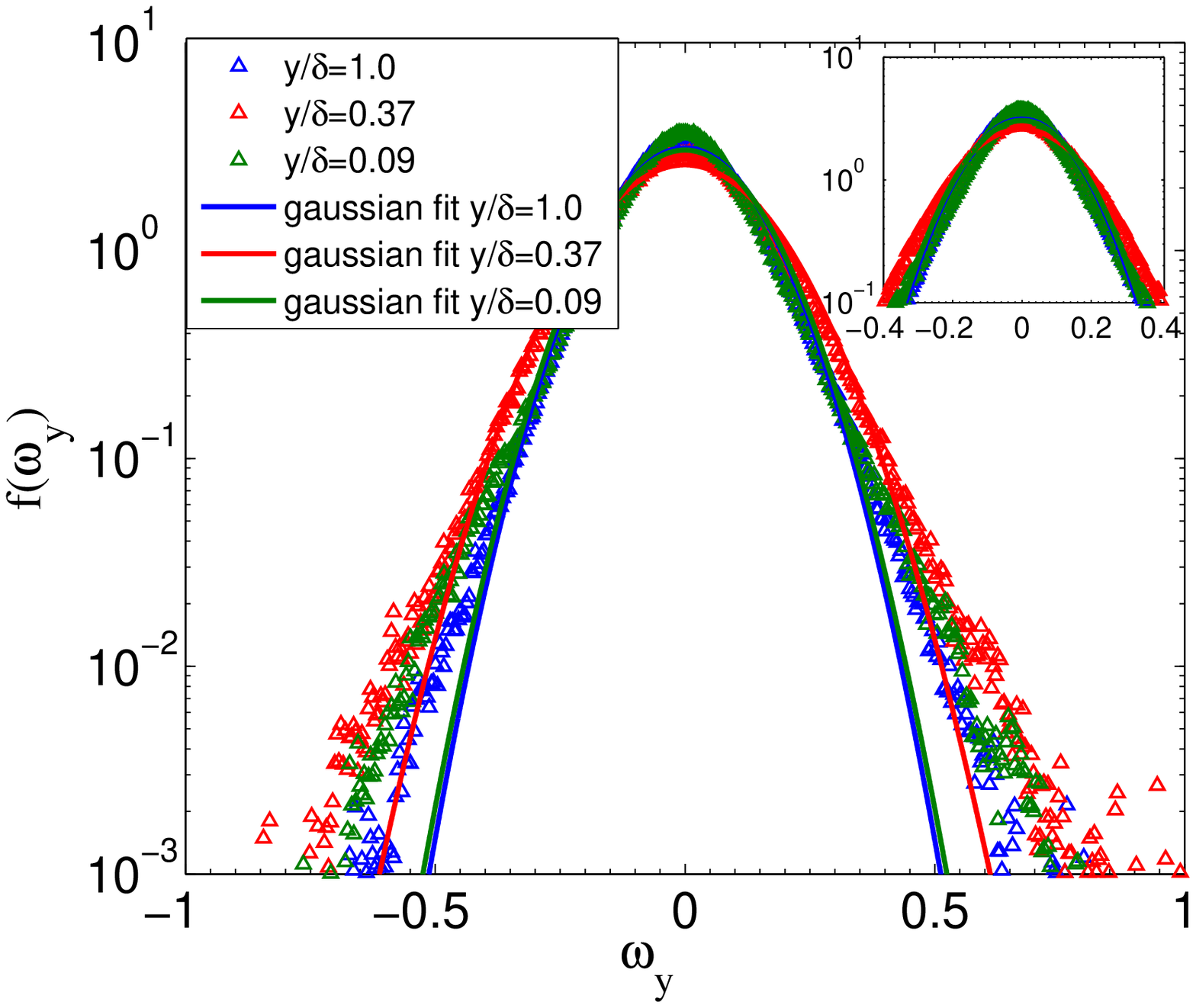}
			\caption*{(b)}
		
			\includegraphics[width=0.45\textwidth]{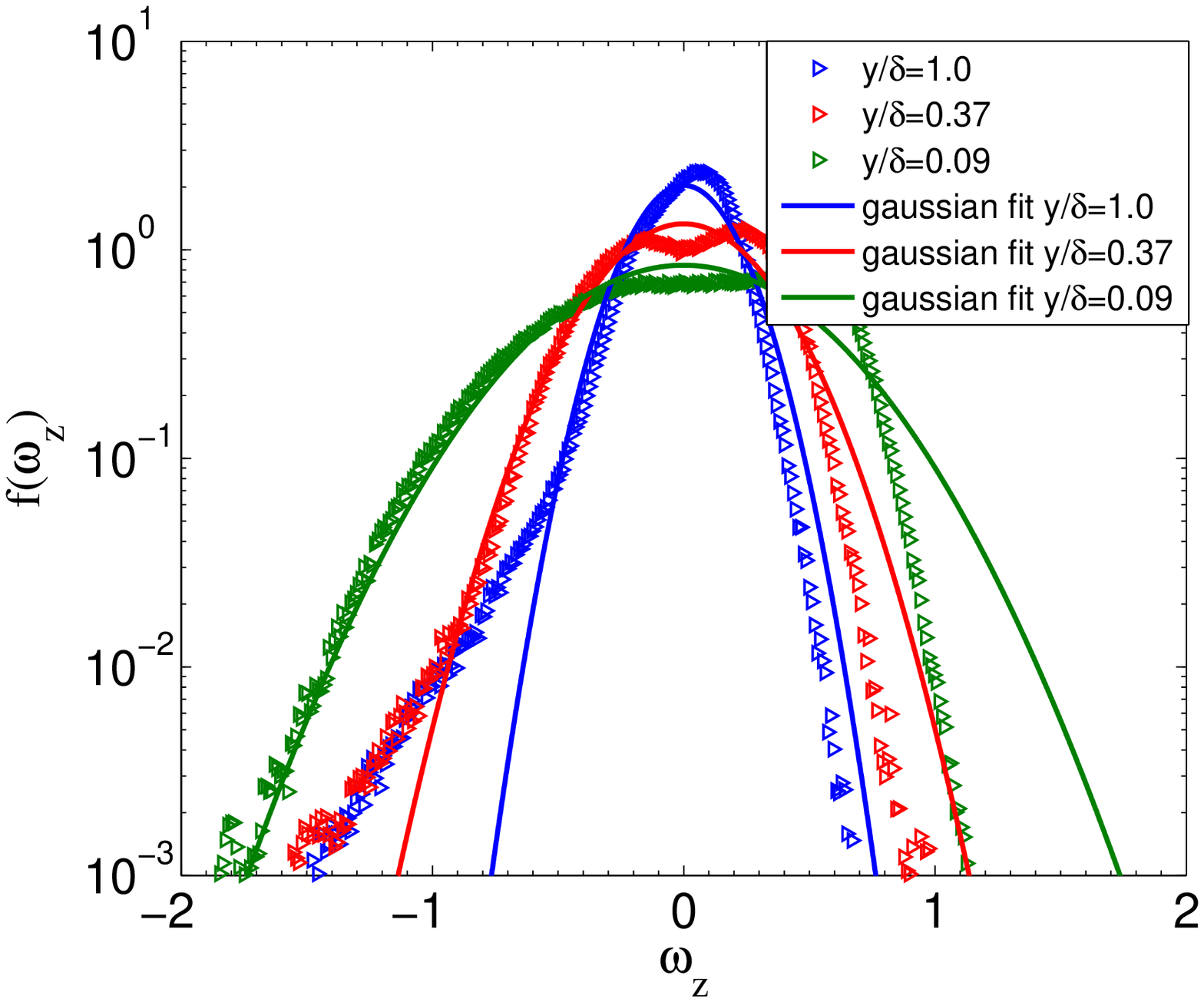}
			\caption*{(c)}
				\caption{Components of particle rotational velocity distributions  (a) $f(\omega_x)$, (b) $f(\omega_y)$, and (c) $f(\omega_z)$ at different wall normal position ($y/\delta$) of the channel.}
			\label{f(omega_i)}     
		\end{figure}

		This section describes the Particle Rotational Velocity Distribution $f(\omega_i)$ in three different directions. All the distribution functions are plotted at three different wall normal locations in the channel as shown in figures \ref{f(omega_i)} (a) to (c). Followings are the  traits of  fluctuating rotational velocity statistics.
		\begin{figure*}[!]
			\centering
			\includegraphics[width=1.0\textwidth]{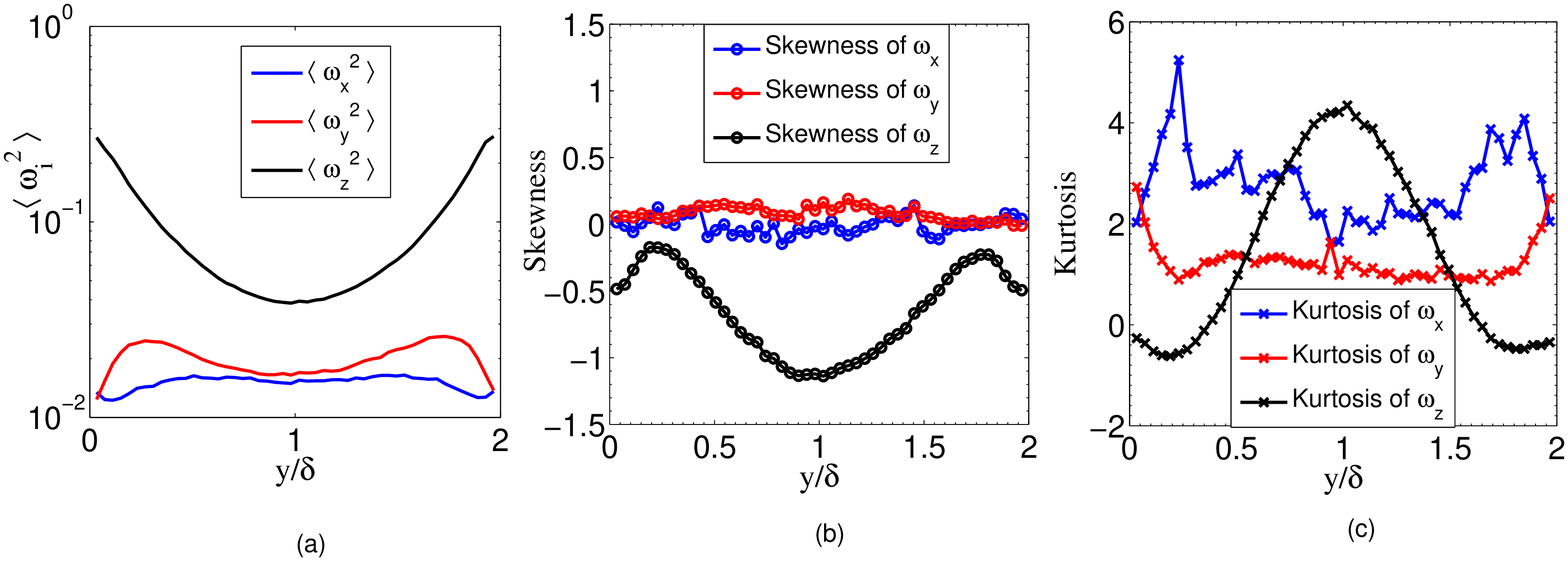}
			\caption{Moments of rotational velocity distributions; (a) second moment of rotational velocity fluctuation, (b) skewness, (c) kurtosis.}
			\label{rotational_velocity_moments}
		\end{figure*}
	
      	Figures \ref{f(omega_i)} (a), (b)  show that $f(\omega_x)$ and $f(\omega_y)$ follow Gaussian distribution at all the wall normal locations at 
      	lower values. Deviation is observed at the tail of the distribution functions. 
	    There is a distinct deviation from the Gaussian distribution for the principal component of angular velocity $f(\omega_z)$. At the center of the channel 
	    there is a long tail with higher negative fluctuation, which is mainly 
	    induced by the cross stream motion of the particles from higher negative (clock wise) angular velocities. The distribution near the wall shows most probable distribution at the positive angular velocity fluctuation and also a very sharp decay at positive fluctuations. 
	The properties of $f(\omega_i)$ e.g. the spread, the skewness are presented by the higher moments of rotational velocity  statistics shown in figure \ref{rotational_velocity_moments}. $\langle \omega_z^2\rangle$ is almost one order 
	of magnitude higher than the $f(\omega_x)$ and $f(\omega_y)$. Unlike $\langle \omega_z^2\rangle$, the variation of $\langle \omega_x^2\rangle$ and $\langle \omega_y^2\rangle$ across the channel-width is non-monotonic in nature.
	$f(\omega_x)$ and $f(\omega_y)$ are symmetric in nature with almost zero skewness across the channel as shown in figure \ref{rotational_velocity_moments} (b). 
	$f(\omega_z)$ shows mild negative skewness accross the channel. 
	The positive excess Kurtosis values in figure \ref{rotational_velocity_moments}(c) reflect significant deviation from corresponding normal distributions with longer tails evident in figures \ref{f(omega_i)} (a) to (c). However, a small negative excess Kurtosis is observed for $f(\omega_z)$  at the wall region (\ref{f(omega_i)} (c)), which corresponds to the sharper decay in the distribution at positive fluctuations.

		\subsection{Effect of Roughness on Particle Fluctuating Statistics}
		\label{section_roughness}
		In order to  understand the effect of roughness mechanistically, it is necessary to  analyze the  rotational and linear velocity and acceleration statistics. 
		Therefore, this section encompasses the effect of roughness on the translational and rotational velocity and acceleration distribution functions as well as their moments in the form of Variance, Skewness and Kurtosis.
		\subsubsection{Particle Acceleration And Velocity Distribution}
\begin{figure}[!]
		\includegraphics[width=1.0\linewidth]{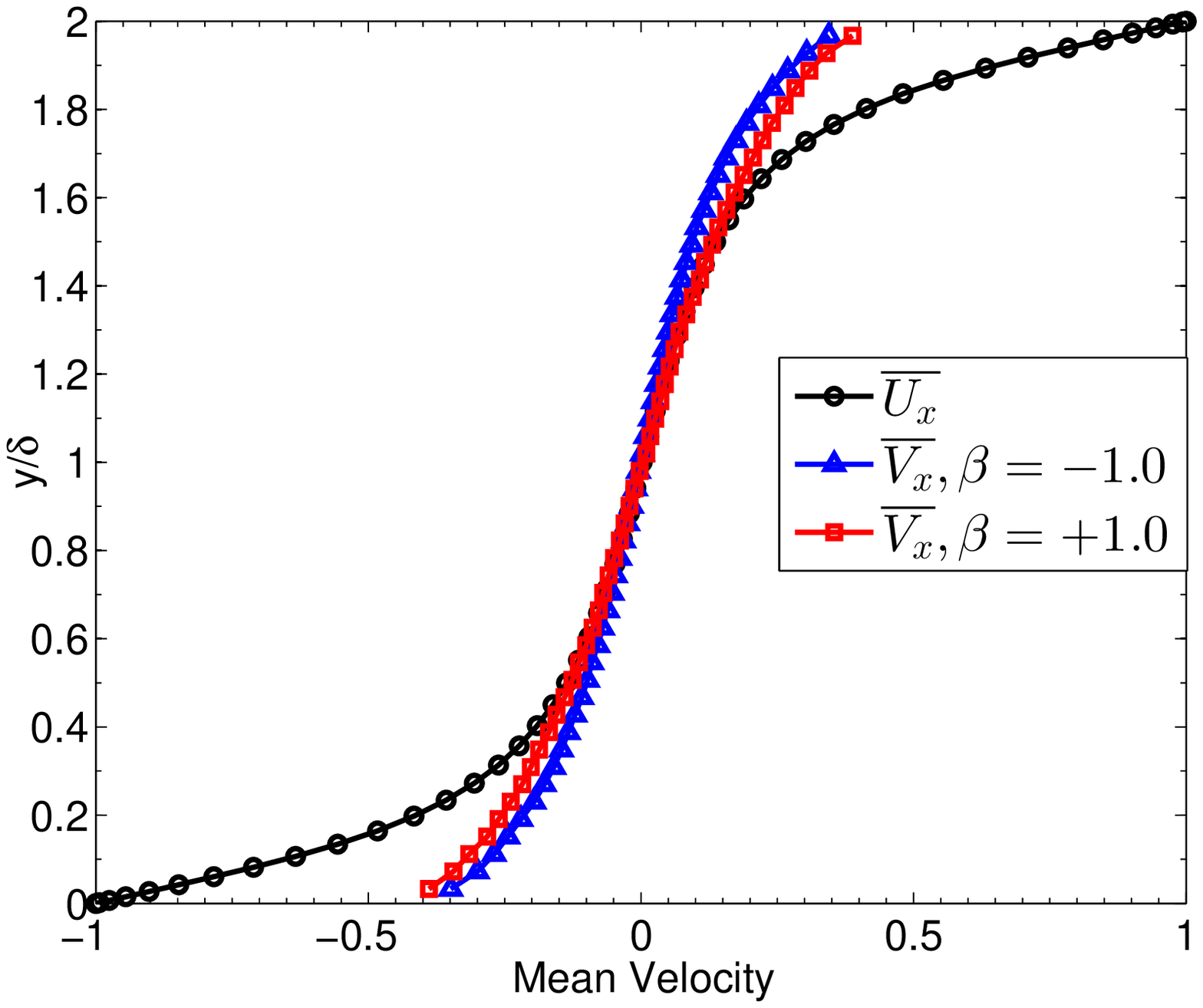}
			\caption{ Mean Velocity of fluid (black), velocity profile for smooth particles (blue), and velocity profile for rough particles (red) as a Function of Channel-width ($y/\delta$)}
			\label{mean(v_x)}
		\end{figure}

\begin{figure}[h!]
			\centering
			\includegraphics[width=0.5\textwidth]{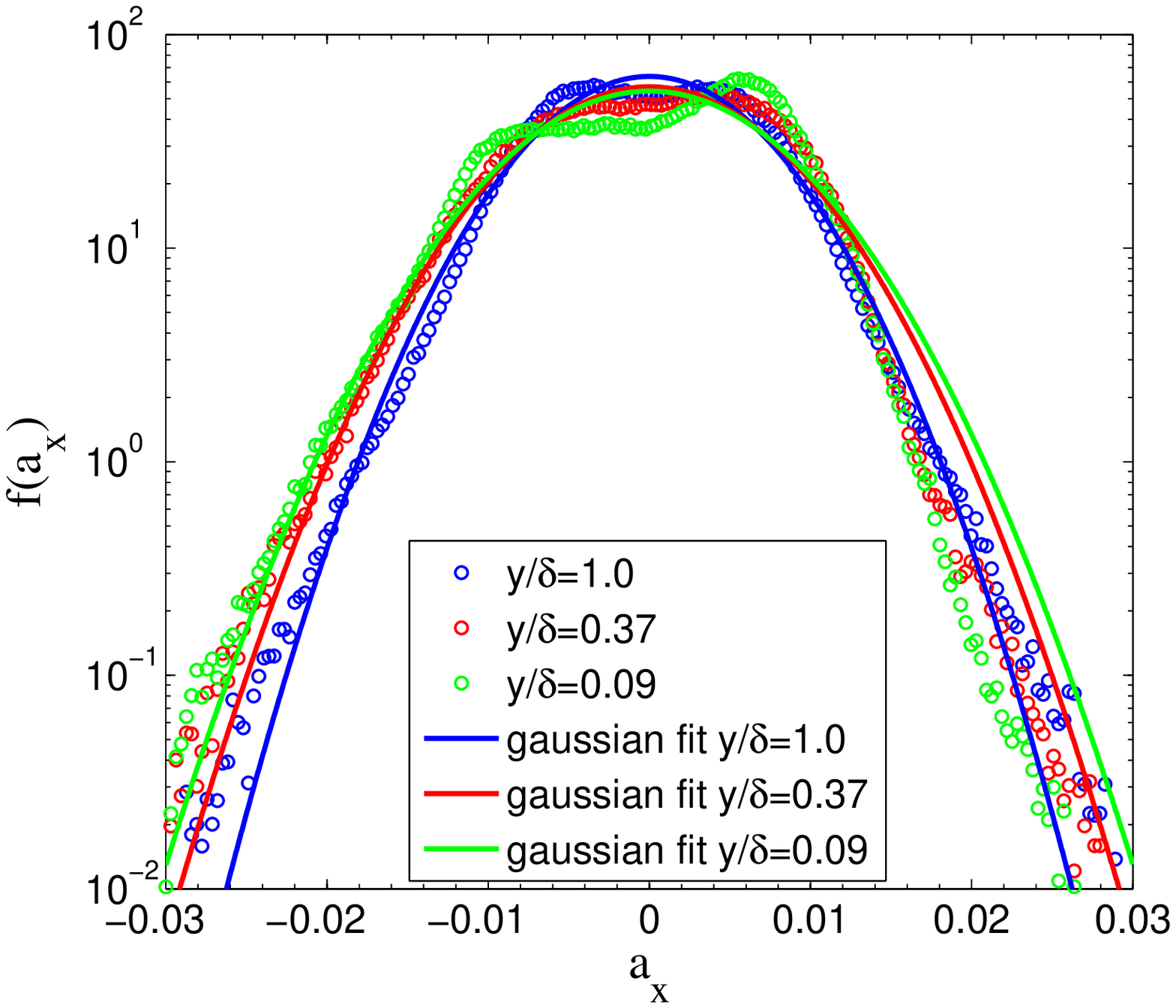}
			\caption*{(a)}
			\includegraphics[width=1.0\linewidth, height=5.5cm]{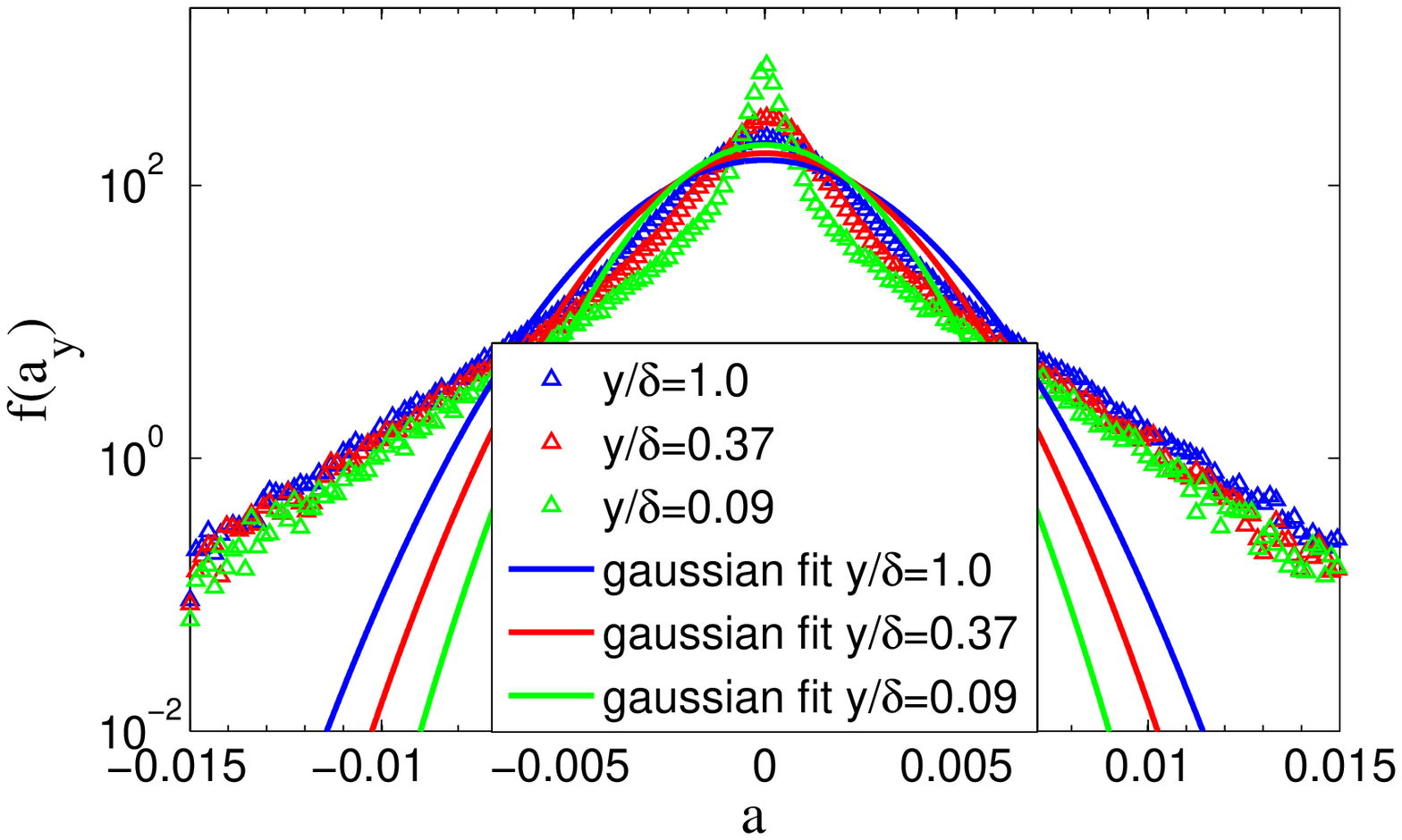}
			\caption*{(b)}
			\includegraphics[width=1.0\linewidth, height= 5.5cm]{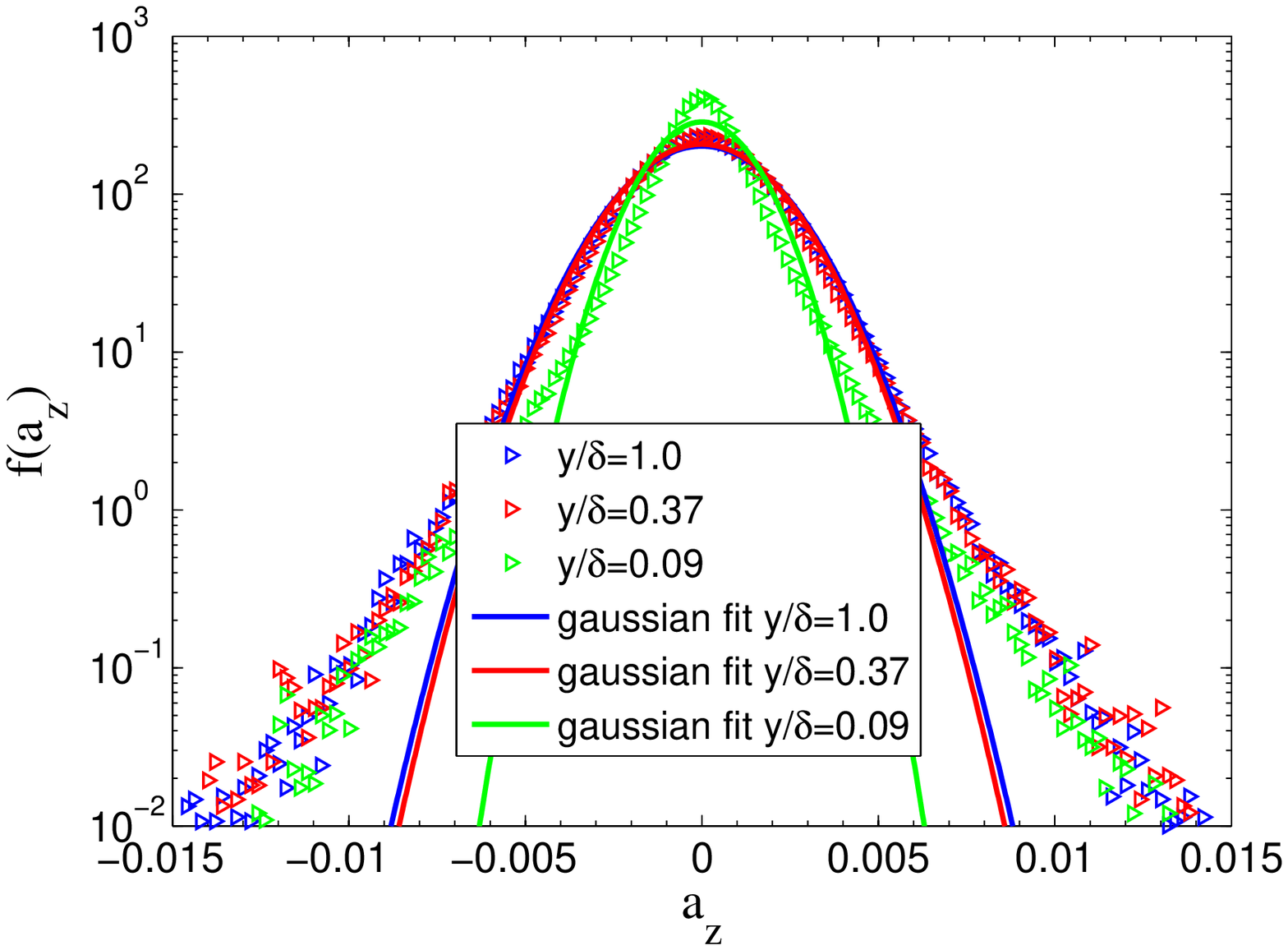}
			\caption*{(c)}
			%
			\caption{Components of particle acceleration distributions (a) $f(a_x)$, (b) $f(a_y)$, and (c) $f(a_z)$ at different wall normal positions ($y/\delta$) of the channel for $\beta=+1.0$.}
			\label{rough_f(a_i)}     
		\end{figure}

\begin{figure}[h!]
			\centering
			\includegraphics[width=1.0\linewidth, height=6.5cm]{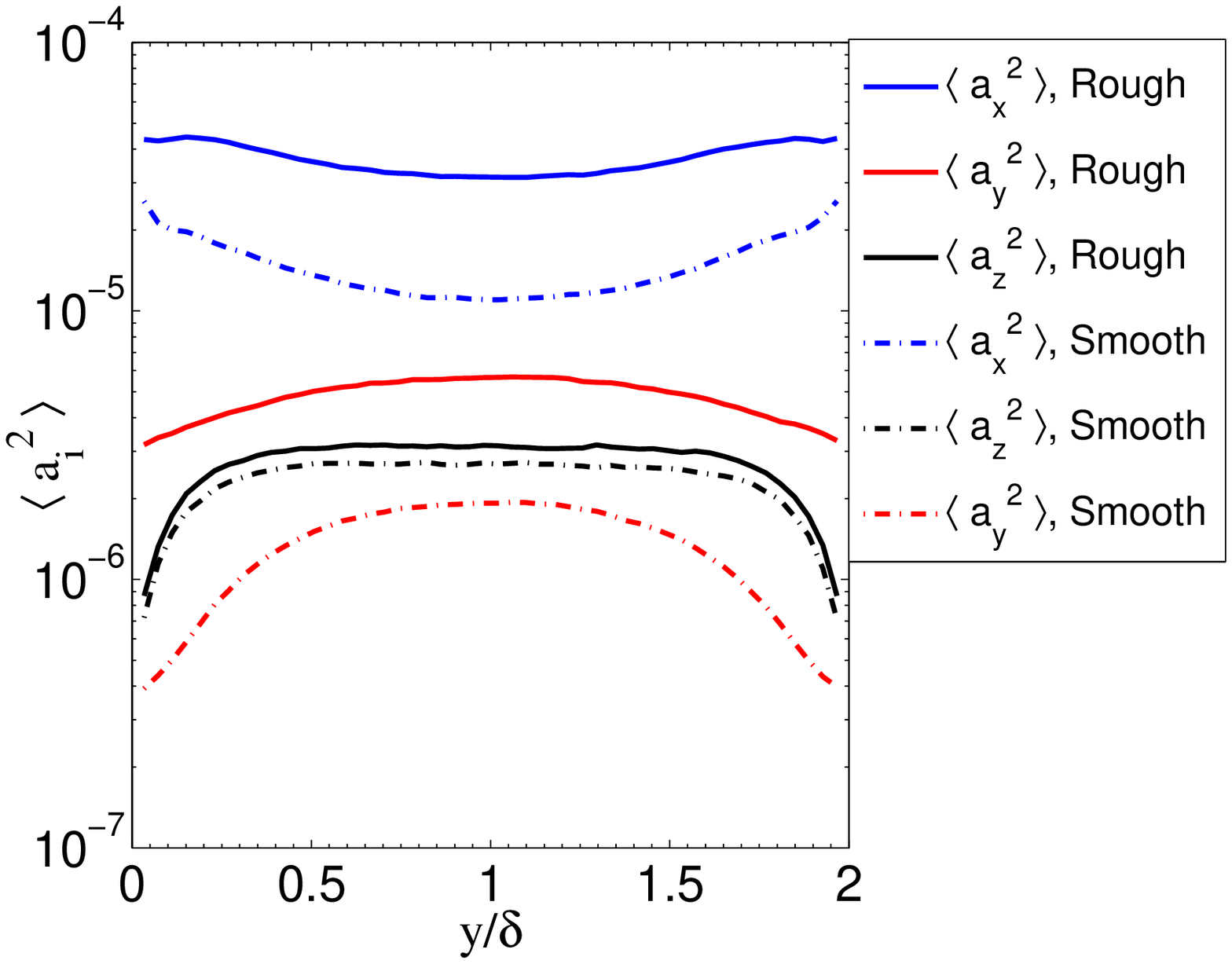}
			\caption{\label{rough_variance_acc}Effect of rough collisions on variance of acceleration}
			\includegraphics[width=1.0\linewidth, height=5.5cm]{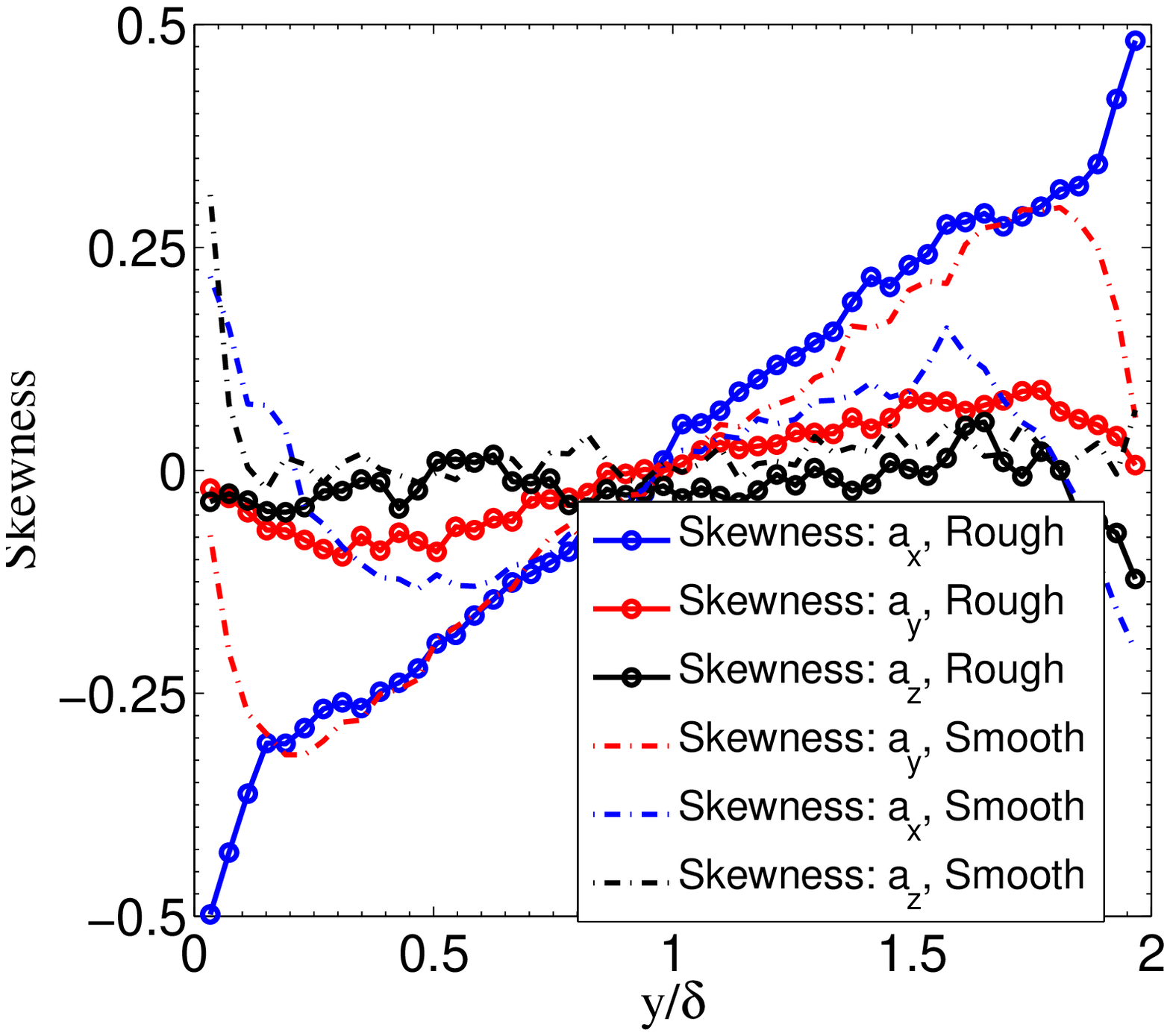}
			\caption{\label{rough_skewness_acc}Effect of rough collisions on skewness of acceleration}
			\includegraphics[width=1.0\linewidth, height= 5.5cm]{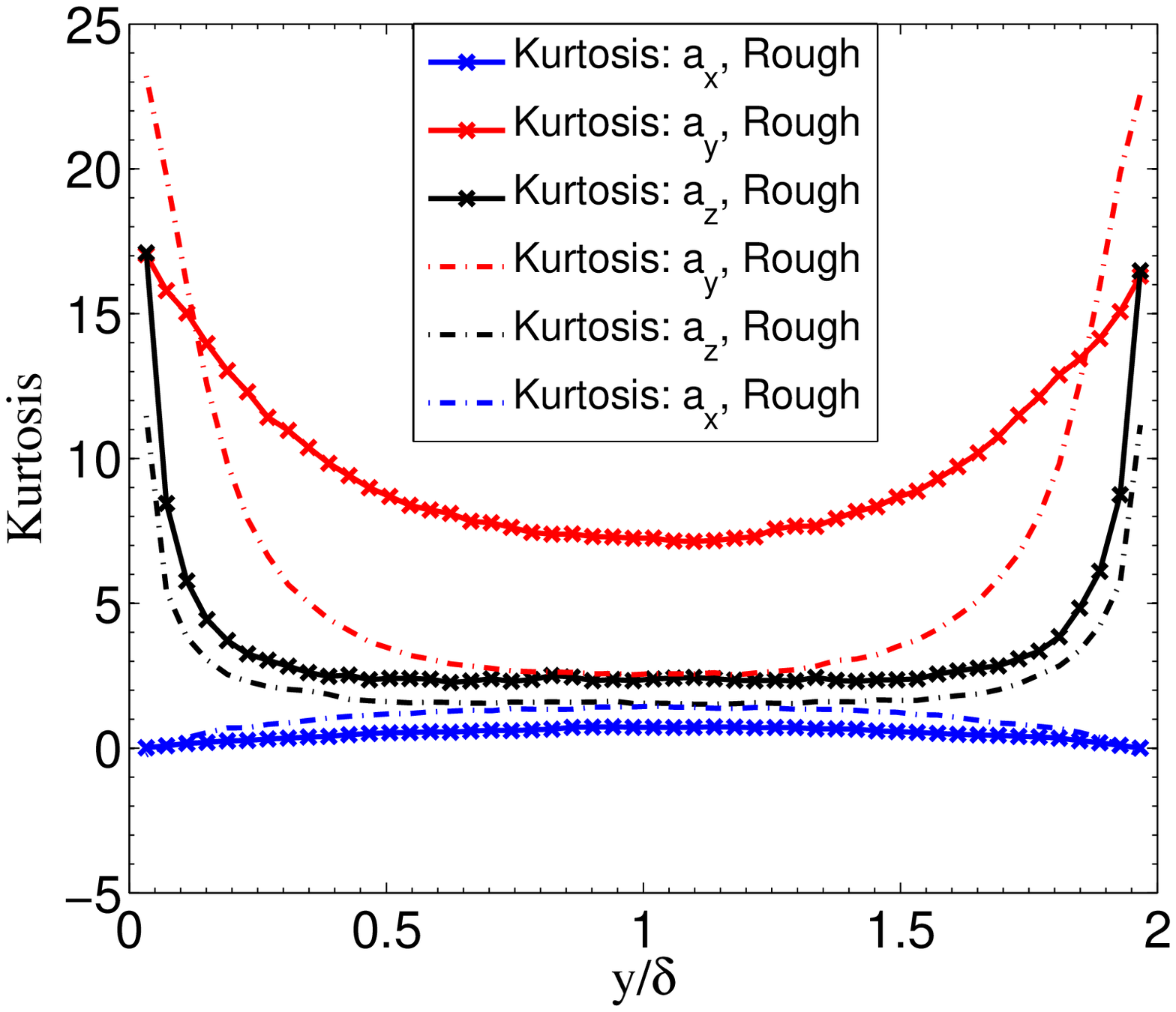}
			\caption{\label{rough_kurtosis_acc}Effect of rough collisions on kurtosis of acceleration}
		\end{figure}

\begin{figure}[h!]
			\centering
			\includegraphics[width=0.5\textwidth ]{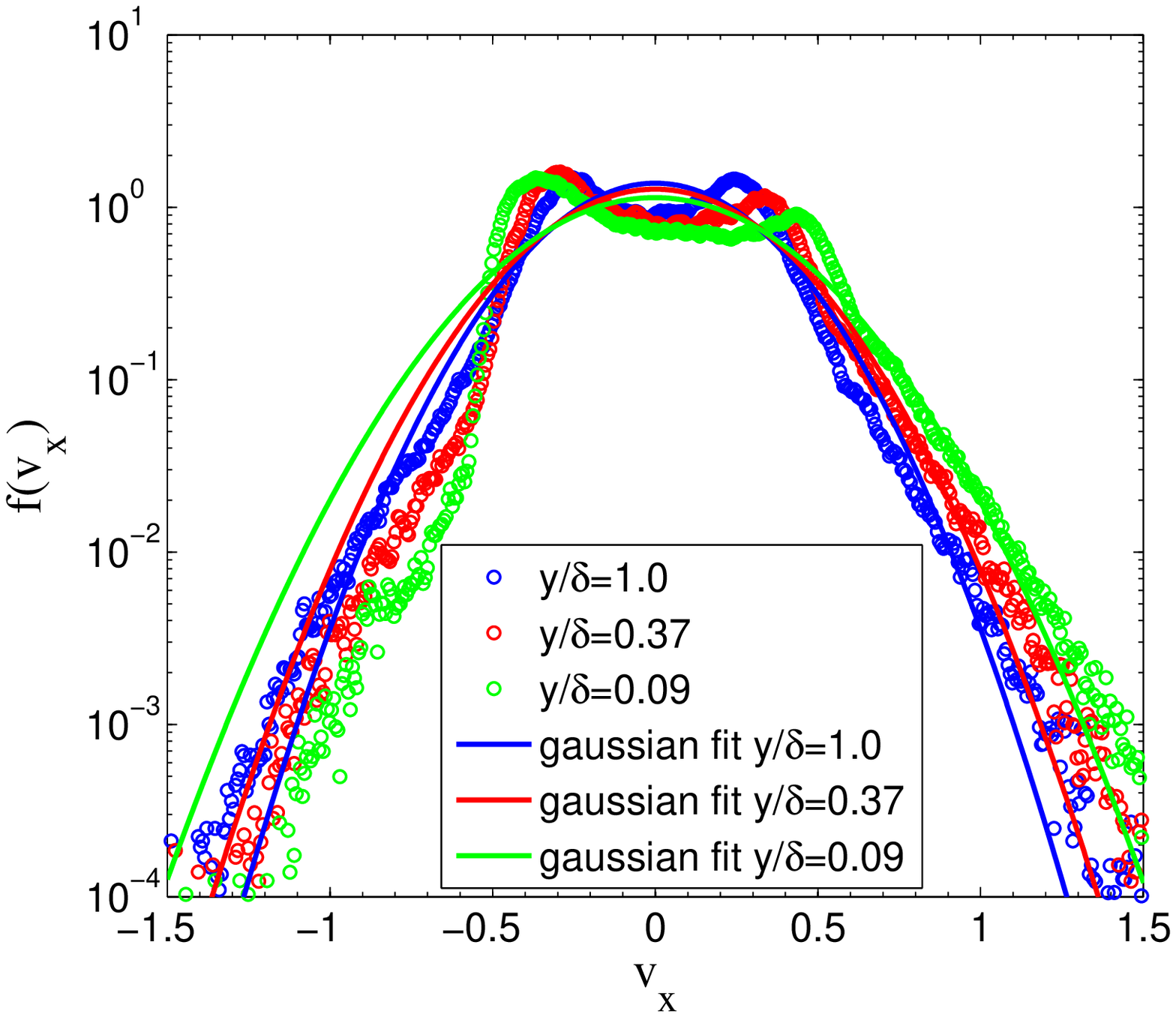}
			\caption*{(a)}
			\includegraphics[width=1.0\linewidth, height=5.5cm]{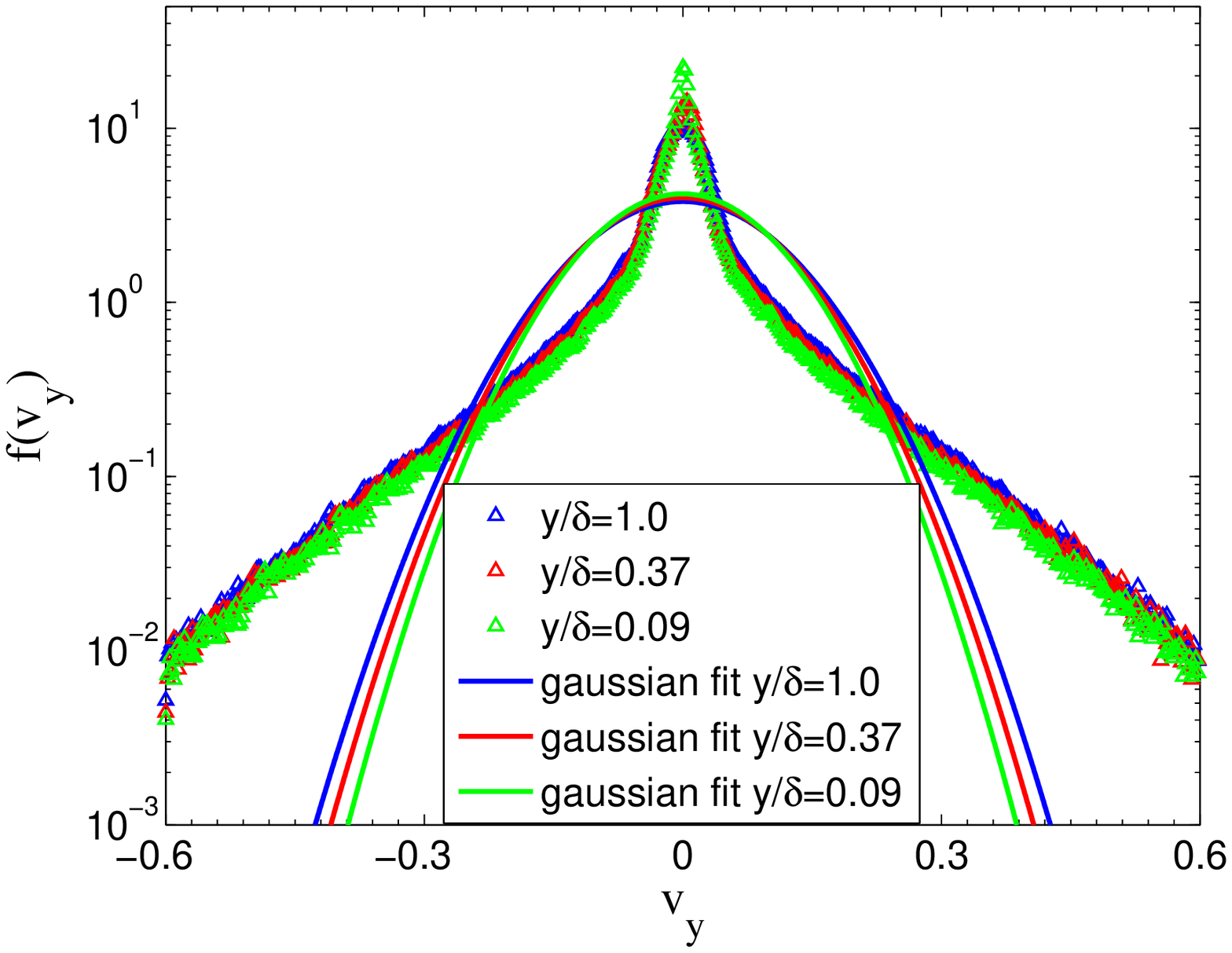}
			\caption*{(b)}
			\includegraphics[width=1.0\linewidth, height=5.5cm]{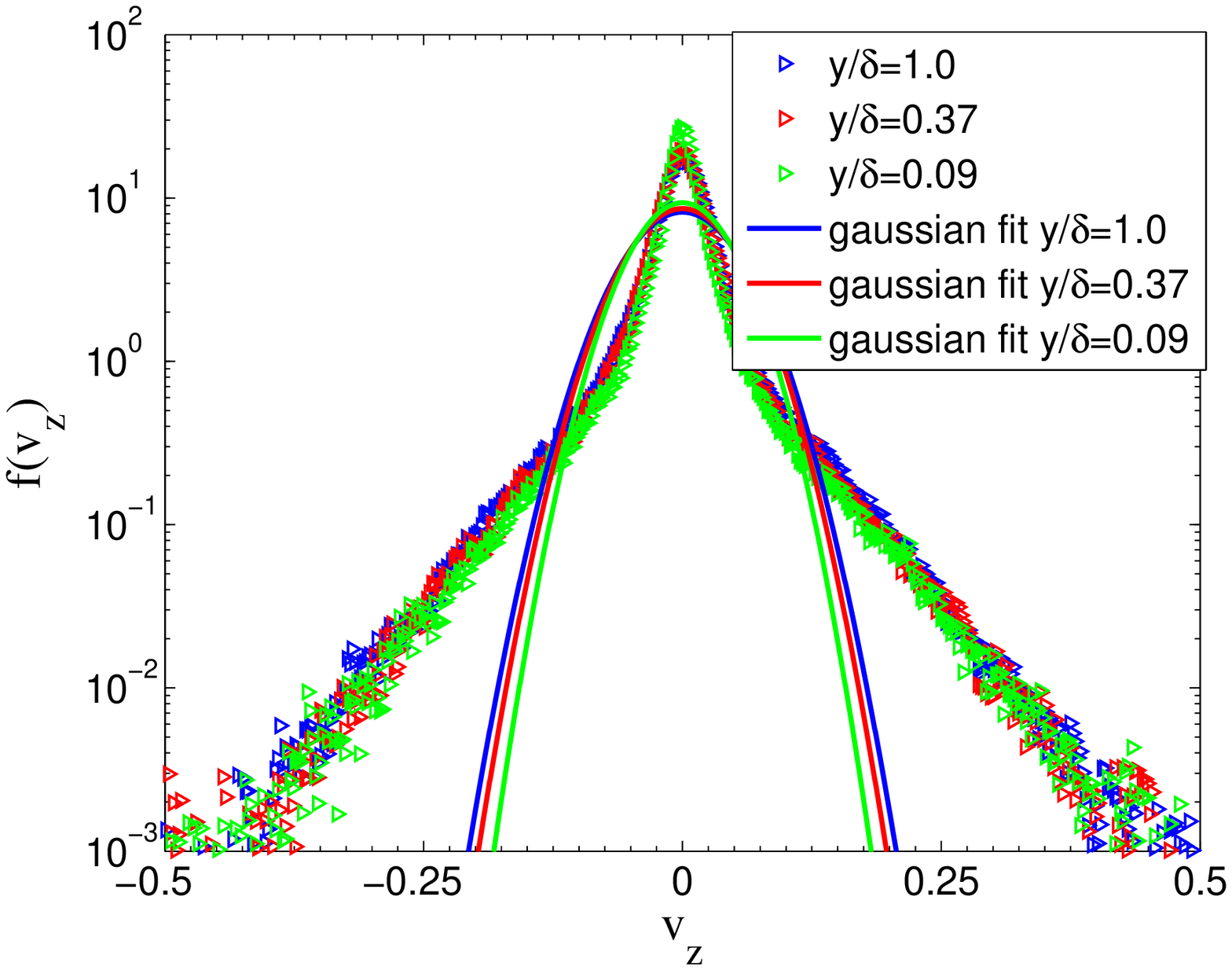}
			\caption*{(c)}
			\caption{Components of particle velocity distributions (a) $f(v_x)$, (b) $f(v_y)$, and (c) $f(v_z)$ at different wall normal positions ($y/\delta$) of the channel for $\beta=+1.0$.}
			\label{rough_f(v_i)}     
		\end{figure}

\begin{figure}[!]
			\centering
			\includegraphics[width=1.0\linewidth]{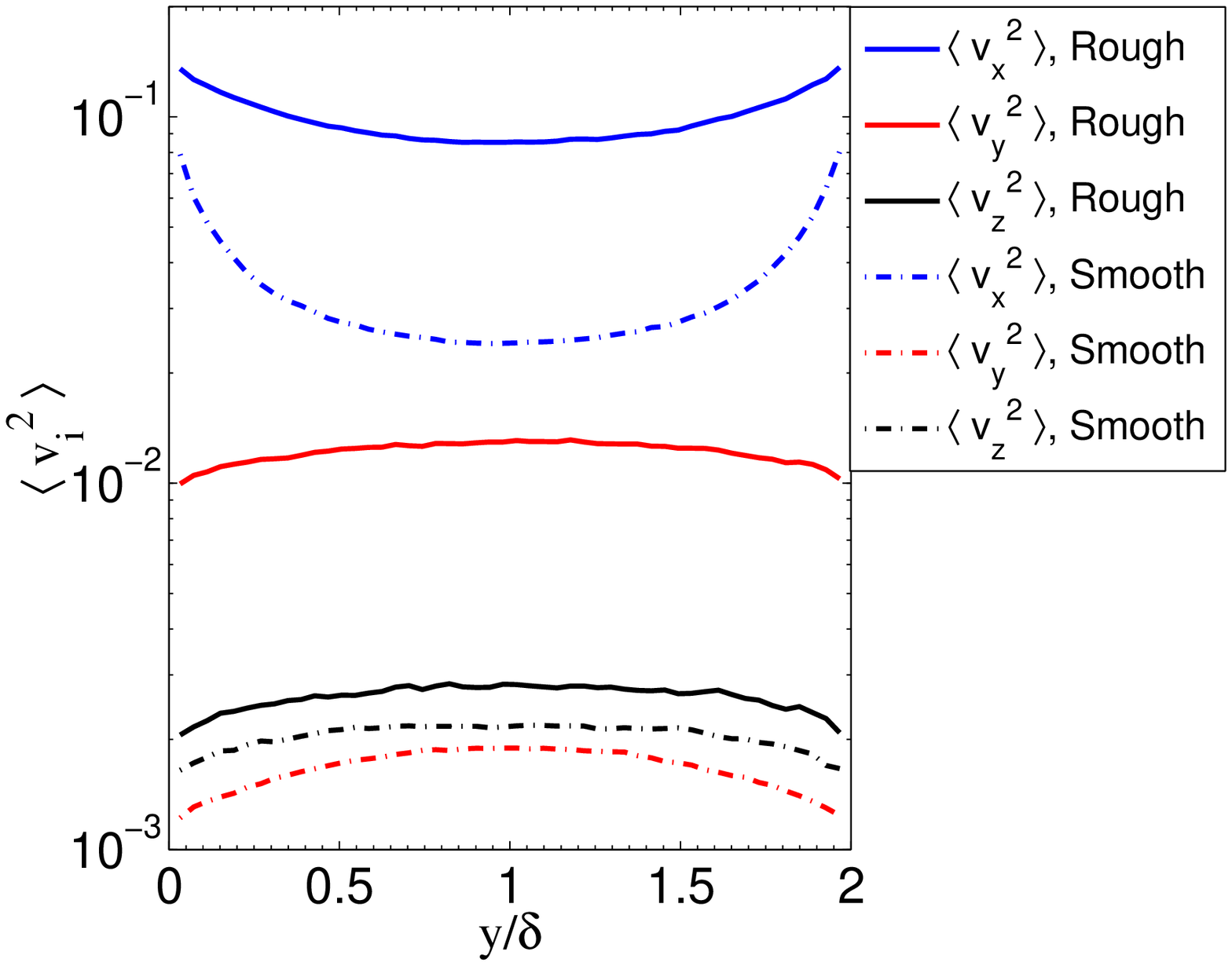}
			\caption{\label{rough_variance_vel}Effect of rough collisions on variance of velocity}
			\includegraphics[width=1.0\linewidth]{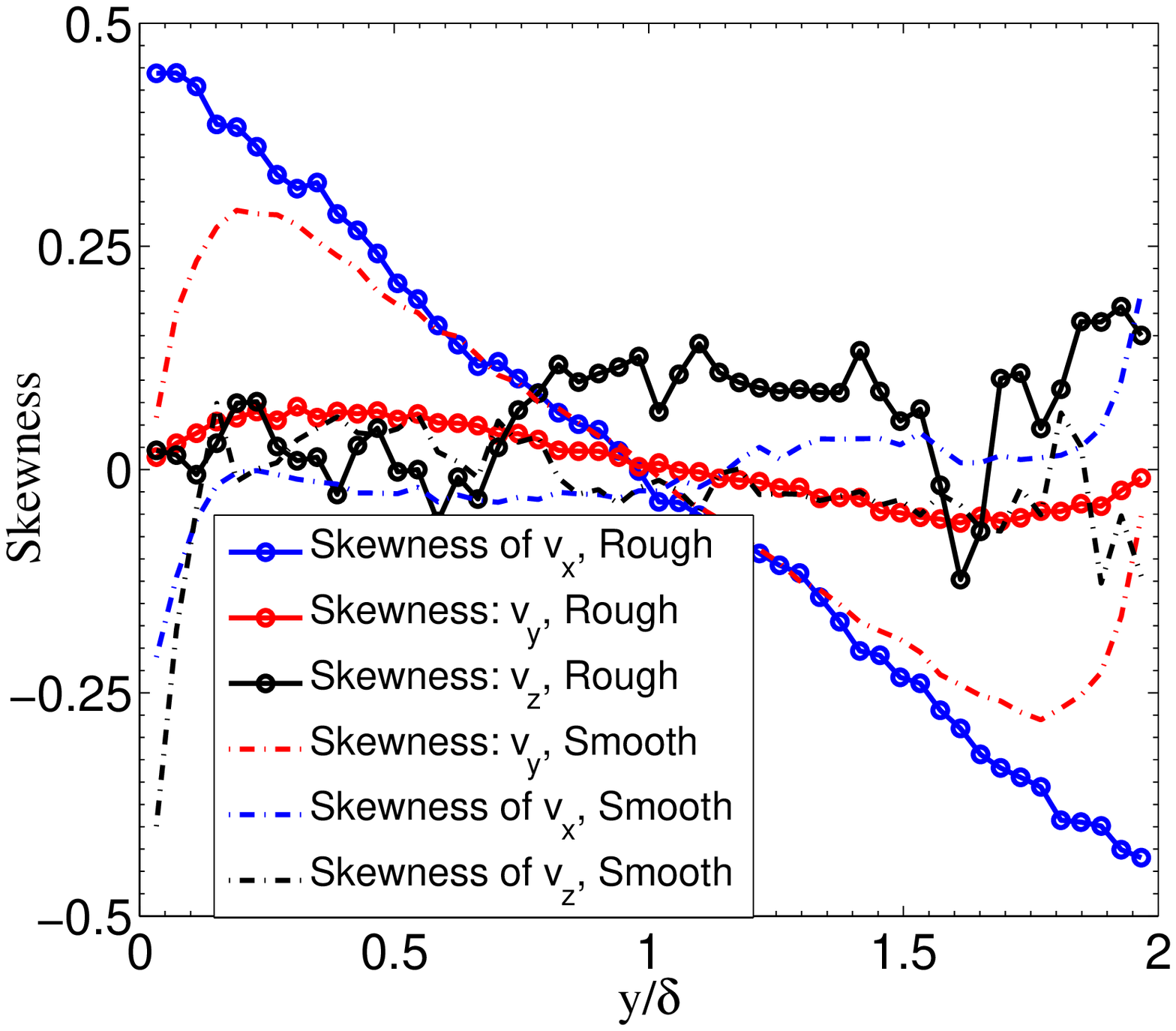}
			\caption{\label{rough_skewness_vel}Effect of rough collisions on skewness of velocity}
			\includegraphics[width=1.0\linewidth]{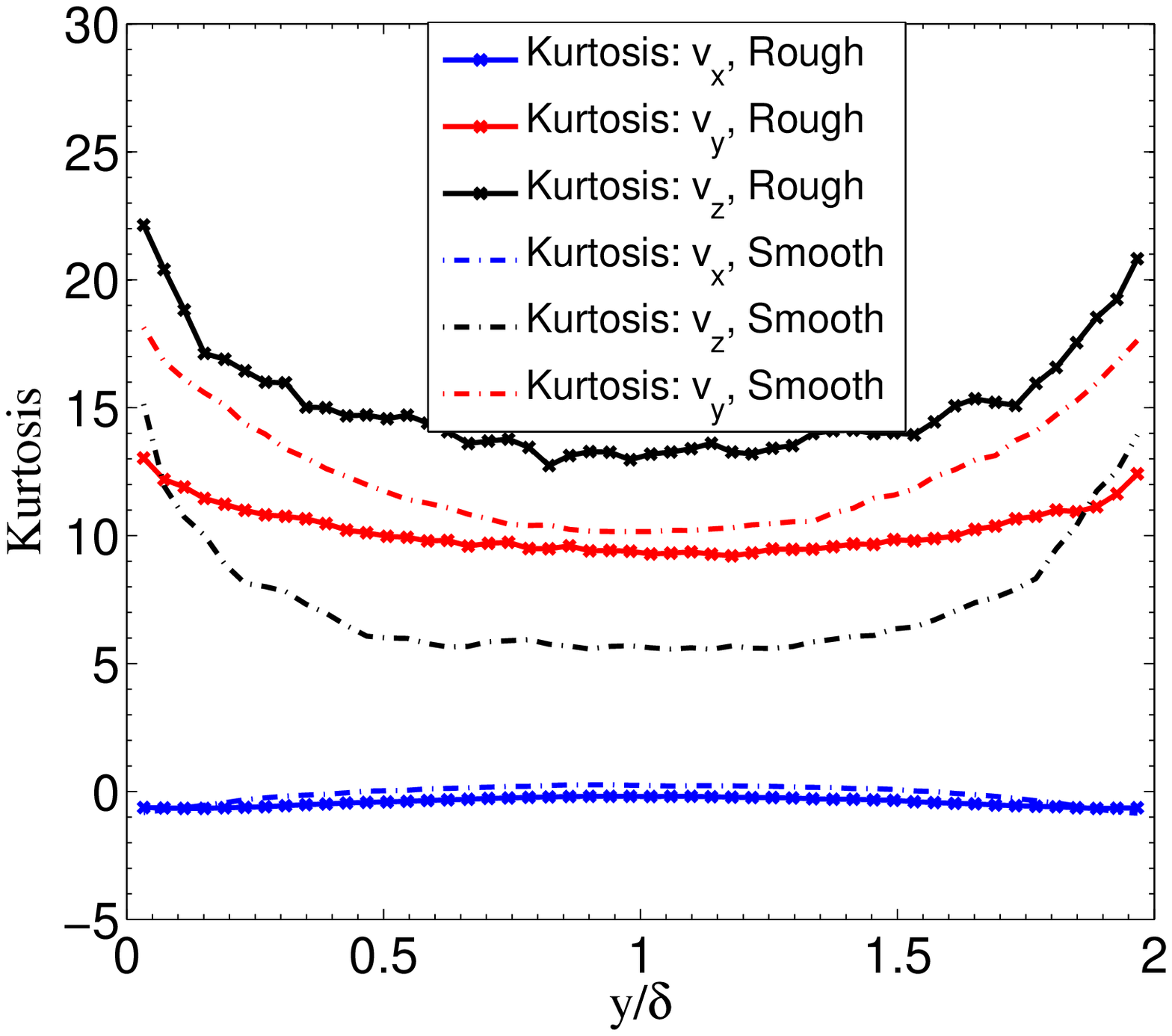}
			\caption{\label{rough_kurtosis_vel}Effect of rough collisions on kurtosis of velocity}
		\end{figure}

		Roughness factor  affect the translational acceleration and velocity distribution primarily through wall-particle collision. The average inter-particle collision times are 110 and 193 non-dimensional ($\delta/U_{wall}$) units  for rough and smooth particles respectively; thus indicating that in presence of roughness factor inter-particle collisions occur 1.74 times more often. The wall-particle collisions happen to be about 2.5 times more often in presence of roughness as the average wall-particle collision times are 32.25 and 80.65 $\delta/U_{wall}$ units for rough and smooth particles respectively. It is observable that for both the cases wall-particle collision occur more frequently than inter-particle collision. A few pronounced effects are observed in the Particle  Acceleration and Velocity fluctuations as shown in  figures \ref{rough_f(a_i)} - \ref{rough_kurtosis_vel}. It is to be noted that the 
		following results are obtained at the limit of completely rough collisions i.e. $\beta=+1$. Therefore besides local fluid velocity (figure \ref{mean(v_x)}), particle-particle or particle-wall interactions play important role in controlling  the dynamics of the particles.
	\\	The tails of $f(a_x)$ an $f(v_x)$ deviates slightly from the corresponding Gaussian fits at the channel center and at the intermediate location. But in case of distribution near the wall, the deviation is more prominent.  Unlike smooth collisions, the high probability low magnitude zone of the fluctuations qualitatively differ from Gaussian nature with  the existence of weakly bi-modal nearly symmetrical peaks (except very near the wall) at lower values of acceleration and velocity fluctuations as shown in figure \ref{rough_f(a_i)} (a) and \ref{rough_f(v_i)} (a). The flat distribution at the peak may originate due to increase is collision frequency and roughness induced change in tangential velocity at the contact point. Due to high frequency of collision and and high level of anisotropy 
		in streamwise and other component of fluctuation, momentum transfer happens from streamwise to other components of fluctuations results in slowly decaying long tail in the distribution function of cross stream velocities and acceleration fluctuations as shown in figures \ref{rough_f(a_i)} (b) , \ref{rough_f(a_i)} (c), \ref{rough_f(v_i)} (b), and \ref{rough_f(v_i)} (c). The decay in $f(v_y)$ is comparatively more slower than that in $f(v_z)$,  which indicates that the probability of higher velocity fluctuation is more in case of $f(v_y)$, which originates because of the 
		roughness induced momentum transfer from the moving wall.
		The effect of particle roughness on the second moment of acceleration and velocity fluctuations are shown in figures \ref{rough_variance_acc} and \ref{rough_variance_vel} respectively. Roughness increases the variance of velocity and acceleration distribution. There is a marginal increase in z-direction but in other two directions, variances increases significantly.
		The deviation of acceleration and velocity distribution form the Gaussian has been quantified in terms of skewness and kurtosis as shown in figures \ref{rough_skewness_acc}, \ref{rough_kurtosis_acc}, \ref{rough_skewness_vel} and \ref{rough_kurtosis_vel}. A similar asymmetry is also reported in case of Granular Poiseuille Flow in presence of rough walls \cite{vijayakumar2007velocity, alam2010velocity, alam2010velocity}. The pdf deviates from  Gaussian nature and the peak shifts from zero in the near wall region. In the present simulations skewness is observed at the positive fluctuation when the local mean velocity is in the 
		negative direction and vice versa. The reason for weak bi-modality is as follows.  When the particles move from the center towards the wall (we may consider the lower wall moving with -ve velocity) it induces a positive fluctuation and once it collides with the wall, due to rough collision particle will acquire more negative velocity and it will induce a negative fluctuation in the PDF. The peaks corresponds to the most probable distribution indicates the cross stream migration 
		of the particles plays more important role in deciding the pdf of low velocity fluctuations. The trend in skewness is exactly opposite for the distribution of velocity and acceleration as shown in figures \ref{rough_skewness_acc} and \ref{rough_skewness_vel}.


		\subsubsection{Particle Rotational Acceleration and Velocity Distribution}
\begin{figure}[!]
			\includegraphics[width=1.0\linewidth, height=6cm]{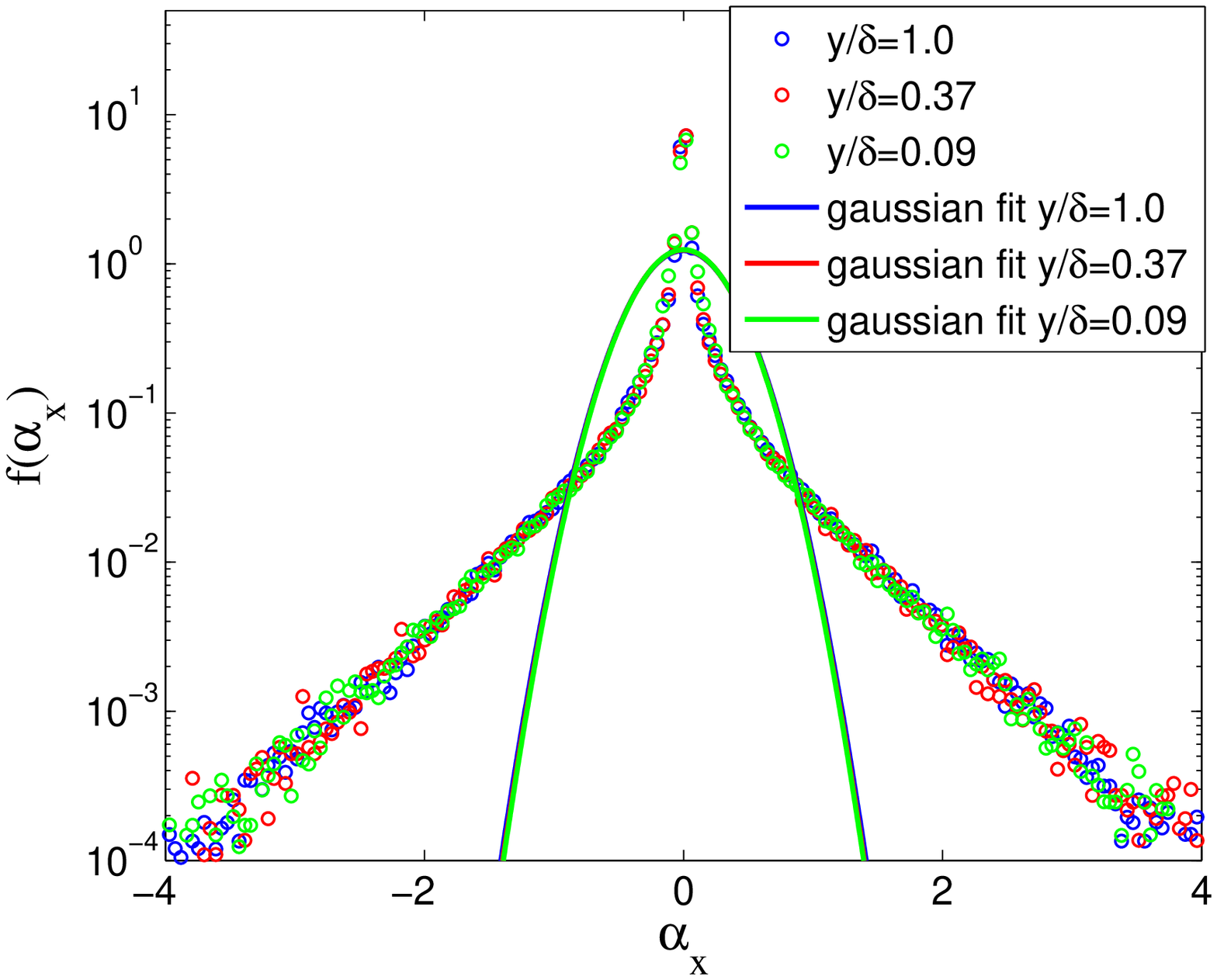}
			\caption*{(a)}
			\includegraphics[width=1.0\linewidth, height=6cm]{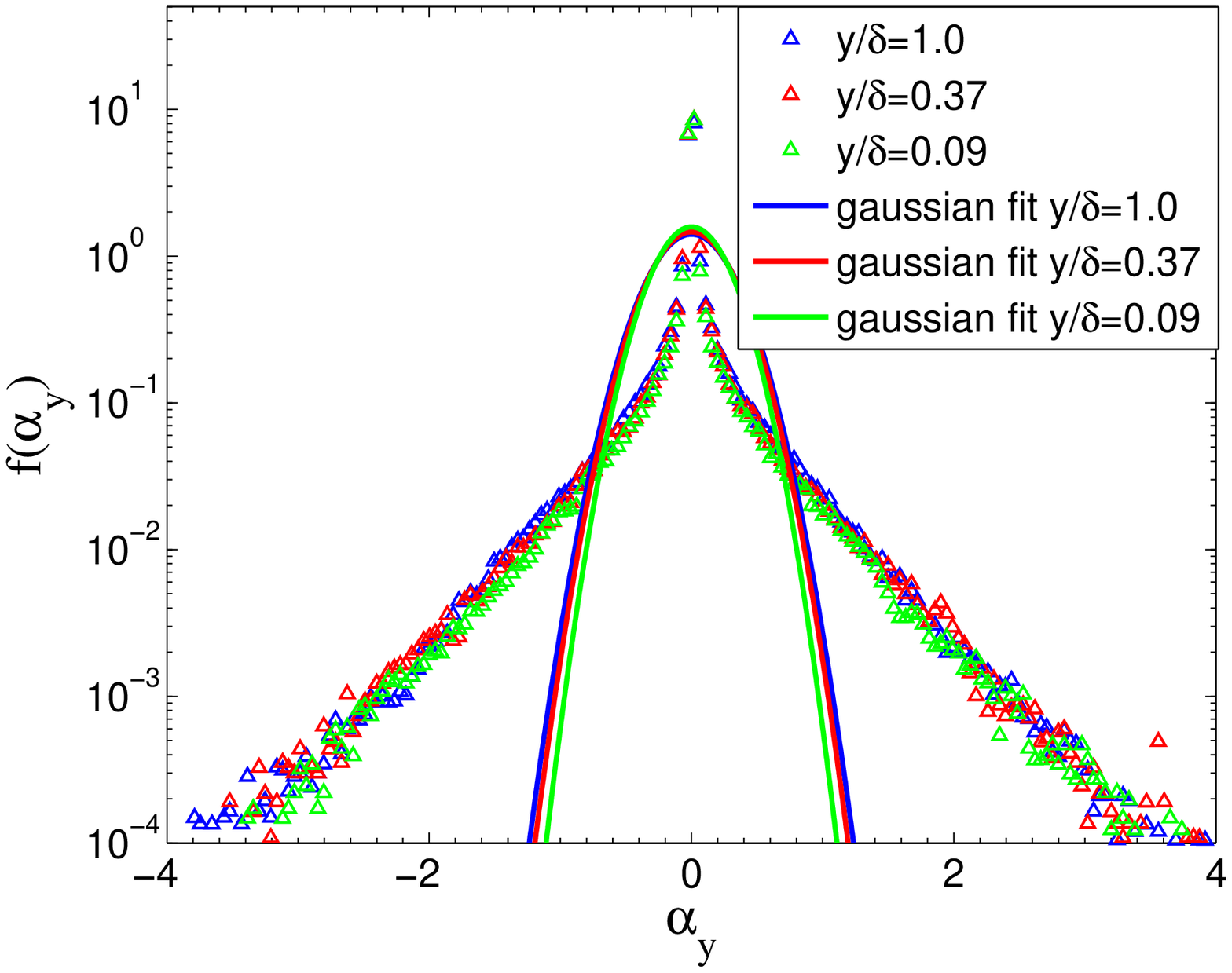}
			\caption*{(b)}
			\centering
			\includegraphics[width=1.0\linewidth]{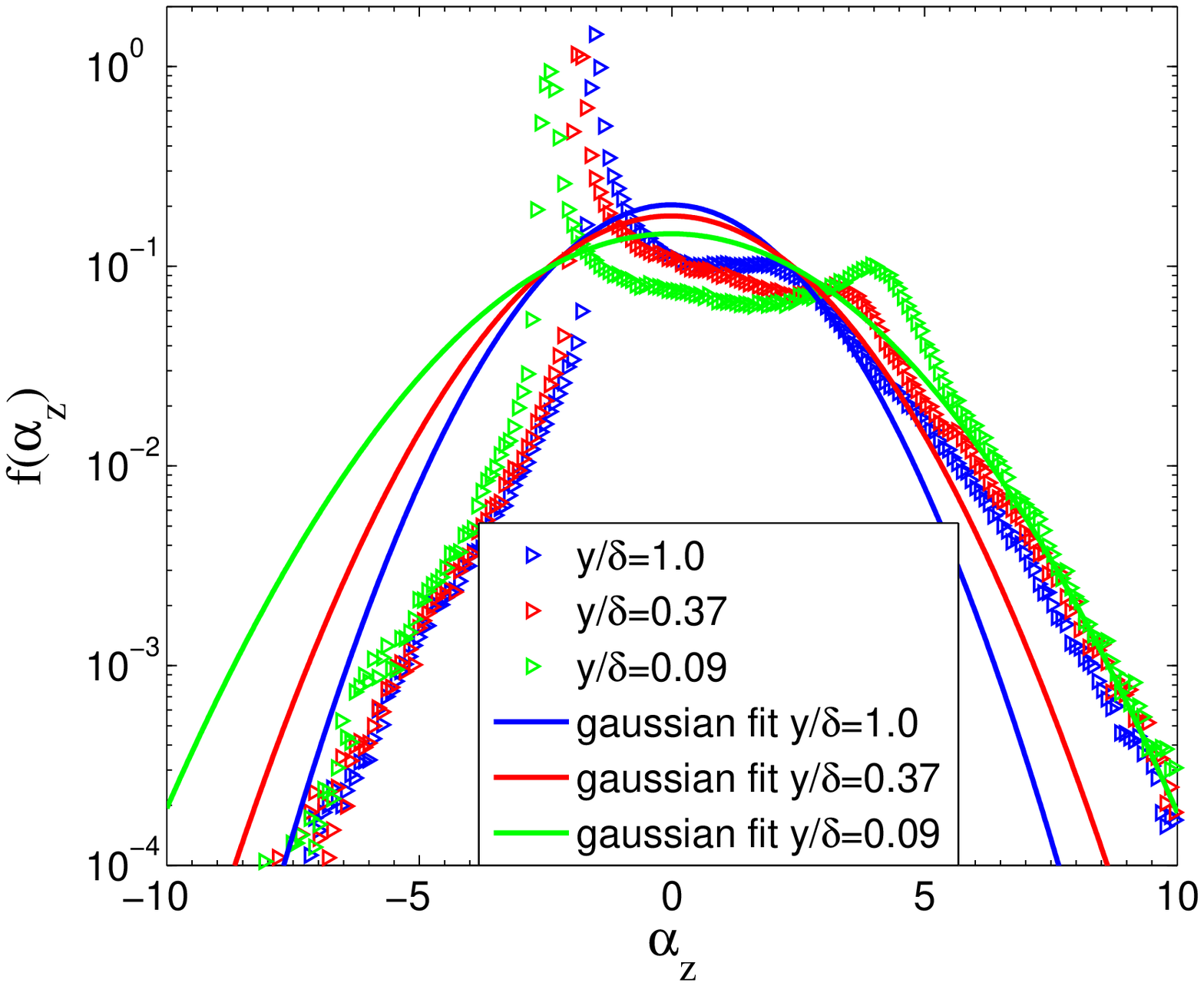}
			\caption*{(c)}
			\caption{Components of particle rotational acceleration distributions (a) $f(\alpha_x)$, (b) $f(\alpha_y)$, and (c) $f(\alpha_z)$ at different wall normal positions ($y/\delta$) of the channel for $\beta=+1.0$.}
			\label{rough_f(alpha_i)}     
		\end{figure}

\begin{figure}[!]
			\centering
			\includegraphics[width=1.0\linewidth, height=6.5cm]{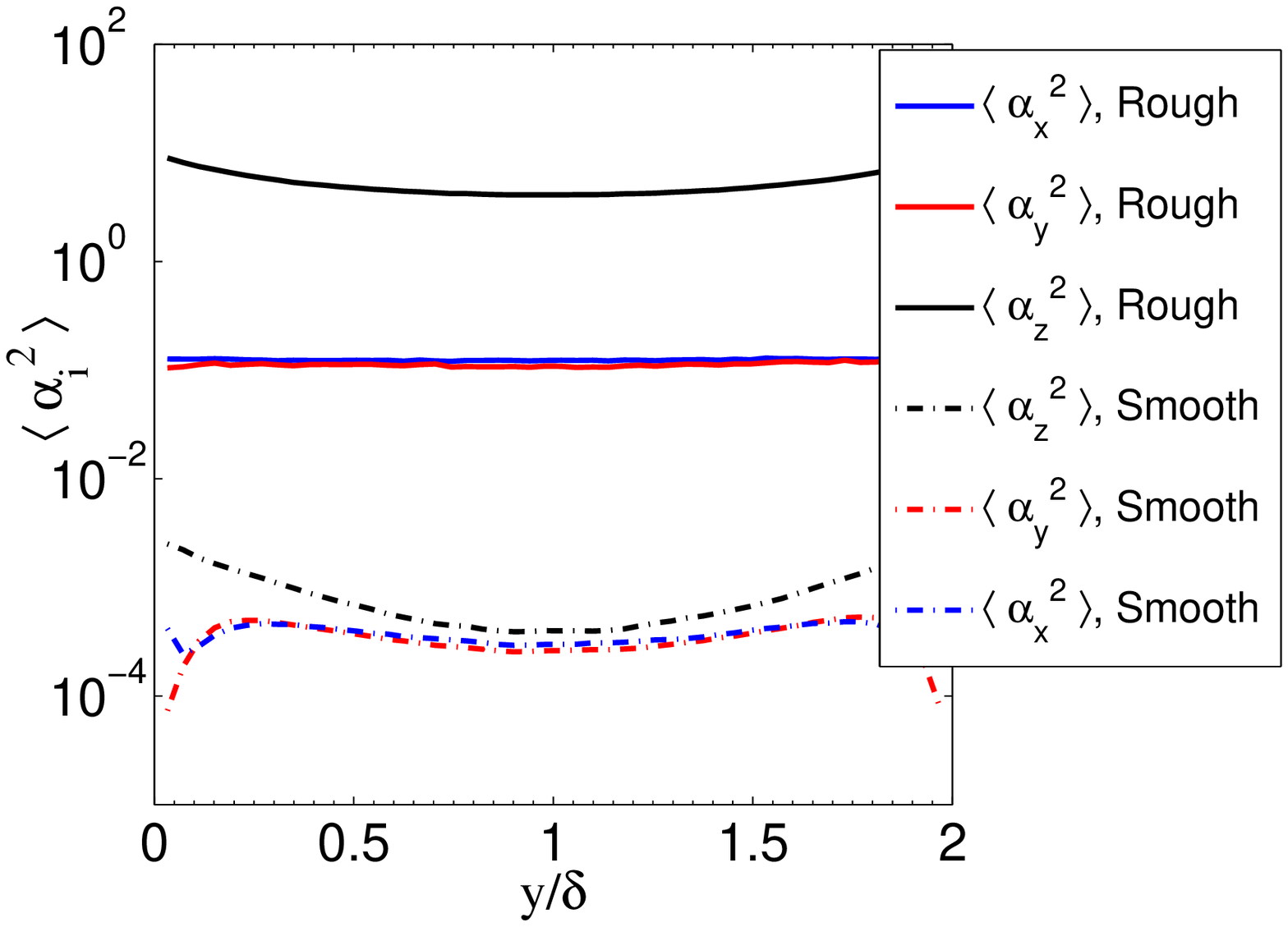}
			\caption{\label{rough_variance_rot_acc}Effect of rough collisions on variance of rotational acceleration}
			\includegraphics[width=1.0\linewidth, height=6.5cm]{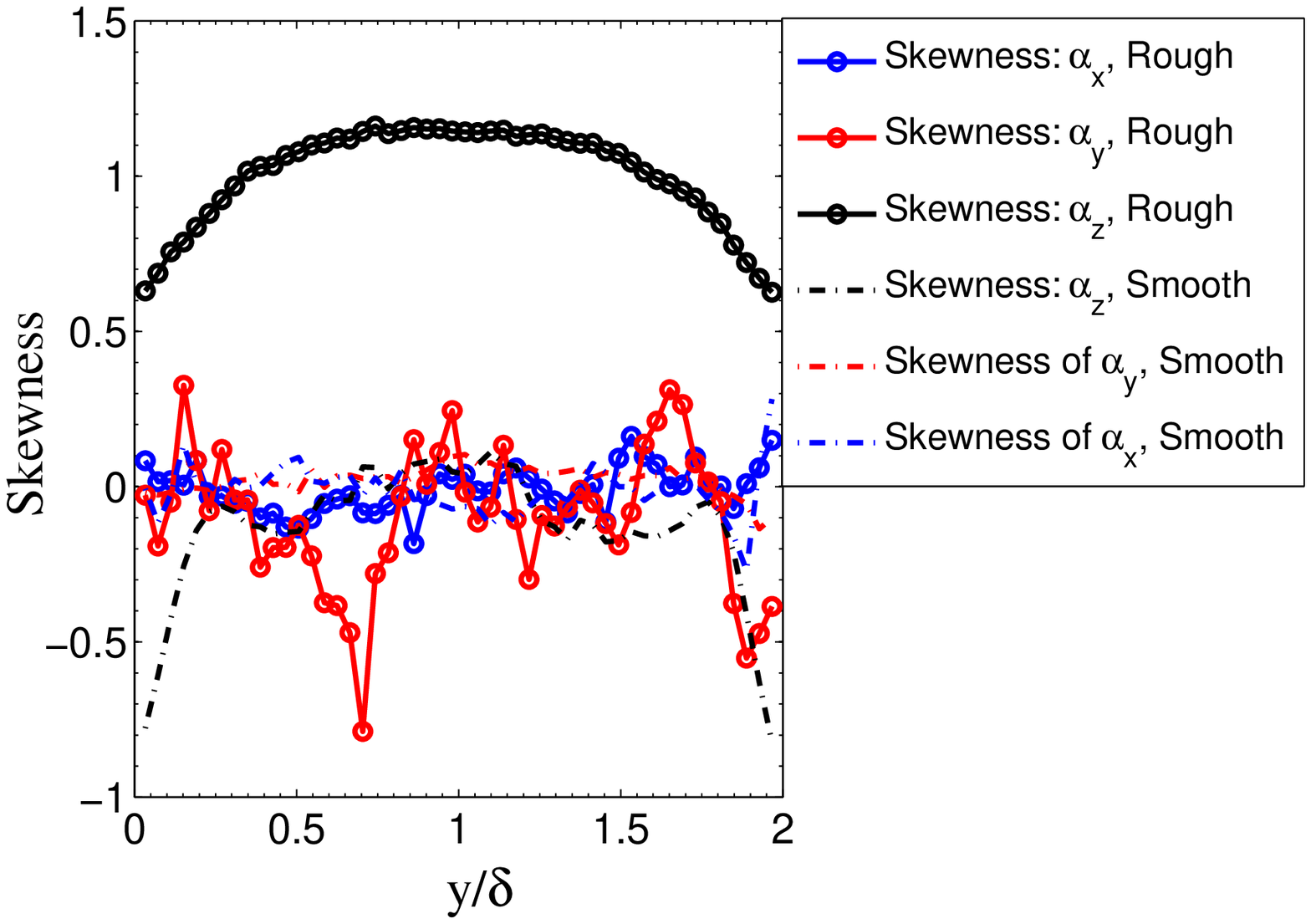}
			\caption{\label{rough_skewness_rot_acc}Effect of rough collisions on skewness of rotational acceleration}
			\includegraphics[width=1.0\linewidth, height= 6.5cm]{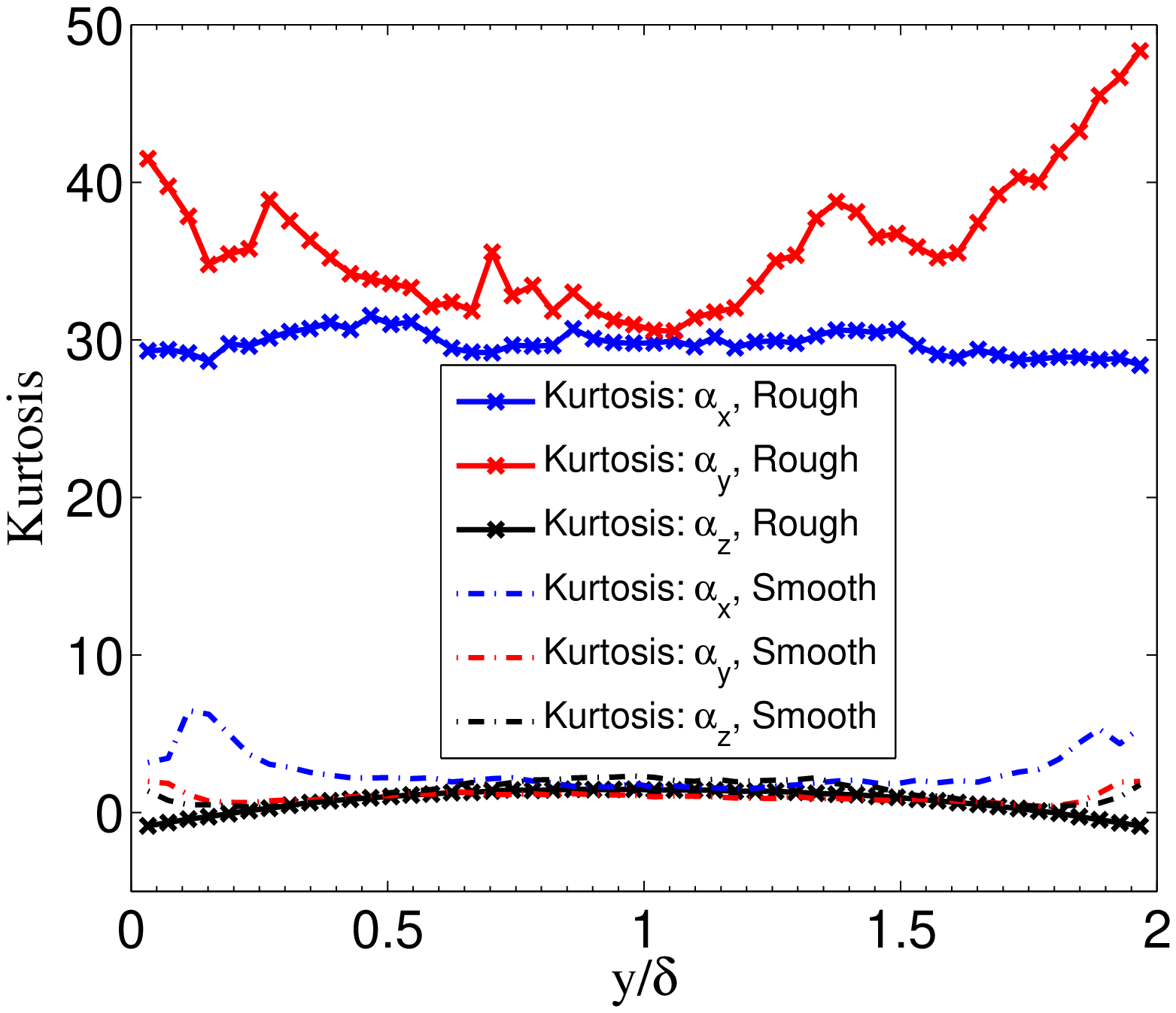}
			\caption{\label{rough_kurtosis_rot_acc}Effect of rough collisions on kurtosis of rotational acceleration}
		\end{figure}

\begin{figure}[!]
			\includegraphics[width=1.0\linewidth, height=5.5cm]{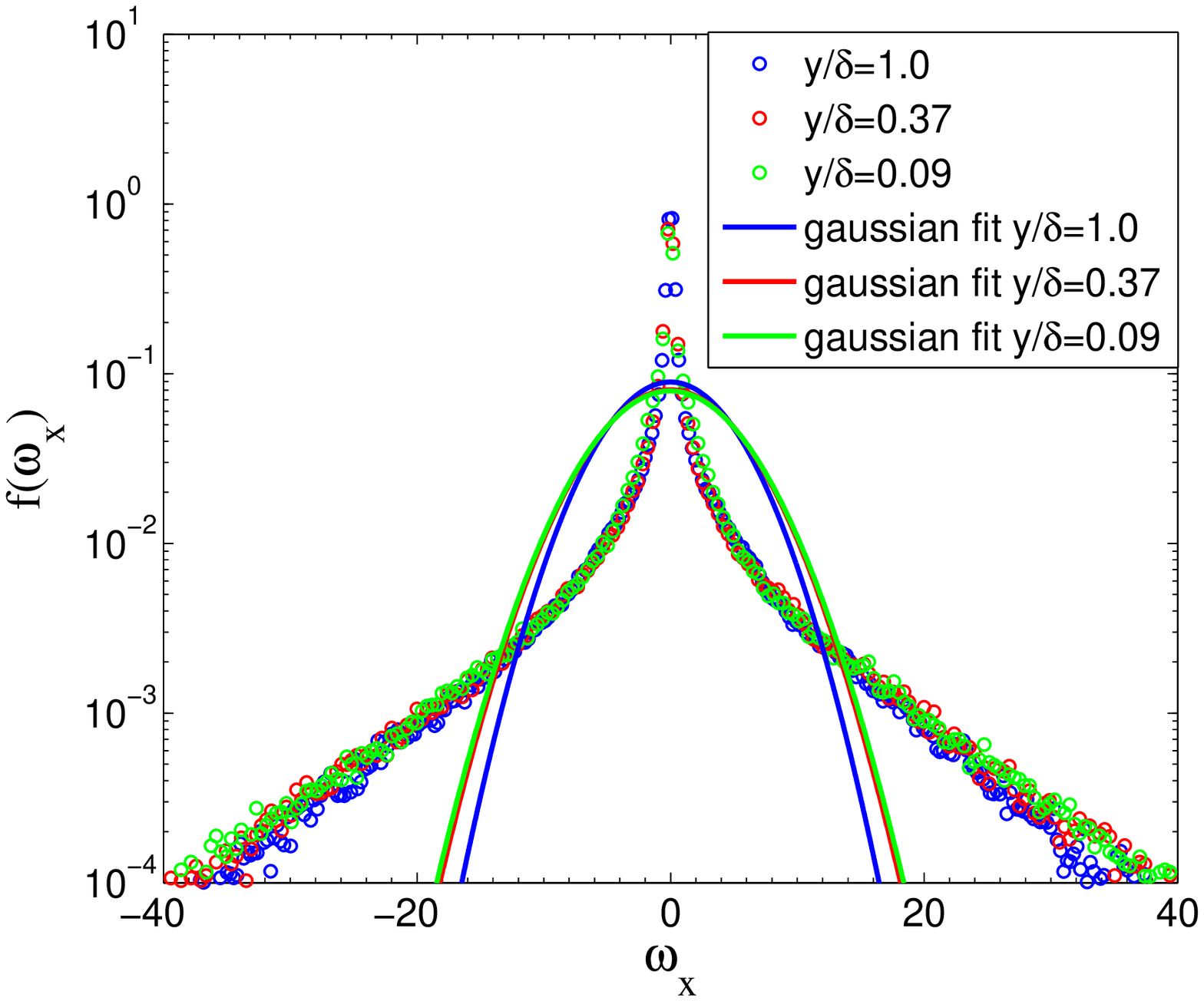}
			\caption*{(a)}
			\includegraphics[width=1.0\linewidth, height=5.5cm]{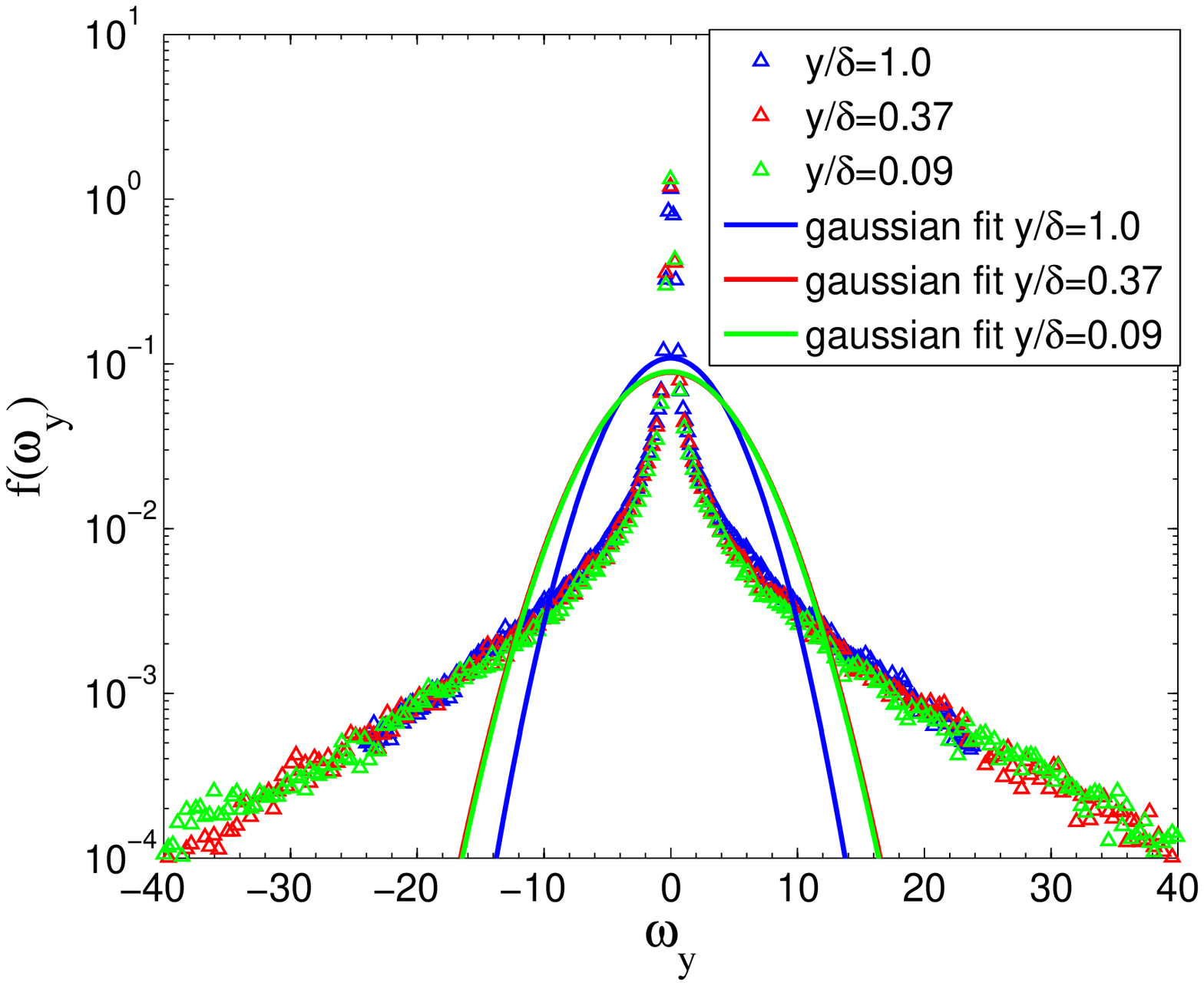}
			\caption*{(b)}
			\includegraphics[width=1.0\linewidth]{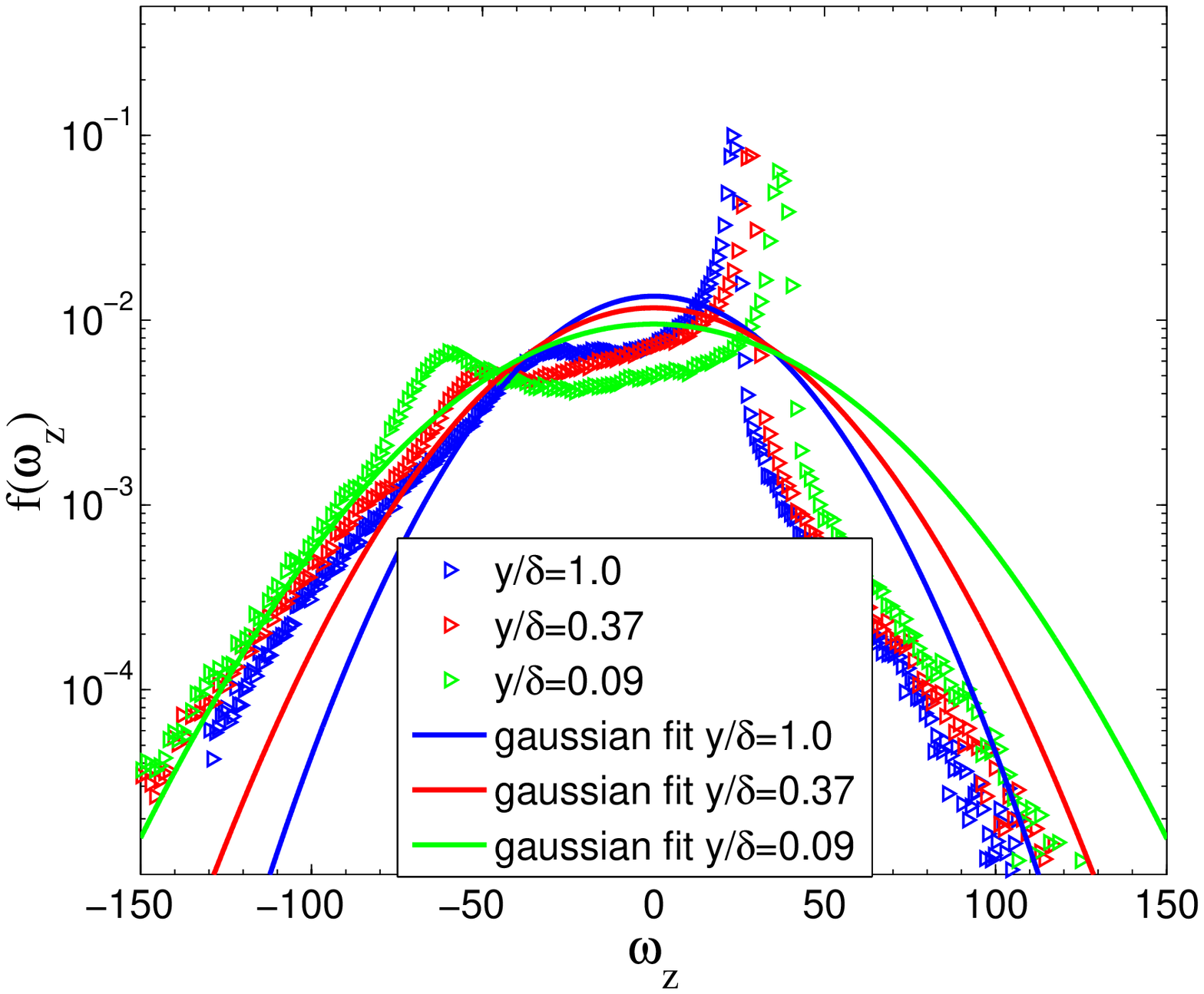}
			\caption*{(c)}
			\caption{Components of particle rotational velocity distributions (a) $f(\omega_x)$, (b) $f(\omega_y)$, and (c) $f(\omega_z)$ at different wall normal positions ($y/\delta$) of the channel for $\beta=+1.0$.}
						\label{rough_f(omega_i)}     
		\end{figure}

\begin{figure}[!]
			\centering
			\includegraphics[width=1.0\linewidth, height=6.5cm]{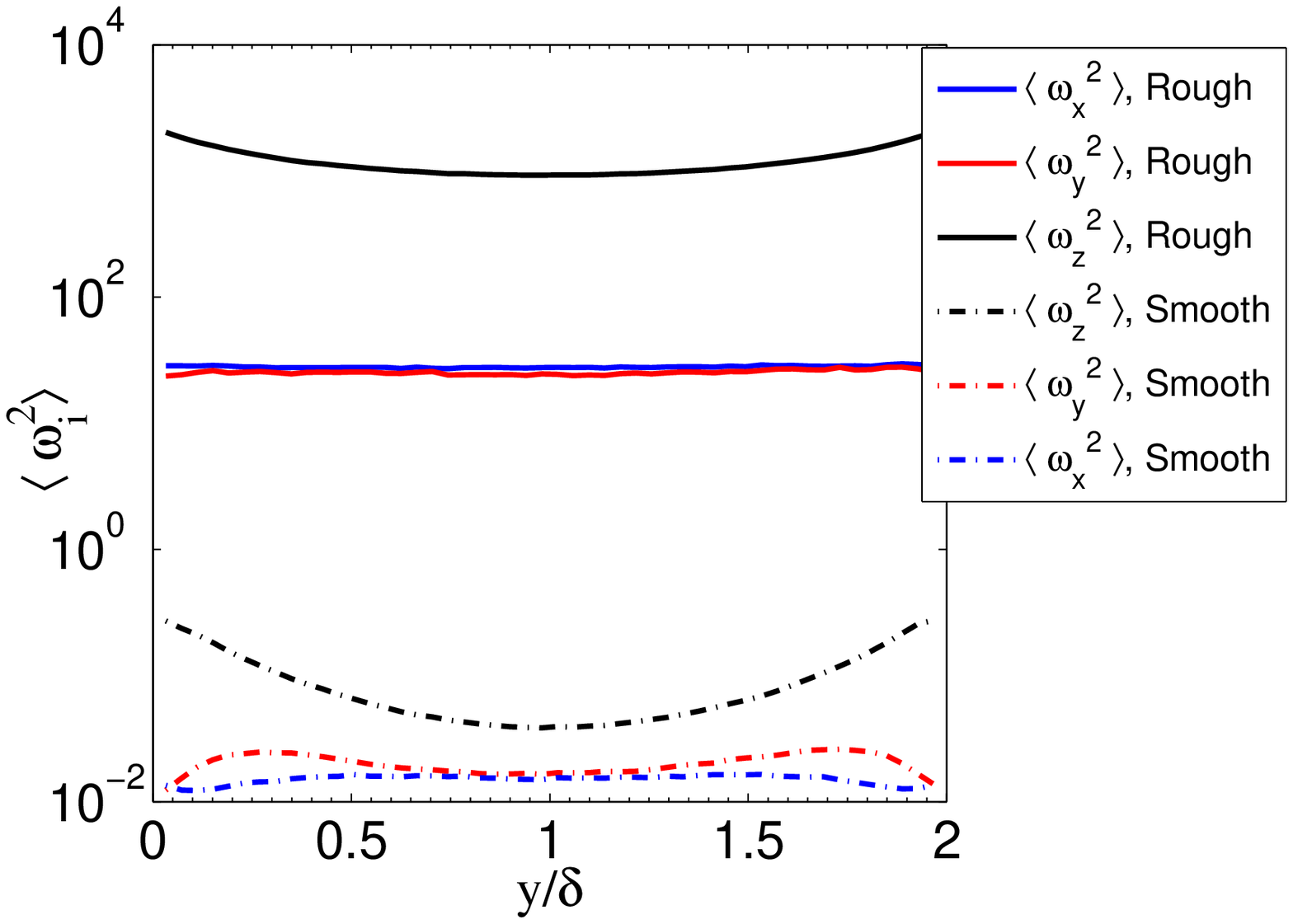}
			\caption{\label{rough_variance_rot_vel}Effect of rough collisions on variance of rotational velocity}
			\includegraphics[width=1.0\linewidth, height=6.5cm]{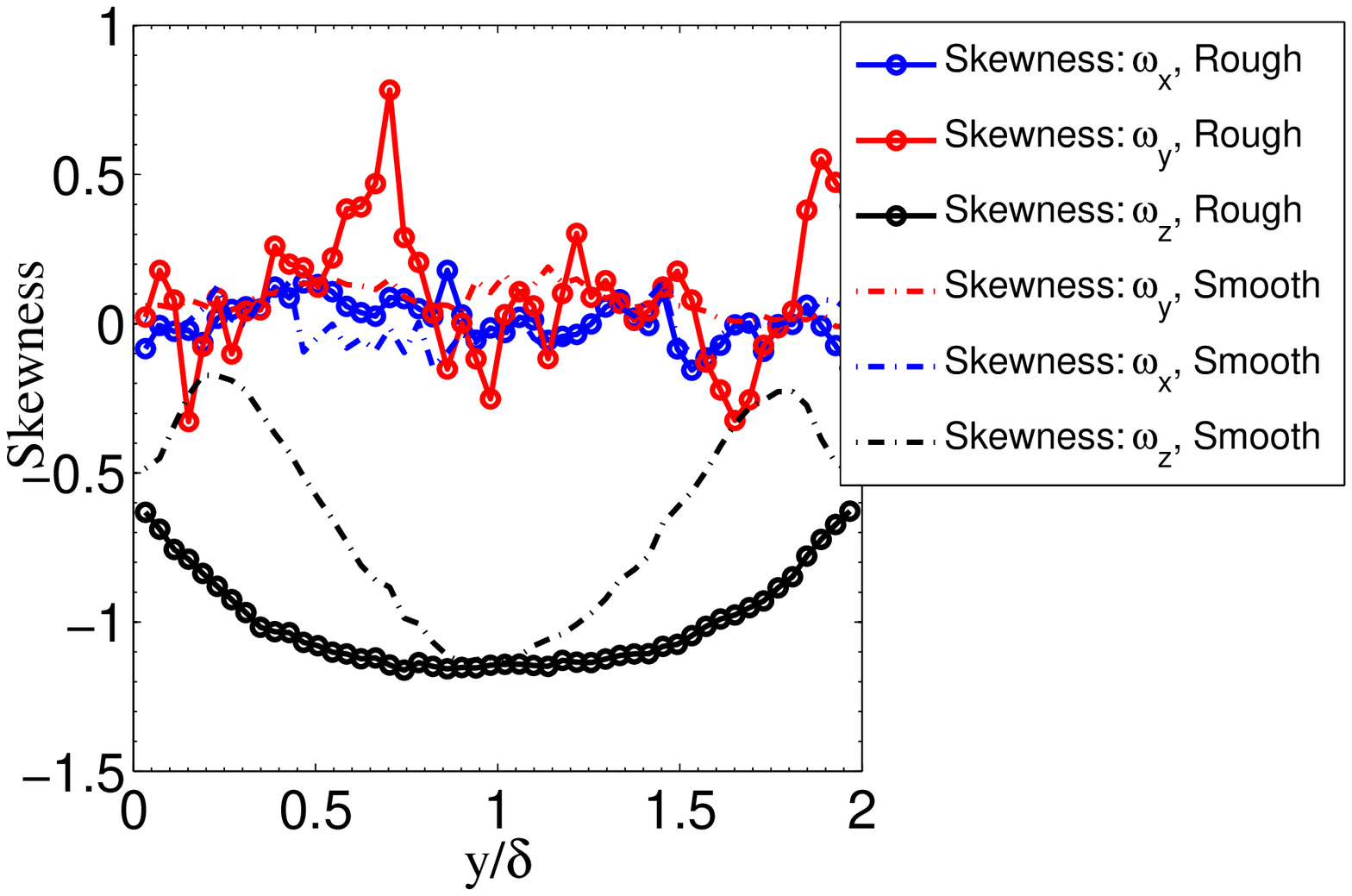}
			\caption{\label{rough_skewness_rot_vel}Effect of rough collisions on skewness of rotational velocity}
			\includegraphics[width=1.0\linewidth, height= 6.5cm]{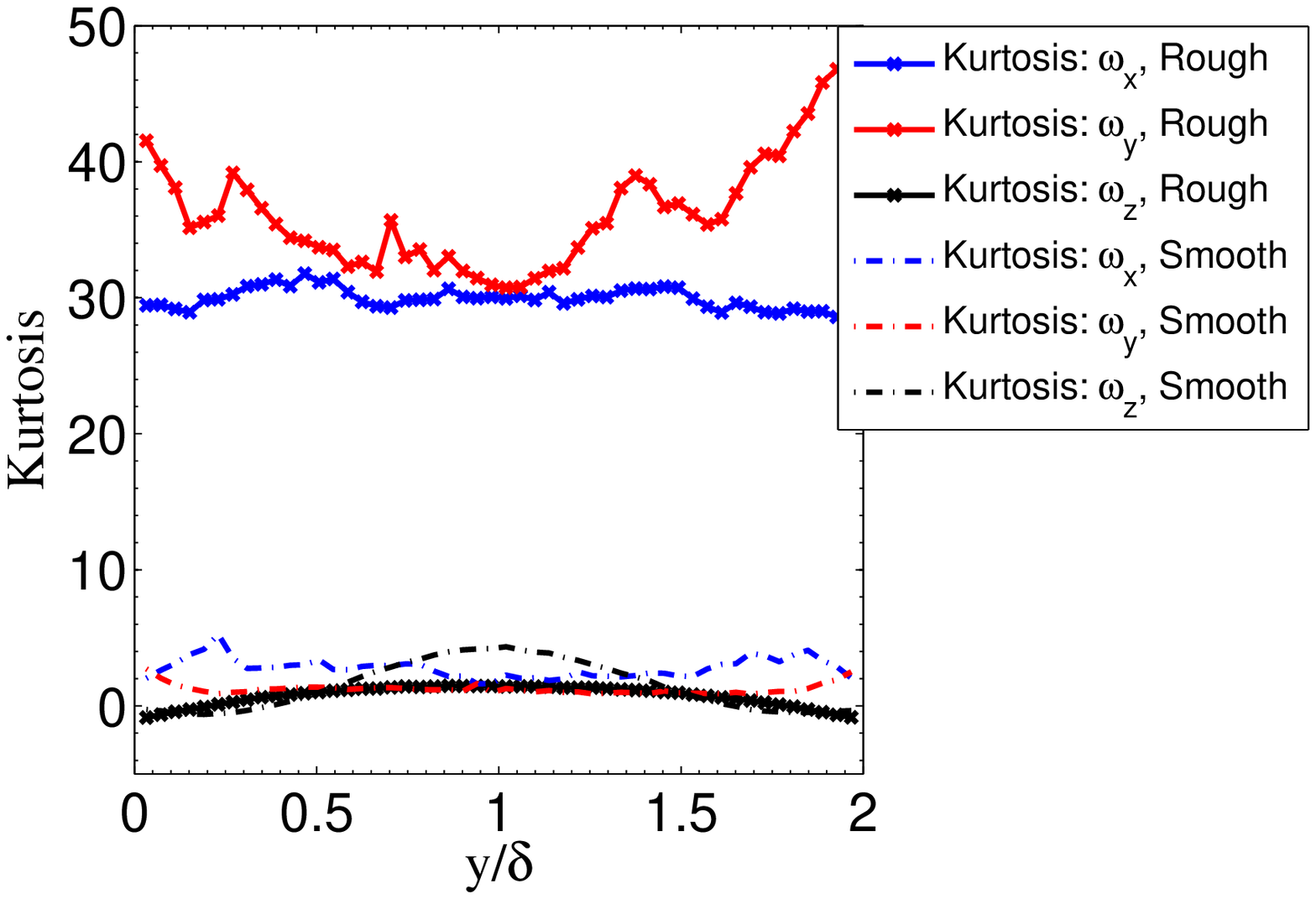}
			\caption{\label{rough_kurtosis_rot_vel}Effect of rough collisions on kurtosis of rotational velocity}
		\end{figure}

		Surface roughness of the particle directly affect the angular acceleration and velocity statistics. Due to rough collisions, the mean-angular velocity gets distributed in all the directions increasing the angular acceleration and 
		angular velocity fluctuations across the channel-width. Both the distribution functions, their skewness and flatness are presented in figures \ref{rough_f(alpha_i)} to \ref{rough_kurtosis_rot_vel}. 
		
		Angular acceleration and velocity distribution of stream wise and wall normal 
		components show a significant deviation from the Gaussian distribution with 
		slowly decaying tail for all the wall normal positions (figures \ref{rough_f(alpha_i)} (a), (b), \ref{rough_f(omega_i)} (a), and (b)). Appearance of the long tail in the distribution functions 
		is also reflected by the higher kurtosis values in figures \ref{rough_kurtosis_rot_acc} and \ref{rough_kurtosis_rot_vel}.

	The tails of $f(\alpha_z)$ and $f(\omega_z)$ deviates from the corresponding Gaussian fits. Unlike smooth collisions, the high probability low magnitude zone of the fluctuations qualitatively differ from Gaussian nature. All these distributions have single sharp peaks with non-zero value accompanied with a valley about zero magnitude as shown in figures \ref{rough_f(alpha_i)} (c) and \ref{rough_f(omega_i)} (c). 
	In case of $f(\alpha_z)$, the peaks appear  at negative fluctuations with a probability almost one order of magnitude higher than the corresponding Gaussian peaks. There are also a longer positive tails at positive values of fluctuations.
	
	Plots of $f(\omega_z)$, show the peaks at positive fluctuations accompanied with longer negative tails. This is due to the fact that, in particle equation of motion net particle acceleration fluctuation and particle acceleration fluctuation originates due to particle rotational velocity fluctuation have opposite signs. Skewness of all the 
	distribution functions are shown figures \ref{rough_skewness_rot_acc} and \ref{rough_skewness_rot_vel} at different wall normal location of the channel.
	Skewness is mainly observed in case of principal component of angular acceleration and 
	angular velocity fluctuations.	
	The  increase  in  fluctuations  due  to  particle surface roughness  is  observed  through  the  increased variance of rotational acceleration and rotational velocities  as shown in figures \ref{rough_variance_rot_acc} and \ref{rough_variance_rot_vel}.  This increase in variance for z-component is approximately four orders of magnitude but for the other two cases these are about two orders of magnitude higher than those originate from smooth particle-particle and particle-wall interactions. 

	
	 The  asymmetry and  sharp peak in the distribution function at the positive angular velocity fluctuation (figure \ref{rough_f(omega_i)}) (c) happens as an outcome of the rough collisions between wall and the particles. Peak at the low positive fluctuation appears due to cross stream particle migration from the zone of lower angular velocity whereas negative fluctuations in the p.d.f. are mainly contributed by the particles migrates to the zone of interest after colliding with the rough wall. Distribution function for angular acceleration shows an exactly opposite trend to that of the angular velocity distribution function, which is shown in figure \ref{rough_f(alpha_i)} (c). To summarize, roughness plays an important role on the 
	 variance of the particle linear and angular velocity fluctuations. Tail of the distribution functions decays more slowly in presence of rough interaction. It has 
	 been observed that in presence of roughness, frequency of inter-particle collision is 1.7 times  and wall-particle collision is 2.5 times more than that occurs in the absence of roughness.

		\section{Modelling  fluctuating torque on the particle}
		\label{section_rot_acc}
\begin{figure}[!]
			\centering
			\includegraphics[width=1.0\linewidth]{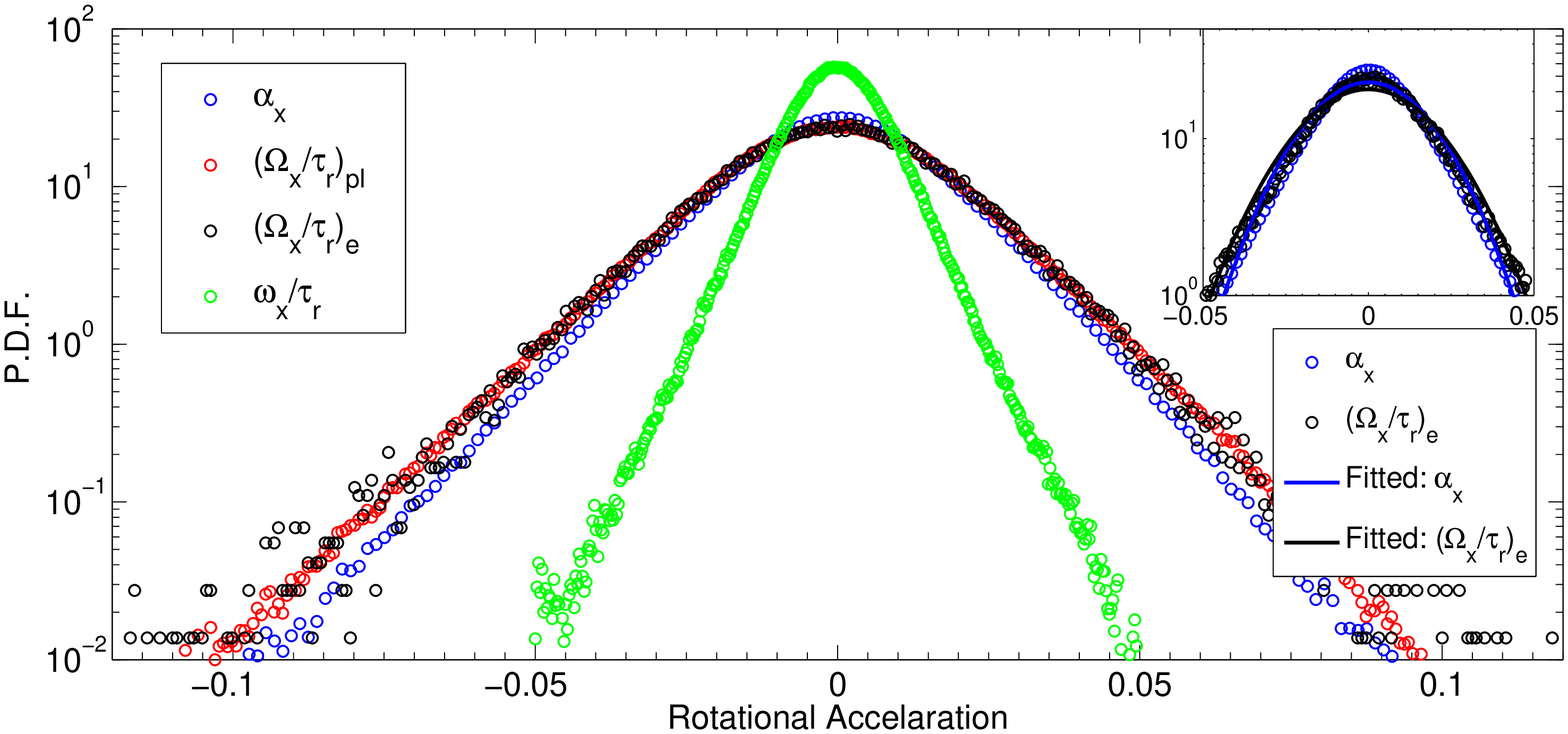}
			\caption*{(a)}
			\includegraphics[width=1.0\linewidth]{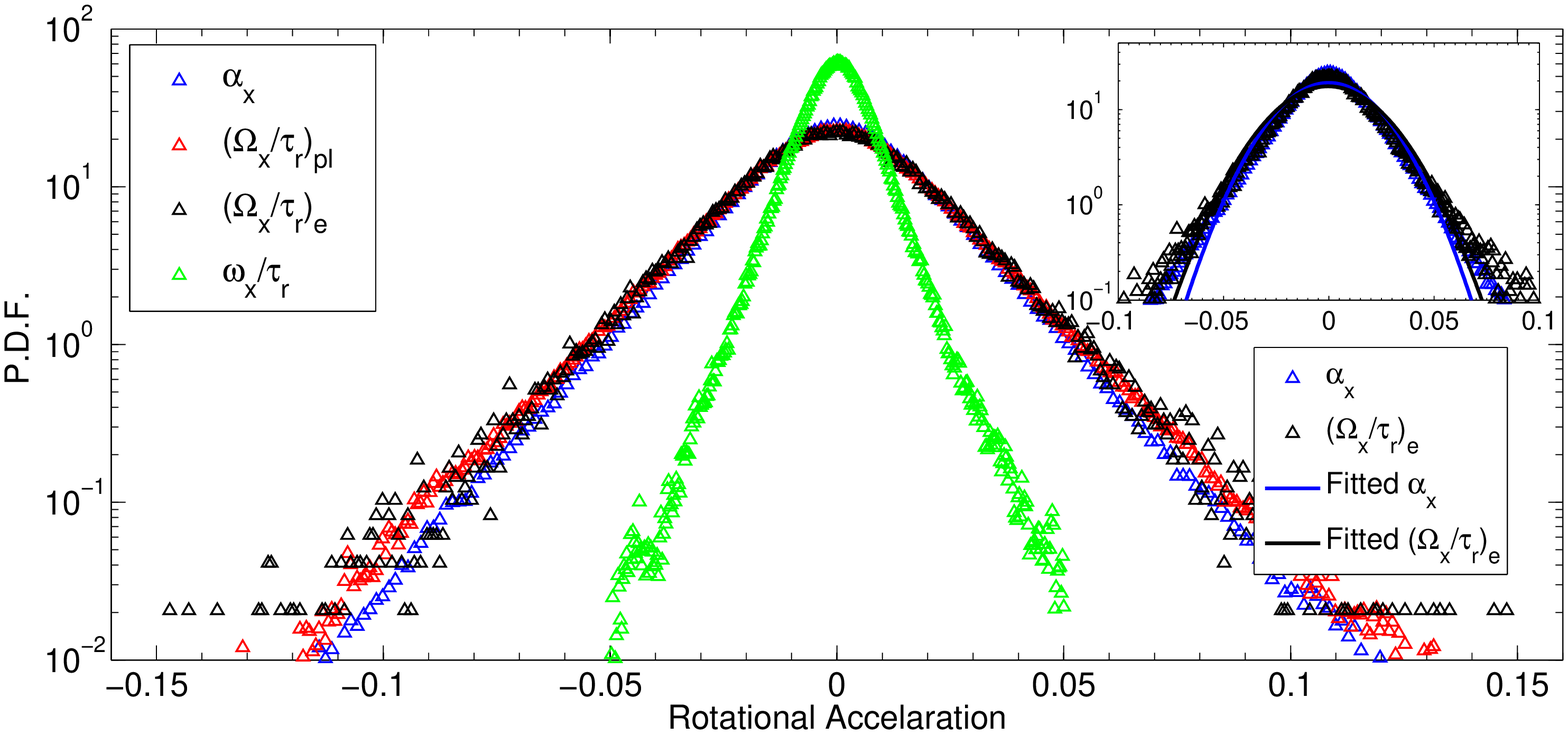}
			\caption*{(b)}
			\includegraphics[width=1.0\linewidth]{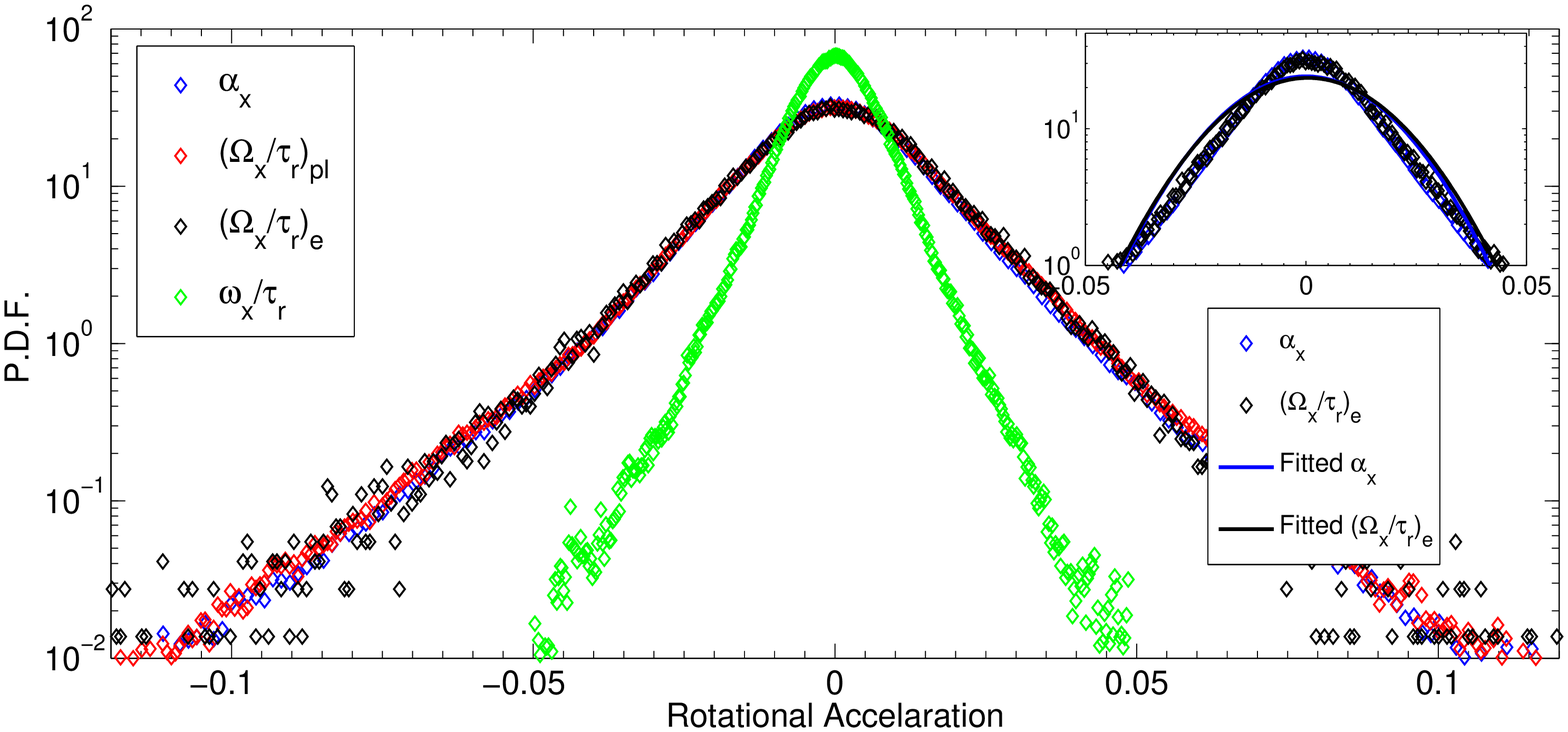}
			\caption*{(c)}
			\caption{Rotational acceleration distribution of x-component ($f(\alpha_x)$) at three different positions in Couette:
			 (a) $y/\delta=1.0$, (b) $y/\delta=0.37$ and (c)  $y/\delta=0.09$}
			\label{rot_acc_x}
		\end{figure}

\begin{figure}[!]
			\centering
			\includegraphics[width=1.0\linewidth]{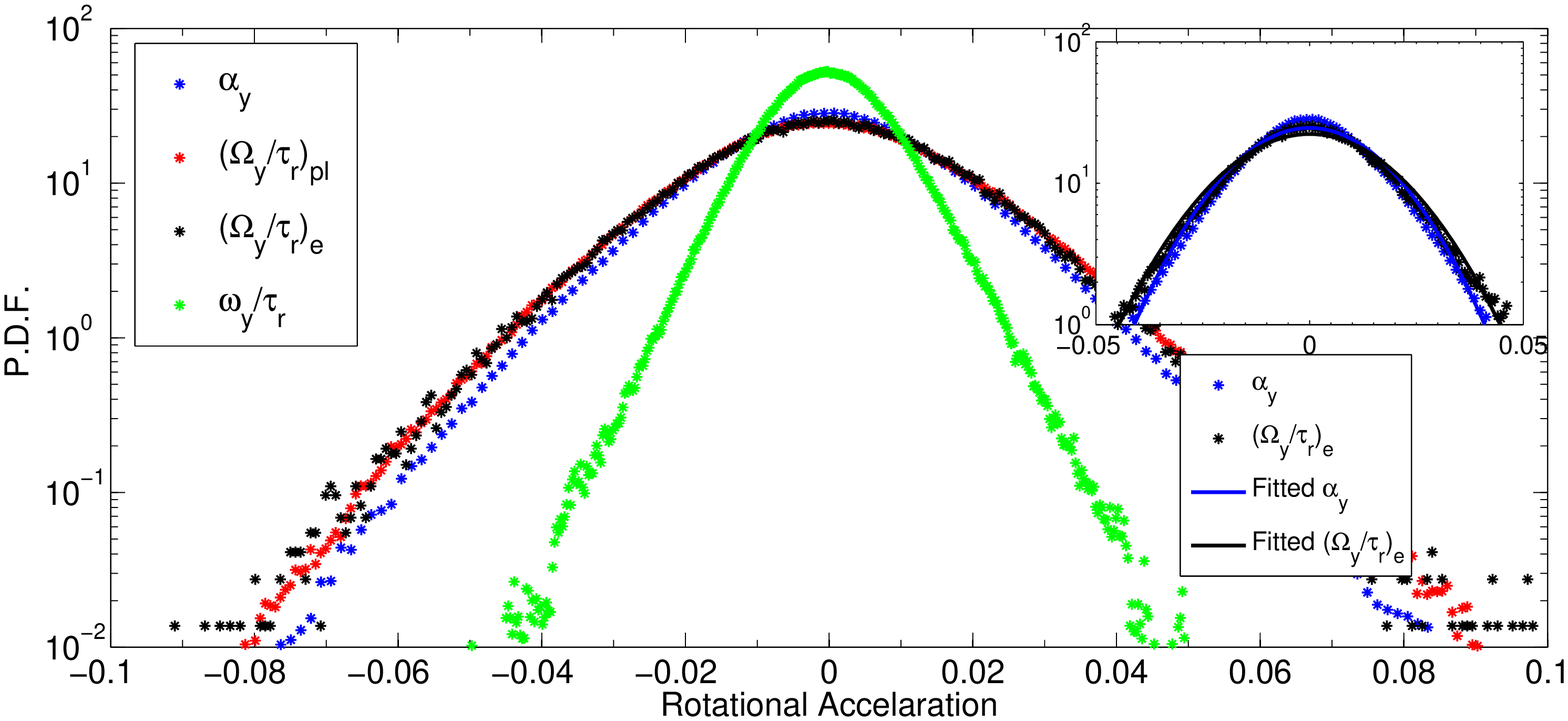}
			\caption*{(a)}
			\includegraphics[width=1.0\linewidth]{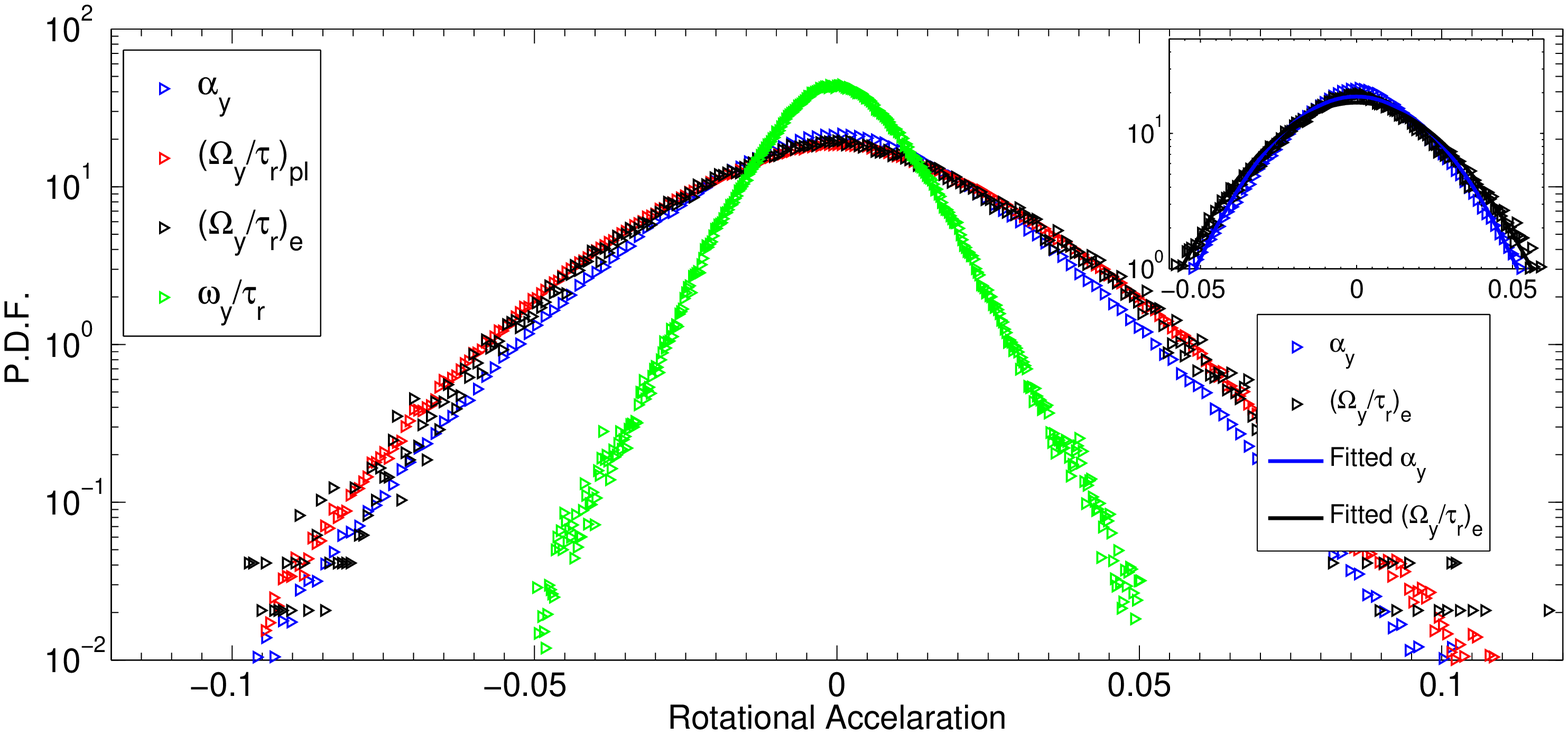}
			\caption*{(b)}
			\includegraphics[width=1.0\linewidth]{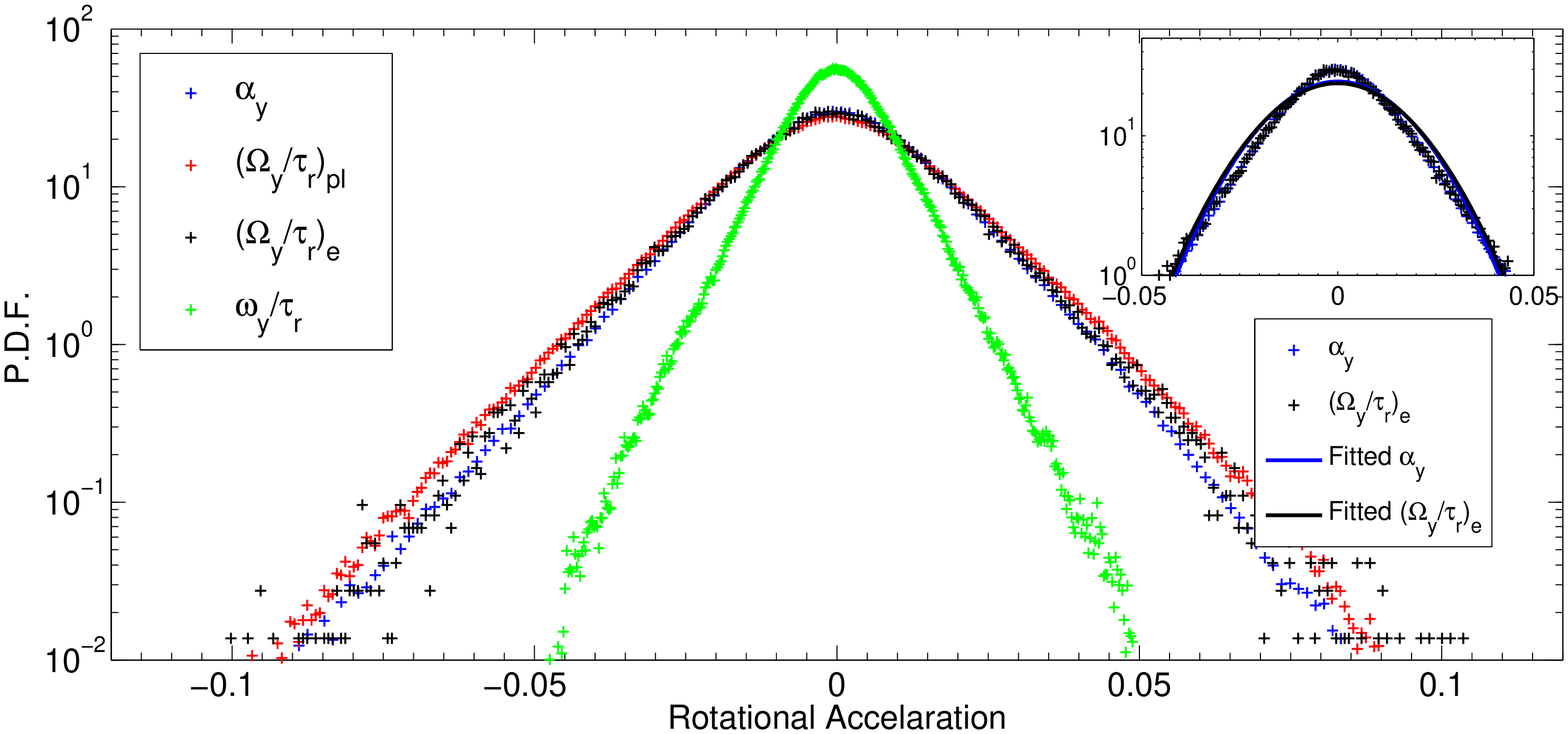}
			\caption*{(c)}
			\caption{Rotational acceleration distribution of y component ($f(\alpha_y)$) at three different positions in the Couette:
			 (a) $y/\delta=1.0$, (b) $y/\delta=0.37$ and (c)  $y/\delta=0.09$}
			\label{rot_acc_y}
		\end{figure}

\begin{figure}[!]
			\centering
			\includegraphics[width=1.0\linewidth]{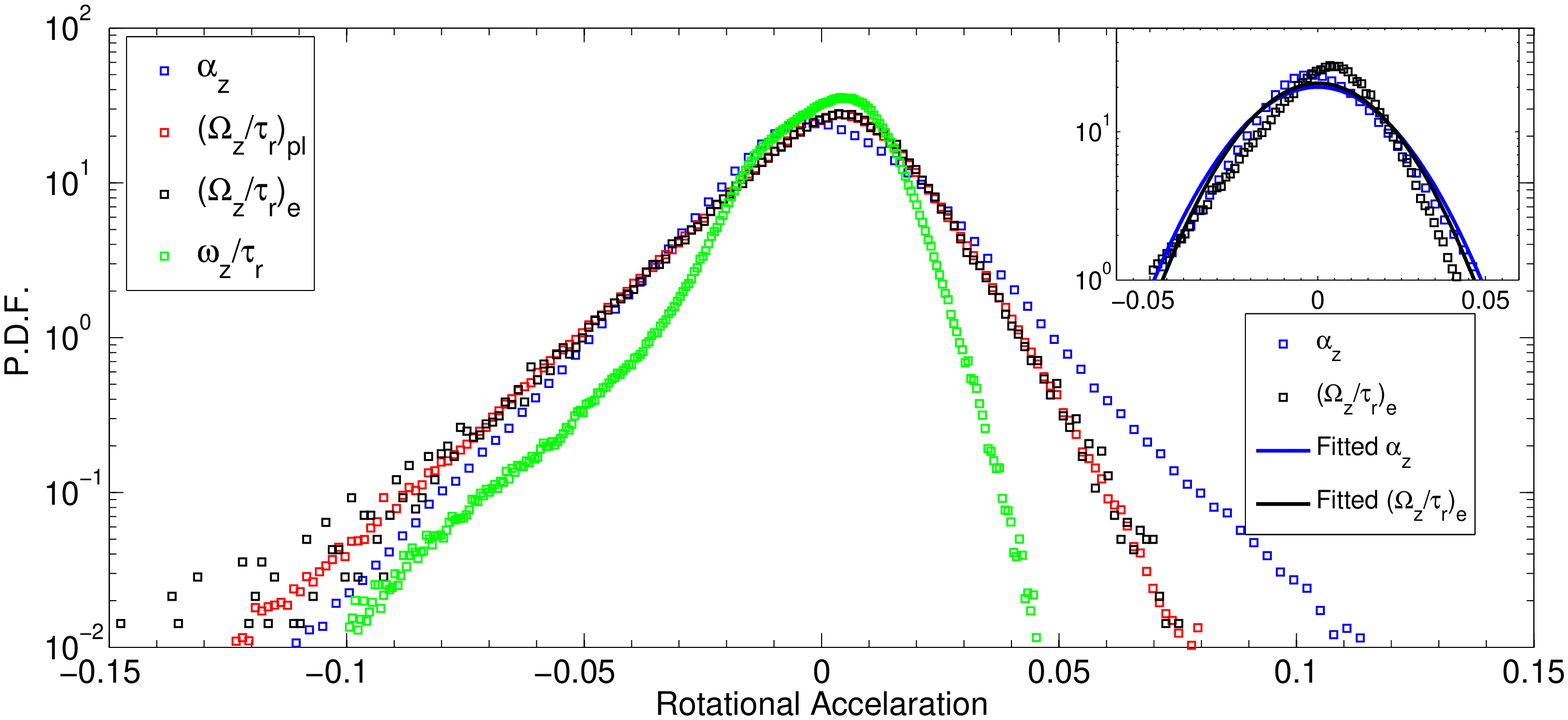}
			\caption*{(a)}
			\includegraphics[width=1.0\linewidth]{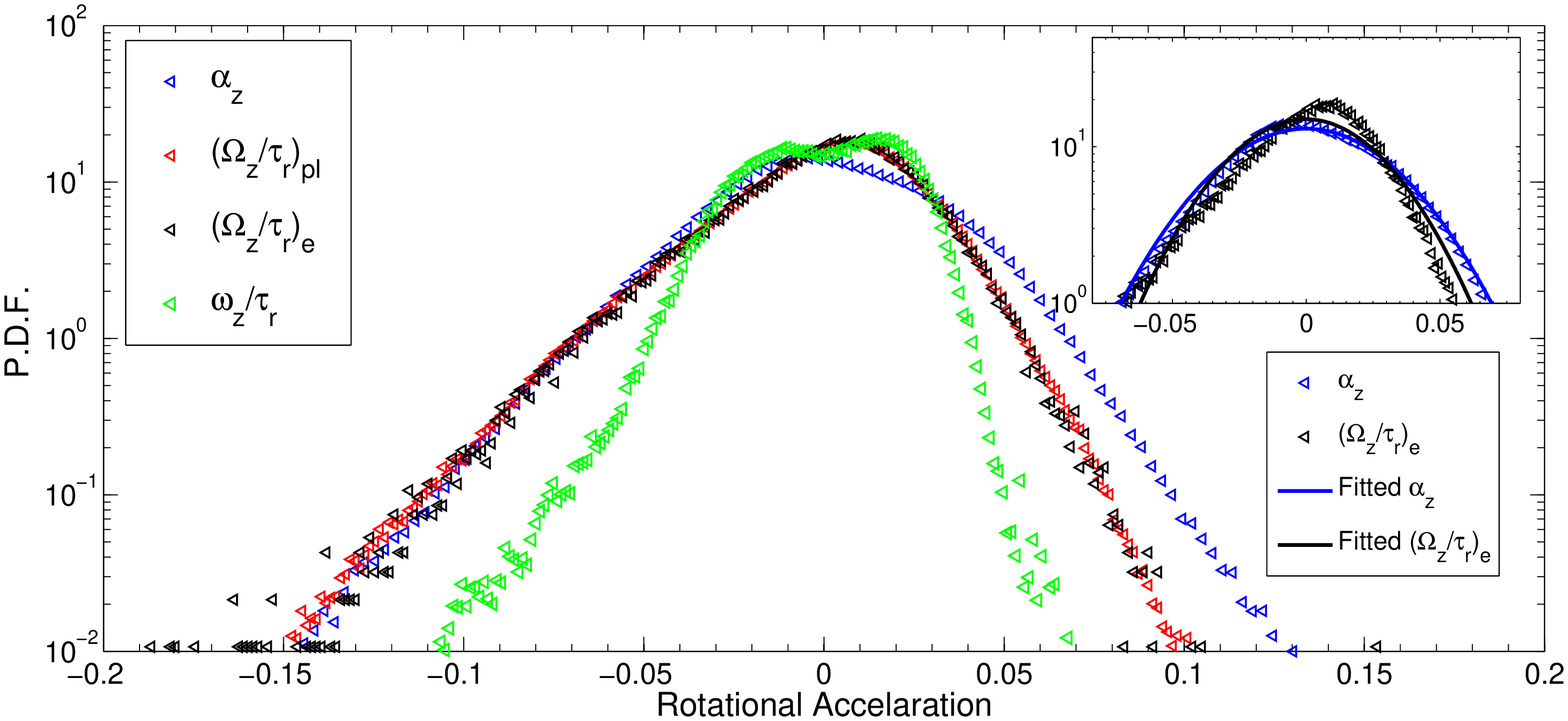}
			\caption*{(b)}
			\includegraphics[width=1.0\linewidth]{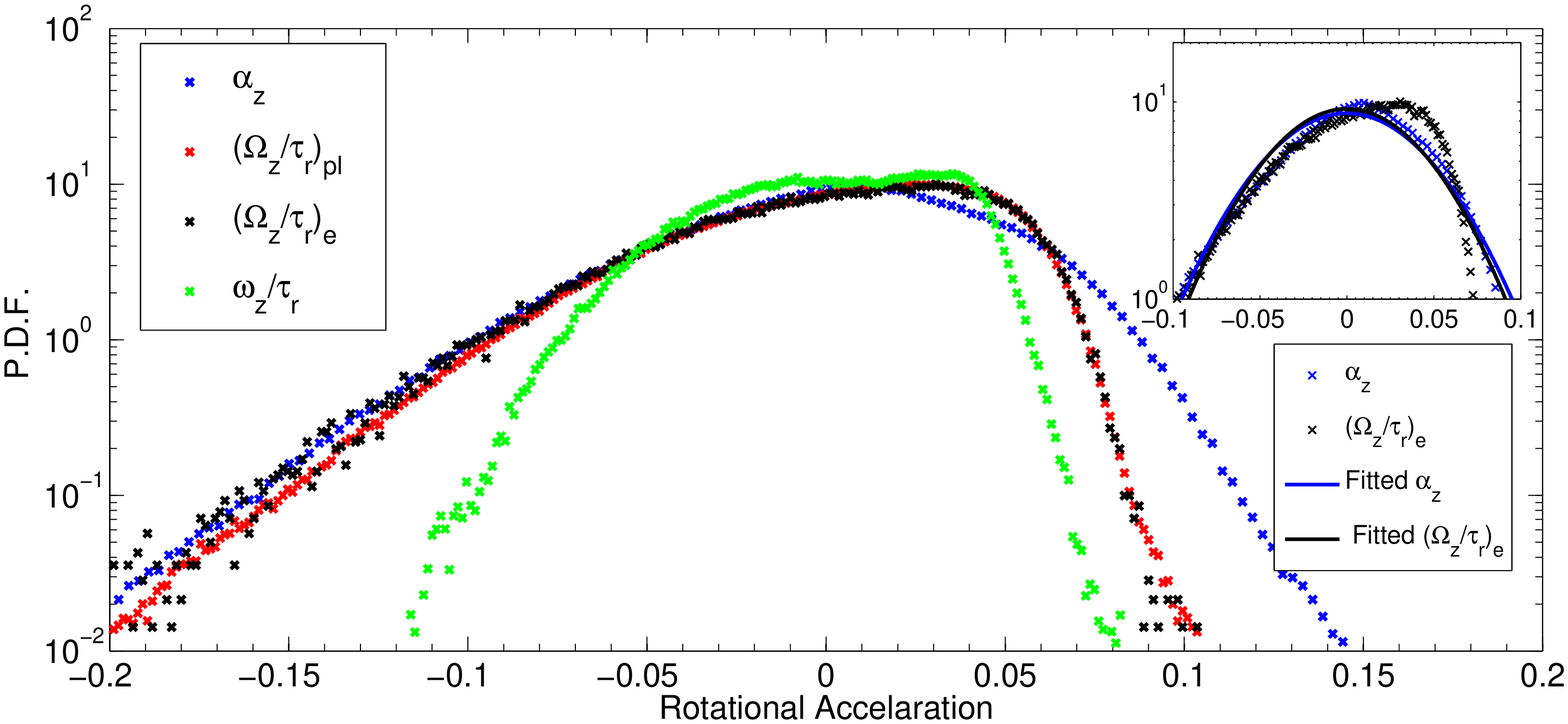}
			\caption*{(c)}
			\caption{Rotational acceleration distribution of z-component ($f(\alpha_z)$) at three different positions in the Couette:
			 (a) $y/\delta=1.0$, (b) $y/\delta=0.37$ and (c)  $y/\delta=0.09$}
			\label{rot_acc_z}
		\end{figure}

		The particle phase equation of angular momentum can be represented by the equation \ref{rot_acc_eq}. The net angular acceleration acting on a particle $\alpha_i$ can be equated with particle rotational drag that arises due to the net slip between  particle angular velocity $\omega_i$ and  local fluid angular velocity, given by, $\varOmega_i=-\frac{1}{2}\epsilon_{ijk}\frac{\partial}{\partial x_j}u_k$.  Local fluid angular velocity is computed at the particle center of mass position, $x_{pi}$.
		On decomposition of the rotational field variables in to a mean and a fluctuating part denoted by over-bar ($\overline{.}$) and ($.\prime$) respectively, we arrive at the expression where the fluctuating angular acceleration of the particle $\alpha_i'$ is expressed as a linear combination of fluid and particle rotational velocities $\varOmega_i'$ and $\omega_i'$ respectively.     
		\begin{equation}\label{rot_acc_eq}
		\begin{split}
		\alpha_i&=\frac{\varOmega_i(x_{pi})-\omega_i}{\tau_r}\\
		\overline{\alpha_i}+\alpha_i'&= \frac{\overline{\varOmega_i}(x_{pi})-\overline{\omega_i}}{\tau_r}+\frac{\varOmega_i'(x_{pi})-\omega_i'}{\tau_r} \\
		\alpha_i'&=\frac{\varOmega_i'(x_{pi})-\omega_i'}{\tau_r} \\
		\alpha_i'&=\frac{\varOmega_i'(x_{pi})}{\tau_r}-\frac{\omega_i'}{\tau_r}
		\end{split}
		\end{equation}
		Here, $\tau_r$ represents the rotational viscous relaxation time of the particle. This section contains the comparative study of the particle rotational acceleration distribution $f(\alpha_i)$ and the particle acceleration component arising from (i) only fluctuating fluid angular velocity $(\varOmega_i/\tau_r)_{pl}$ computed in the particle- Largrangian frame and (ii) only fluctuating particle angular velocity $\omega_i/\tau_r$  in three different directions at different positions of channel-width $y/\delta$. All the  results are shown in figures \ref{rot_acc_x} (a) to \ref{rot_acc_z} (c). In this comparison the fluid rotational acceleration  $(\varOmega_i/\tau_r)_{e}$ computed in the fluid Eulerian grids are also shown. Hereafter, in all the figures (\ref{rot_acc_x} (a) to \ref{rot_acc_z} (c)), we have omitted the $\prime$ symbol from all the fluctuating variables.

		From figures \ref{rot_acc_x} (a) to \ref{rot_acc_z} (c)  it is observed 
		that the distribution function originates from the particle angular velocity fluctuation  ($f(\omega_i/\tau_r)$) does not resemble the distribution function 
		of net particle angular acceleration fluctuation ($f(\alpha_i)$). On the other hand, the distribution function of rotational accelerations arising out of fluid rotational velocity fluctuations computed in particle-Lagrangian frame $f(\varOmega_i/\tau_r)_{pl}$ quite resemble the net particle rotational acceleration distribution functions $f(\alpha_i)$ in all three directions. However, 
		in case of acceleration fluctuation in z-direction, we observe a deviation at positive values of the fluctuations. In that zone pdf of particle acceleration fluctuation decays at a slower rate compared to the fluctuations originate from the fluid angular velocity fluctuations. 
		The probability of particle acceleration originates from the fluid rotational angular velocity  distribution computed in the Eulerian (fluid) Grids $f(\varOmega_i/\tau_r)_{e}$ is similar in nature to $f(\varOmega_i/\tau_r)_{pl}$ and deviate only in the tail region.
		Gaussian fitting of $f(\alpha_i)$ and $f(\varOmega_i/\tau_r)_{e}$ is performed and is shown up to one and a half to two decades lower from the peak frequency in the inset of each of the plots. The Guassian functions fit quite well for $f(\alpha_x)$, $f(\alpha_y)$, $f(\varOmega_x/\tau_r)_{e}$ and $f(\varOmega_y/\tau_r)_{e}$. The Gaussian fits deviate marginally along the z direction especially in case of $f(\varOmega_z/\tau_r)_{e}$. A detailed quantification can be found in \ref{JSD} table \ref{table:4}.

		
		\subsection{The Jensen-Shannon divergence based similarity assessment}\label{JSD}
		Jensen-Shannon Divergence (JSD) is a method of quantifying the similarity between two probability distribution functions. JSD is a modified symmetrized  version of Kullback-Liebler (KL) divergence. The KL divergence quantifies the difference between two distribution functions and  is expressed in terms of the relative Shannon-entropy of the two distributions \cite{lin1991divergence}. 
		Let $P(x)$ and $Q(x)$ be two continuous probability distribution functions of the random variable x in probability space $\chi$.
		The Kullback-Liebler divergence of Q with respect to P is represented as $D(P||Q)$ is given by\\
		\begin{equation}
		D(P||Q)=\int_{-\infty}^{\infty}P(x)log_{2}\left(\frac{P(x)}{Q(x)}\right)dx 
		\label{KL}
		\end{equation}
		Jensen-Shannon Divergence is obtained by symmetrized and smoothened KL divergence as follows:
		\begin{equation}
		JSD(P||Q)=\frac{1}{2}D(P||M)+\frac{1}{2}D(Q||M)
		\end{equation}
		where $M=\frac{1}{2}(P+Q)$
		When 2 is used as the base of logarithm in KL divergence, the upper bound of JSD for two probabilities become 1. Hence, 
		$0\leq JSD(P||Q) \leq 1$. When two distributions are identical, $JSD(P||Q)$ value becomes 0.  
		\begin{table}[h!]
			\centering
			\begin{tabular}{||c | c | c | c ||} 
				\hline
				Position & $f(\frac{\varOmega_x}{\tau_r})_{pl}$ & $f(\frac{\varOmega_x}{\tau_r})_{e}$ & $f(\frac{\omega_x}{\tau_r})$ \\ [0.5ex] 
				\hline\hline
				$y/\delta\approx$0.09 & 0.0005 & 0.0013 & 0.1080 \\ 
				$y/\delta\approx$0.37 & 0.0016 & 0.0023 & 0.161 \\
				$y/\delta\approx$1.0 & 0.0034 & 0.0035 & 0.1173 \\ [1ex] 
				\hline
			\end{tabular}
			\caption{Jensen-Shannon divergence of the rotational acceleration components $f(\frac{\varOmega_x}{\tau_r})_{pl}$,$f(\frac{\varOmega_x}{\tau_r})_{e}$ and $f(\frac{\omega_x}{\tau_r})$, computed with respect to net particle rotational acceleration $f(\alpha_x)$ }
			\label{table:1}
		\end{table}
		\begin{table}[h!]
			\centering
			\begin{tabular}{||c | c | c | c ||} 
				\hline
				Position & $f(\frac{\varOmega_y}{\tau_r})_{pl}$ & $f(\frac{\varOmega_y}{\tau_r})_{e}$ & $f(\frac{\omega_y}{\tau_r})$ \\ [0.5ex] 
				\hline\hline
				$y/\delta\approx$0.09 & 0.002 & 0.0005 & 0.082 \\ 
				$y/\delta\approx$0.37 & 0.005 & 0.004 & 0.1214 \\
				$y/\delta\approx$1.0 & 0.0038 & 0.0032 & 0.0944 \\ [1ex] 
				\hline
			\end{tabular}
			\caption{Jensen-Shannon divergence of the rotational acceleration components $f(\frac{\varOmega_y}{\tau_r})_{pl}$,$f(\frac{\varOmega_y}{\tau_r})_{e}$ and $f(\frac{\omega_y}{\tau_r})$, computed with respect to net particle rotational acceleration $f(\alpha_y)$ }
			\label{table:2}
		\end{table}
		\begin{table}[h]
			\centering
			\begin{tabular}{||c | c | c | c ||} 
				\hline
				Position & $f(\frac{\varOmega_z}{\tau_r})_{pl}$ & $f(\frac{\varOmega_z}{\tau_r})_{e}$ & $f(\frac{\omega_z}{\tau_r})$ \\ [0.5ex] 
				\hline\hline
				$y/\delta\approx$0.09 & 0.0158 & 0.0159 & 0.0674 \\ 
				$y/\delta\approx$0.37 & 0.0163 & 0.0189 & 0.0691 \\
				$y/\delta\approx$1.0 & 0.0105 & 0.0114 & 0.0573 \\ [1ex] 
				\hline
			\end{tabular}
			\caption{Jensen-Shannon divergence of the rotational acceleration components $f(\frac{\varOmega_z}{\tau_r})_{pl}$,$f(\frac{\varOmega_z}{\tau_r})_{e}$ and $f(\frac{\omega_z}{\tau_r})$, computed with respect to net particle rotational acceleration $f(\alpha_z)$ }
			\label{table:3}
		\end{table}
		\\From tables \ref{table:1}, \ref{table:2} and \ref{table:3} it is clear that the extent of similarity of $f(\alpha_i)$ to $f(\frac{\omega_i}{\tau_r})$ is quite lesser compared to that of $f(\frac{\varOmega_i}{\tau_r})_{pl}$ and $f(\frac{\varOmega_i}{\tau_r})_{e}$. In general $f(\alpha_i)$ is more similar to the fluid angular velocity distribution at particle Lagrangian point. But it is not feasible to assess the particle Lagrangian position during modelling the system. Therefore we have also computed JSD for the pdf of fluid angular velocity fluctuation at fluid Eulerian grid points. It is observed that the JSD values for $f(\frac{\varOmega_i}{\tau_r})_{e}$ is ~ $10^{-2}$ to $10^{-3}$. Therefore $f(\alpha_i)$ can be well approximated as $f(\frac{\varOmega_i}{\tau_r})_{e}$.
		\\ Following the above discussion and our observation that principal component of fluid vorticity ($\Omega_z$) is non-Gaussian in nature, it is tempting to perform a JSD analysis to comment on the probability of particle angular acceleration from the fluid vorticity as a Gaussian distribution function.
		\begin{table}[h!]
			\centering
			\begin{tabular}{||c | c | c | c ||} 
				\hline
				Direction & Position & $JSD(f(\alpha_i)||f(\frac{\varOmega_i}{\tau_r})_{e})$ & $JSD(f(\alpha_i)||g({\frac{\varOmega_i}{\tau_r}})_{e}))$ \\ [0.5ex] 
				\hline\hline
				\multirow{3}{*}{x} &$y/\delta\approx$0.09 & 0.0013 & 0.0191 \\ 
				&$y/\delta\approx$0.37 & 0.0023 & 0.0145\\
				&$y/\delta\approx$1.0  & 0.0035 & 0.0115 \\ [1ex]  \hline
				\multirow{3}{*}{y} &$y/\delta\approx$0.09 & 0.0005 & 0.0082 \\ 
				&$y/\delta\approx$0.37 & 0.0040 & 0.0068 \\
				&$y/\delta\approx$1.0  & 0.0032 & 0.0080 \\ [1ex]  \hline
				\multirow{3}{*}{z} &$y/\delta\approx$0.09 & 0.0159 & 0.0072 \\ 
				&$y/\delta\approx$0.37 & 0.0189 & 0.0071 \\
				&$y/\delta\approx$1.0  & 0.0114 & 0.0074 \\ [1ex] \hline
			\end{tabular}
			\caption{Jensen-Shannon Divergence of the Gaussian Fitted Fluid Rotational Acceleration Computed in Eulerian Grids $g({\frac{\varOmega_i}{\tau_r}})_{E}$ With Respect to Net Particle Rotational Acceleration $f(\alpha_i)$ }
			\label{table:4}
		\end{table}
		\\Table \ref{table:4} gives a comparison of the JSD values of  $f(\frac{\varOmega_i}{\tau_r})_e$ and its gaussian fit $g(\frac{\varOmega_i}{\tau_r})_e$ with respect to $f(\alpha_i)$. It is observable that the guassian fit $g(\frac{\varOmega_i}{\tau_r})_e$ gives marginally lesser JSD values along z direction and greater JSD values along x direction with respect to the actual $f(\frac{\varOmega_i}{\tau_r})_e$. Although overall the JSD values remain at a magnitude very near to zero O($10^{-2}-10^{-3}$) in the interval of [0 1]. From the above plots and statistics, it is evident that $f(\alpha_i)$ can be approximated as the gaussian fit of $f(\frac{\varOmega_i}{\tau_r})_e$. Hence the particle accelaration distribution function can be well approximated by a gaussian fit of the distribution function of the fluid phase angular velocity $\varOmega_i$ rescaled by $\tau_r$ computed in fluid-phase Eulerian grids.  
		\subsection{Rotational Acceleration and Velocity Correlations for Smooth Elastic Collisions}
		\label{correlation_decay}
\begin{figure}
			\includegraphics[width=8cm, height= 6cm]{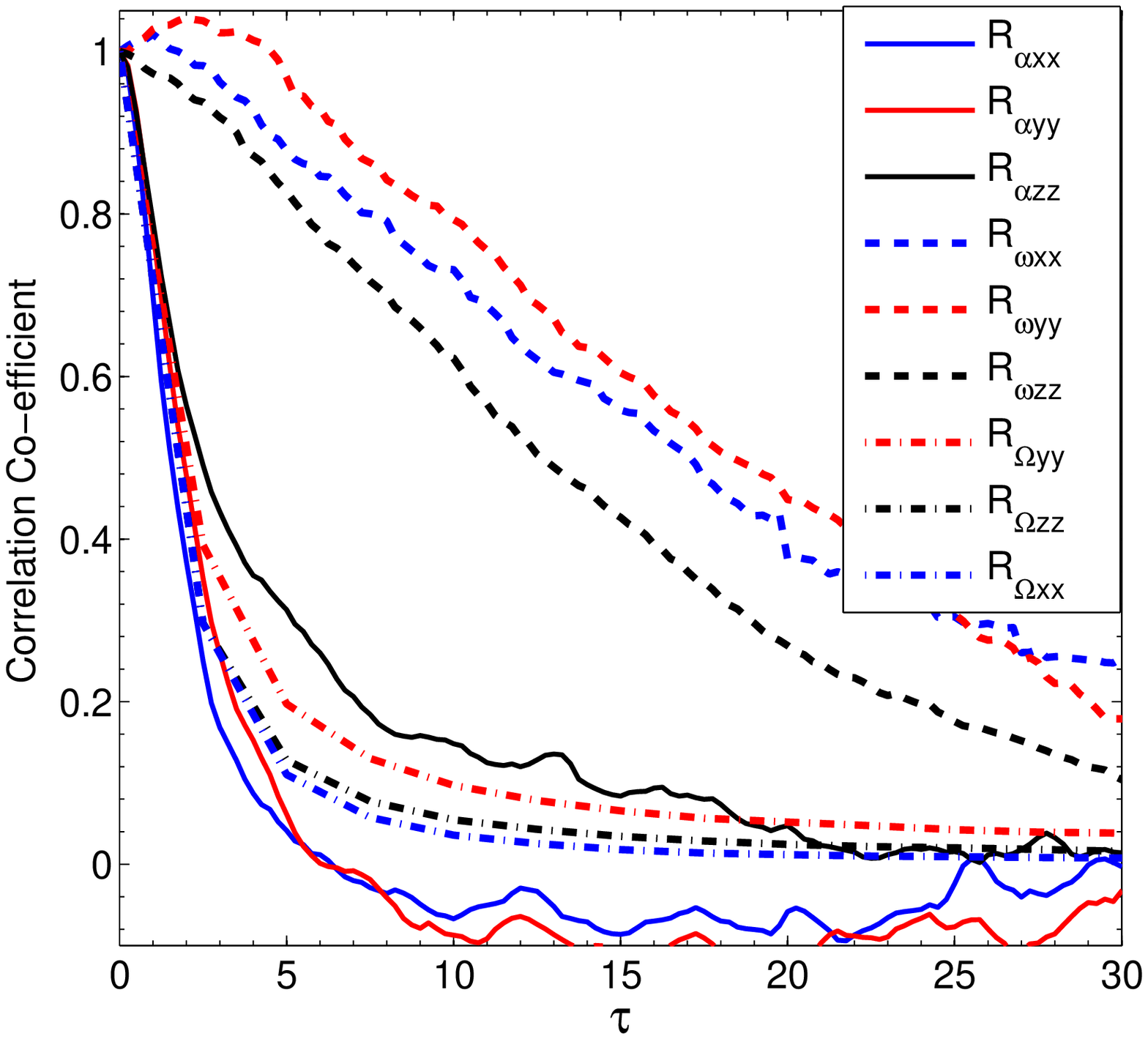}
			\caption*{(a)}
			\includegraphics[width=8cm, height= 6cm]{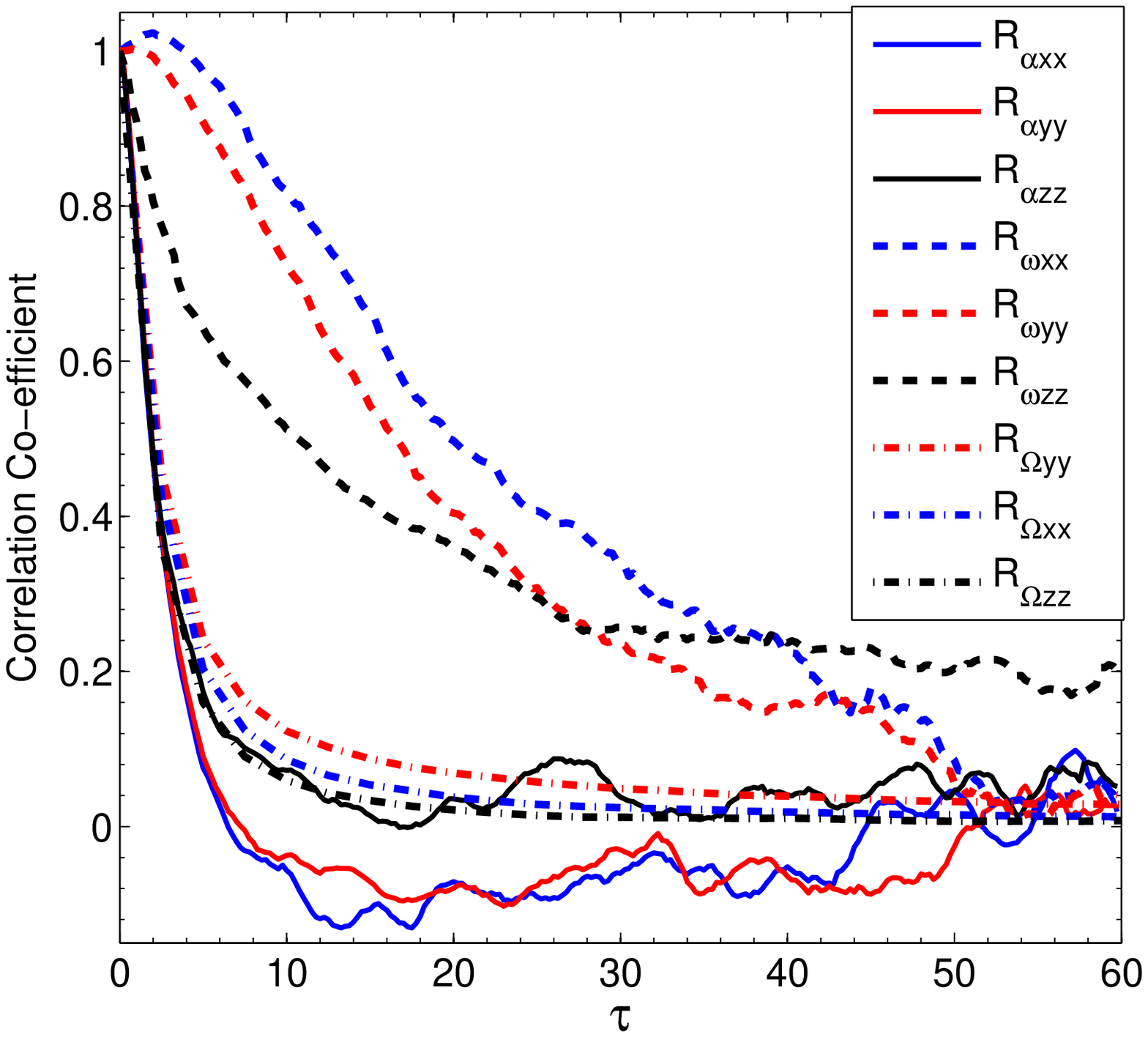}
			\caption*{(b)}
			 \caption{ Decay of Particle Rotational Acceleration Correlation and Rotational Velocity at Two Different Positions of Couette: (a) $y/\delta=0.2\mbox{-}0.4$, (b) $y/\delta=0.6\mbox{-}1.0$}
			 \label{R(alpha)}
		\end{figure}

		The analysis in the previous section suggests that the major constituent of the particle distribution function which originates from the fluid angular velocity can be well aproximated as a Gaussian distribution function in an Eulerian reference frame. To develop a fluid and particle phase decoupled model, it is also important to analyse the decorrelation pattern for particle angular accelaration, particle angular velocity and fluid angular velocity. \\
		Figure \ref{R(alpha)} (a) and (b) show the temporal correlation of particle angular velocity, particle angular acceleration and fluid angular velocity at fluid Eulerian grid. It is to be noted that the last term is related to the decorrelation of particle angular acceleration which originates from fluid angular velocity and can be represented as the ratio of fluid angular velocity to the particle angular relaxation time ($\Omega'_i/\tau_r$). In all the cases we have computed the decorrelation time by integrating the correlation coefficient up to a large time. The results are tabulated in Table\ref{table:5}. It is observed that the particle angular acceleration decays similarly as the angular fluid velocity fluctuation; indicates that the fluctuating torque on the particle can be modeled by the particle angular acceleration fluctuation due to angular velocity fluctuation of the fluid. 
		
			\begin{table}[h!]
			\begin{tabular}{||c | c | c | c ||} 
				\hline
				Direction & Statistics & Time-scale, y=0.6-1.0 & Time-scale, y=0.2-0.4 \\ [0.5ex] 
				\hline\hline
				\multirow{3}{*}{x} & $R_{\alpha ii}$ & 0.2823 & 0.6937 \\ 
				& $R_{\omega ii}$ & 24.9922 & 21.3135\\
				& $R_{\varOmega ii}$  & 6.6262 & 3.3033\\   \hline
				\multirow{3}{*}{y} & $R_{\alpha ii}$ & 0.8921 & 0.1774 \\ 
				& $R_{\omega ii}$ & 21.0774 & 21.573 \\
				& $R_{\varOmega ii}$  & 11.8408 & 9.8230 \\   \hline
				\multirow{3}{*}{z} & $R_{\alpha ii}$ & 3.7767 & 4.9558 \\ 
				& $R_{\omega ii}$ & 20.6116 & 14.603 \\
				& $R_{\varOmega ii}$  & 4.5533 & 4.5524 \\   \hline
			\end{tabular}
			\caption{Time-Scales Computed from the Corresponding Rotational Statistics  }
			\label{table:5}
		\end{table}
		Particle rotational correlations are calculated in particle-Lagrangian reference frame. So while computing the statistics in a sampling bin, particles from the bin do move out to the adjacent zones whereas particles from the neighbouring zones do move into the bin. This cross-stream movement induces noise in the correlation tails figure \ref{R(alpha)} (a) and (b). Consequently integrating over the entire domain of correlation length gives erroneous results. An approximation over the time-length cut-off is necessary. Here the rotational velocity auto-correlations $R_{\omega ii}$ are integrated over $\tau=60$ non-dimensional times and $R_{\alpha ii}$ over $\tau=30$ non-dimensional times. For heavy inertial particles fluid phase rotational accelaration can be approximated as $\alpha_{f i}=\varOmega_i/\tau_r$ and hence the properties of covariance yields $R_{\alpha_f ii}=R_{\varOmega ii}$.     
		From table \ref{table:5} the following are observed:
		\begin{itemize}
			\item Time-scale of particle rotational velocity de-correlation $T_{\omega i}$is $O(10^1-10^2)$ magnitude higher than Time-scale of particle rotational acceleration de-correlation $T_{\alpha i}$. 
			\item Fluid Eulerian rotational acceleration decorrelation time $T_{\varOmega}$ is $O(10^0-10^1)$ in magnitude with respect to the particle rotational accelaration de-correlation time-scale $T_{\alpha i}$, but quite less than particle rotational velocity decorrelation time-scale $T_{\omega i}$.   
		\end{itemize}
		\section{Conclusion}
		In the present study we have analysed the translational and angular velocity and acceleration statistics of the particle phase in detail using Direct Numerical Simulation (DNS). The particles are considered to be with high Stokes number. The particle volume fraction is low such that presence of particle does not significantly modify the turbulence. Therefore one-way Direct Numerical Simulation has been performed for the present study.
		The inter-particle and wall-particle interactions were considered to be perfectly smooth and elastic. The fluctuating velocity, vorticity field, and mean velocity gradient of the turbulent fluid act as the primary source of particle velocity and angular velocity fluctuations rather than collisional transport.
		The important conclusions of the detail analysis of the velocity and acceleration distribution functions are as follows. The streamwise linear velocity fluctuation can be approximated as Gaussian distribution except very near the wall. Distribution in wall-normal and spanwise component of velocity fluctuations show exponentially decaying long tail, which originates mainly due to collisional exchange of momentum from the streamwise component. At the center of the Couette all the components of linear acceleration show a Gaussian distribution. The angular acceleration distributions are Gaussian upto two decades at the center of the Couette but asymmetry is observed near the wall. It is also observed that particle roughness plays a very important role on deciding the shape of the distribution function. From the particle equation of rotational motion, we arrive at the expression where the fluctuating angular acceleration $\alpha_i'$ of the particle is expressed as the ratio of a linear combination of fluctuating rotational velocities of particle ($\omega_i'$) and fluid angular velocity ($\varOmega_i'$) to the particle rotational relaxation time $\tau_r$ (for convenience all the primed superscript are dropped in the notations for fluctuations).
		 The analysis was done using comparative study of the  particle rotational acceleration distribution $f(\alpha_i')$,  with the distributions of particle acceleration component arising from (i) the fluctuating fluid angular velocity in the particle-Largrangian frame $f((\varOmega'_i/\tau_r)_{pl})$, (ii) fluctuating particle angular velocity $f(\omega'_i/\tau_r)$, and (iii) the fluid rotational acceleration $f((\varOmega'_i/\tau_r)_{e})$ computed in the fluid Eulerian grids.  \textbf{Jensen-Shannon Divergence}, which is a \textbf{relative Shannon-entropy based statistic}, was used to quantify the similarity of (i),(ii) and (iii) with $f(\alpha_i')$. The rigourous statistical analysis led to the conclusion that $f(\alpha_i')$ can be well approximated by the gaussian fit of the distribution function of the ratio of fluid phase angular velocity $\varOmega'_i$ to $\tau_r$ and computed in fluid Eulerian grids.
		From the sub-section \ref{correlation_decay} it can be substantiated that the time-scale of particle angular acceleration fluctuation and hence the net fluctuating torque is very small than that of particle rotational velocity fluctuation. Even if the particle rotational acceleration fluctuation is approximated with the ratio of Eulerian fluid phase rotational velocity to $\tau_r$, the consequent time-scale remains substantially low. Under these circumstances the effect of fluid fluctuating torque on net particle rotational acceleration, in between two successive collisions, can be modelled as Gaussian Random White Noise.
		In section \ref{section_roughness} it is argued that the nature of distribution function changes in presence of roughness factor $\beta$ due to collisional re-distribution of mean translational and rotational kinetic energy along with the increase of fluctuations (mean-square values). As the change in the nature of the distribution functions is purely a collisional effect, the approximation of particle rotational acceleration, as mentioned above, in between two successive collisions, can be assumed to be unaffected by rough collisions. Hence, it is noteworthy to mention that the statistics reported in section \ref{section_rot_acc} and sub-section \ref{correlation_decay} are not required to be analysed in presence of rough collision before arriving at the applicability of the above Gaussian White-Noise approximation. The detail methodology of Langevin type modeling of the fluctuating torque due to fluid vorticity fluctuation and the resulting dynamics of the particle phase will be addressed in our future communication.

			\begin{acknowledgments}
				We wish to acknowledge the financial support of SERB, DST, Government of INDIA.
			\end{acknowledgments}
		
		\section*{Data Availability Statement}
		The data that support the findings of this study are available
		from the corresponding author upon reasonable request.
		
		\bibliography{arxiv_manuscript.bib}
		
	\end{document}